\documentclass{aa}
\usepackage{graphicx,caption} 
\usepackage{txfonts}
\usepackage{placeins}

\makeatletter
\renewcommand*\aa@pageof{, page \thepage{} of \pageref*{LastPage}}
\usepackage{hyperref}
\hypersetup{
 colorlinks=true,
 allcolors=[rgb]{0,0,0.8},
 pdftitle={The SRG/eROSITA all-sky survey: Cosmology constraints from cluster abundances in the western Galactic hemisphere},
 pdfauthor={Ghirardini et al.},
 }

\newcommand{\erosita}{eROSITA\xspace}
\newcommand{\planck}{\emph{Planck}\xspace}

\newcommand{\chandra}{\emph{Chandra}\xspace}

\newcommand{\eromapper}{\texttt{eROMaPPer}\xspace}
\newcommand{\esass}{\texttt{eSASS}\xspace}
\newcommand{\ext}{\texttt{EXTENT}}
\newcommand{\erass}[1][1]{eRASS1\xspace}
\newcommand{\extlike}{$\mathcal{L}_{ext}$\xspace}
\newcommand{\detlike}{$\mathcal{L}_{det}$\xspace}
\newcommand{\mbproj}{MBProj2D\xspace}
\newcommand{\nh}{\ensuremath{n_{\rm H}}\xspace}
\newcommand{\lcdm}{\ensuremath{\Lambda\mathrm{CDM}}\xspace}

\usepackage{natbib,twoopt}

\makeatletter
\newcommandtwoopt{\citeads}[3][][]{\href{http://adsabs.harvard.edu/abs/#3}%
 {\def\hyper@linkstart##1##2{}%
 \let\hyper@linkend\@empty\citealp[#1][#2]{#3}}}
 \newcommandtwoopt{\citepads}[3][][]{\href{http://adsabs.harvard.edu/abs/#3}%
 {\def\hyper@linkstart##1##2{}%
 \let\hyper@linkend\@empty\citep[#1][#2]{#3}}}
 \newcommandtwoopt{\citetads}[3][][]{\href{http://adsabs.harvard.edu/abs/#3}%
 {\def\hyper@linkstart##1##2{}%
 \let\hyper@linkend\@empty\citet[#1][#2]{#3}}}
 \newcommandtwoopt{\citeyearads}[3][][]%
 {\href{http://adsabs.harvard.edu/abs/#3}
 {\def\hyper@linkstart##1##2{}%
 \let\hyper@linkend\@empty\citeyear[#1][#2]{#3}}}
\makeatother

\begin{document}

\title{The SRG/eROSITA all-sky survey}
\subtitle{Cosmology constraints from cluster abundances in the western Galactic hemisphere}
\author{
V.~Ghirardini\inst{1}, 
E.~Bulbul\inst{1}, 
E.~Artis\inst{1}, 
N.~Clerc\inst{2}, 
C.~Garrel\inst{1}, 
S.~Grandis\inst{3, 5}, 
M.~Kluge\inst{1}, 
A.~Liu\inst{1}, 
Y.~E.~Bahar\inst{1}, 
F.~Balzer\inst{1}, 
I.~Chiu\inst{4}, 
J.~Comparat\inst{1}, 
D.~Gruen\inst{5},
F.~Kleinebreil\inst{3, 8}, 
S.~Krippendorf\inst{5, 6}, 
A.~Merloni\inst{1},
K.~Nandra\inst{1},
N.~Okabe\inst{7}, 
F.~Pacaud\inst{8}, 
P.~Predehl\inst{1},
M.~E.~Ramos-Ceja\inst{1}, 
T.~H.~Reiprich\inst{8}, 
J.~S.~Sanders\inst{1}, 
T.~Schrabback\inst{3, 8}, 
R.~Seppi\inst{1}, 
S.~Zelmer\inst{1},
X.~Zhang\inst{1},
W.~Bornemann\inst{1}, 
H.~Brunner\inst{1}, 
V.~Burwitz\inst{1}, 
D.~Coutinho\inst{1}, 
K.~Dennerl\inst{1}, 
M.~Freyberg\inst{1}, 
S.~Friedrich\inst{1}, 
R.~Gaida\inst{1},
A.~Gueguen\inst{1}, 
F.~Haberl\inst{1}, 
W.~Kink\inst{1}, 
G.~Lamer\inst{9}, 
X.~Li\inst{11}, 
T.~Liu\inst{1}, 
C.~Maitra\inst{1},
N.~Meidinger\inst{1}, 
S.~Mueller\inst{1}, 
H.~Miyatake\inst{13, 14, 15}, 
S.~Miyazaki\inst{12}, 
J.~Robrade\inst{10}, 
A.~Schwope\inst{9}, 
I.~Stewart\inst{1} 
}

\institute{
Max Planck Institute for Extraterrestrial Physics, Giessenbachstrasse 1, 85748 Garching, Germany
\and
IRAP, Université de Toulouse, CNRS, UPS, CNES, F-31028 Toulouse, France
\and
Universit\"at Innsbruck, Institut f\"ur Astro- und Teilchenphysik, Technikerstr. 25/8, 6020 Innsbruck, Austria
\and
Department of Physics, National Cheng Kung University, 70101 Tainan, Taiwan
\and
Universit\"ats-Sternwarte, Faculty of Physics, LMU Munich, Scheinerstr. 1, 81679 M\"unchen, Germany
\and
Arnold Sommerfeld Center for Theoretical Physics, LMU Munich, Theresienstr. 37, 80333 M\"unchen, Germany
\and
Department of Physical Science, Hiroshima University, 1-3-1 Kagamiyama,
Higashi-Hiroshima,Hiroshima 739-8526, Japan
\and
Argelander-Institut f\"ur Astronomie (AIfA), Universit\"at Bonn, Auf dem H\"ugel 71, 53121 Bonn, Germany
\and
Leibniz-Institut f\"ur Astrophysik Potsdam (AIP), An der Sternwarte 16, 14482 Potsdam, Germany
\and
Hamburger Sternwarte, Gojenbergsweg 112, 21029 Hamburg, Germany
\and 
McWilliams Center for Cosmology, Department of Physics, Carnegie Mellon University, Pittsburgh, PA 15213, USA
\and
Subaru Telescope, National Astronomical Observatory of Japan, 650 N Aohoku Place Hilo HI 96720 USA
\and
Kobayashi-Maskawa Institute for the Origin of Particles and the Universe (KMI), Nagoya University, Nagoya, 464-8602, Japan
\and
Institute for Advanced Research, Nagoya University, Nagoya 464-8601, Japan
\and
Kavli Institute for the Physics and Mathematics of the Universe (WPI), The University of Tokyo Institutes for Advanced Study (UTIAS), The University of Tokyo, Chiba 277-8583, Japan
}
\date{\today}

\titlerunning{Cosmology constraints from \erass cluster abundances}
\authorrunning{Ghirardini et al.}

\abstract{The evolution of the cluster mass function traces the growth of linear density perturbations, providing valuable insights into the growth of structures, the nature of dark matter, and the cosmological parameters governing the Universe. The primary science goal of \erosita, on board the \textit{Spectrum Roentgen Gamma (SRG)} mission, is to constrain cosmology through the evolution of the cluster mass function. In this paper, we present a set of cosmological constraints obtained from 5259 clusters of galaxies detected over an area of 12791~deg$^{2}$ in the western Galactic hemisphere of \erosita's first All-Sky Survey (\erass). 
The common footprint region (4968~deg$^{2}$) between the \erosita Survey and Dark Energy Survey (DES), the Kilo-Degree Survey (KiDS), and the Hyper Supreme Camera (HSC) survey is used for calibration of the scaling between X-ray count rate of the clusters and their total mass through measurements of their weak gravitational lensing signal. The 
\erass cluster abundances constrain the \lcdm parameters, namely, the energy density of the total matter to $\Omega_{\mathrm{m}}=0.29^{+0.01}_{-0.02}$ and the normalization of the density fluctuations to $\sigma_{8}=0.88\pm0.02,$ and their combination yields $S_{8}=\sigma_8 ({\Omega_{\mathrm{m}}} / {0.3})^{0.5}=0.86\pm0.01$. These results are  consistent and achieve at a similar precision with state-of-the-art cosmic microwave background (CMB) measurements. Furthermore, the
\erass cosmological experiment places a most stringent upper limit on the summed masses of left-handed light neutrinos to $\sum m_{\nu}< 0.43\mathrm{~eV}$ (95\% confidence interval) from cluster number counts alone. By combining \erass cluster abundance measurements with CMB- and ground-based neutrino oscillation experiments, we measured the summed neutrino masses to be $\sum m_\nu = 0.09_{-0.02}^{+0.04}\mathrm{~eV}$ or $\sum m_\nu = 0.12_{-0.02}^{+0.03}\mathrm{~eV,}$ assuming a normal or inverted mass hierarchy scenario for neutrino eigenstates. The \erass cluster abundances significantly improve the constraints on the dark energy equation of state parameter to $w=-1.12 \pm 0.12$. When $\sum m_{\nu}$ and $w$ are left free, we find consistent results with the concordance \lcdm cosmology. Our results from the first All-Sky Survey improve the cosmological constraints by over a factor of 5 to 9 over the previous cluster surveys, establishing cluster abundance measurements for precision cosmology and setting the stage for deeper \erosita All-Sky Surveys, as well as for future cluster abundance experiments. 
}


\maketitle

\section{Introduction}

The so-called \lcdm concordance cosmological model is currently the most widely accepted theory on the formation and evolution of our Universe, whereby cold dark matter (CDM) and dark energy parameterized by a cosmological constant ($\Lambda$) are the essential ingredients. It is a relatively simple model, where just a handful of cosmological parameters allow for accurate predictions for several observations, for instance: the cosmic microwave background \citep[CMB][]{Smoot1992, Fixsen1996},  large-scale structure \citep{Peebles1980},  abundance of light elements \citep{Walker1991, Cooke2018}, and  accelerated expansion of the Universe \citep{Riess1998, Perlmutter1998}.
In recent years, state-of-the-art experiments have enabled precise measurement of cosmological parameters.
For instance, \planck satellite allowed for detailed constraints on the density field at the CMB epoch \citep{Planck2020}, while deep wide-area surveys, such as
the Dark Energy Survey (DES),  Hyper Supreme-Cam Subaru Strategic Program (HSC Y3) Survey, and  Kilo Degree Survey (KiDS), have 
allowed for constraints to be placed on cosmological parameters combination as the parameter $S_8$ \citep{Abbott2021, Asgari2021, Amon2022, Secco2022, vandenBusch2022, Li2023, Dalal2023, DESKiDS2023}, and the widening of the catalog of known SNe~Ia and Cepheid variables allowed accurate calibration of distance ladder method and precise measurement of $H_0$ \citep{Riess2022}. However, tensions between CMB early-time measurements and large-scale structure late-time experiments have emerged: the so-called $H_0$-tension \citep[e.g.][]{DiValentino2021} and the so-called $S_8$-tension \citep[e.g.][]{Heymans2021}. These tensions indicate that the currently widely preferred cosmological model, the so-called \lcdm, might be revised if new generation experiments increase the statistical significance of these cosmic discordances.
With the discovery of neutrino oscillations between different flavors, electron, muon, and tau, it has become clear that neutrino eigenstates must have non-zero mass \citep{Fukuda1998, Ahmad2002}. This leads to several cosmological implications for CMB and the large-scale structure experiments \citep{Bashinsky2004, Lesgourgues2013, Vagnozzi2017}.
Despite their small size and mass, neutrinos influence large-scale structures' formation by hindering small-scale halos' formation and leaving imprints on the gravitational collapse \citep{Costanzi2013}.

Clusters of galaxies formed from the highest density fluctuations in the early Universe and represent the largest virialized objects that have undergone a gravitational collapse. Their abundance through the halo mass function and spatial distribution traces the growth of linear density perturbations and, therefore, can be utilized as a probe of cosmology. The cluster abundance, $dn / dM$, is particularly sensitive to two cosmological parameters: 
the amplitude of the mass fluctuations on scales of $8 h^{-1} \,$Mpc, namely $\sigma_8$\footnote{See Appendix~\ref{app:sigma8} for detailed definition of $\sigma_8$.},
and the present-day matter density parameter, $\Omega_\mathrm{m}$\footnote{{$\Omega_{\mathrm{m}} = \rho_{\mathrm{m}} / \rho_c$, where $\rho_{\mathrm{m}}$ is the matter density and $\rho_c$ is the critical density of the Universe.}}.
It is also dependent on the dark energy equation of state, $w$, and its evolution because of the growth suppression factor $g(z)$ and the Universe's expansion history \citep[see][for a recent review]{ClercFinoguenov2023}. 
Compared to other probes, such as baryon acoustic oscillations \citep[BAO, e.g.][]{Alam2017}, 
type Ia supernovae
\citep[SNe~Ia, e.g.][]{Scolnic2018, Brout2022}, cosmic microwave background \citep[CMB, e.g.][]{Komatsu2011, Planck2020}, and galaxy and cosmic shear auto- and cross-correlations \citep{Abbott2021, DESKiDS2023}, cluster number counts do not display strong degeneracies among the primary late time cosmological parameters discussed above, while retaining a comparable precision. Furthermore, the sum of the left-handed neutrino masses, $\sum m_\nu$, can also be constrained via cluster abundances, particularly in combination with external probes, as the cosmic neutrinos have a non-negligible effect on the growth of matter perturbations during the matter and dark energy-dominated era \citep{Vikhlinin2009}. In fact, the number of low-mass halos in cluster abundance measurements can shed light on the summed masses of three left-handed neutrino species.

Multiwavelength surveys of clusters of galaxies have been widely utilized to place constraints on the cosmological parameters of the late-time Universe, with sample sizes ranging from tens to thousands of clusters. Among these are the surveys in the X-ray band \citep{Vikhlinin2009, Mantz2015, Garrel2022, Chiu2023a}, in the optical band \citep{IderChitham2020, Abbott2020, Costanzi2021, Lesci2022}, and mm-wave band through the inverse Compton scattering of CMB photons, called Sunyaev-Zel'dovich (hereafter SZ) effect \citep{Planck2014, Planck2016, deHaan2016, Bocquet2019}. The challenge for using cluster abundances in cosmological experiments lies in the reconstruction of the cluster mass distribution at a population level, as individual masses are not directly observable through X-ray, optical, and SZ observations. 
For cluster surveys, many observable mass proxies can be linked to cluster masses, such as
richness, for instance, the number of photometrically assigned member galaxies, the temperature, X-ray luminosity, or gas mass of the intracluster medium (ICM), the X-ray luminosity, the SZ decrement, the tangential reduced shear imprinted by gravitational lensing on the shapes of background galaxies.
These observables can be used to calibrate the observable-mass distributions and thereby reconstruct the cluster mass function.
The observables in the optical, such as richness, suffer from projection effects and contamination, limiting their use in cluster abundance measurements \citep{Abbott2020, IderChitham2020, Costanzi2021}. 
For cluster samples selected in X-rays, the count rate in the soft X-ray band, proportional to the integrated surface brightness, is a convenient observable that correlates tightly with the selection process and scales with halo mass and redshift. Another advantage of the X-ray detection process is that count rate measurements do not suffer from significant projection effects \citep{Garrel2022}. Additionally, the scatter in count rate at a given mass and redshift is dominated by the X-ray emission mechanism, namely, thermal Bremsstrahlung; thus, it is not more sensitive to the underlying cosmology and environment than the scatter in richness.

A well-calibrated scaling relation between the total mass and the observed signal is critical for capitalizing on the statistical potential of large samples of clusters of galaxies. Historically, in their mass calibration, cluster counts experiments have employed X-ray cluster mass measurements relying upon the hydrostatic equilibrium, which assumes that the gravitational potential provided by the dark matter halo is balanced by thermal pressure \citep{Vikhlinin2009, Mantz2010, Planck2016}. However, recent numerical simulations indicate that the contribution of non-thermal pressure due to bulk motions and turbulence can lead to significant biases on the cluster masses measured with these traditional techniques \citep{Lau2009, Rasia2012, Scheck2023}. The use of biased X-ray mass measurements is also reflected in the cosmological parameters obtained from cluster counts \citep{PlanckCollaboration2013}. In the last decade, a significant improvement over this approach has been introduced by integrating the gravitational lensing signal in the cluster mass calibration process of the cosmological studies \citep{Mantz2015, Bocquet2019, Garrel2022, Chiu2023a}. The weak-lensing (WL) effect directly probes the cluster's gravitational potential, thus allowing the masses to be accurately measured without any a priori assumption about their hydrostatic equilibrium properties. It is worth noting that this method is not entirely bias-free; the correlated large-scale structure along the line-of-sight, triaxiality, and assumed underlying dark matter distribution, baryonic feedback, and miscentering of the WL signal could introduce additional bias and scatter on the mass scaling relations \citep{Hoekstra2003, Meneghetti2010, Becker2011, Bahe2012}. However, compared to X-ray measurements, the accuracy of WL measurements is inherently less impacted by hydrodynamical modeling as the majority of the mass is collisionless.
Most importantly, since the WL mass measurements are calibrated with numerical simulations, the bias and systematic uncertainties can be reliably and quantitatively estimated and integrated into the final cluster mass calibration effort \citep{Schrabback2018, Dietrich2019, Grandis2021a, Sommer2022, Zohren2022, Grandis2024}. 

Another critical aspect in applying wide-area survey cluster catalogs in statistical analyses is understanding and adequately accounting for the underlying selection function. Surveys in the X-ray band are prone to significant selection effects, such as Eddington and Malmquist biases \citep{Eddington1913, Malmquist1922}. Additionally, the detected cluster sample often undergoes an optical confirmation process for redshift measurements; optical selection effects may play an essential role in the overall scientific results drawn from the observed cluster population. {A good knowledge of the selection function, namely, the interface between an observed sample and the parent cluster population, is a prerequisite for statistical studies of cluster samples, such as the constraints on cosmology. Comparisons between samples detected at different wavelengths have shed light on our understanding of some of the selection biases \citep{Nurgaliev2017, Rossetti2017, Willis2021, Ramos-Ceja2022}. Thus, such comparisons have been successfully applied in the recent applications of the X-ray selected samples of clusters of galaxies in cosmology experiments \citep[e.g.,][]{Garrel2022}.

\begin{figure*}
\includegraphics[width=\textwidth, clip]{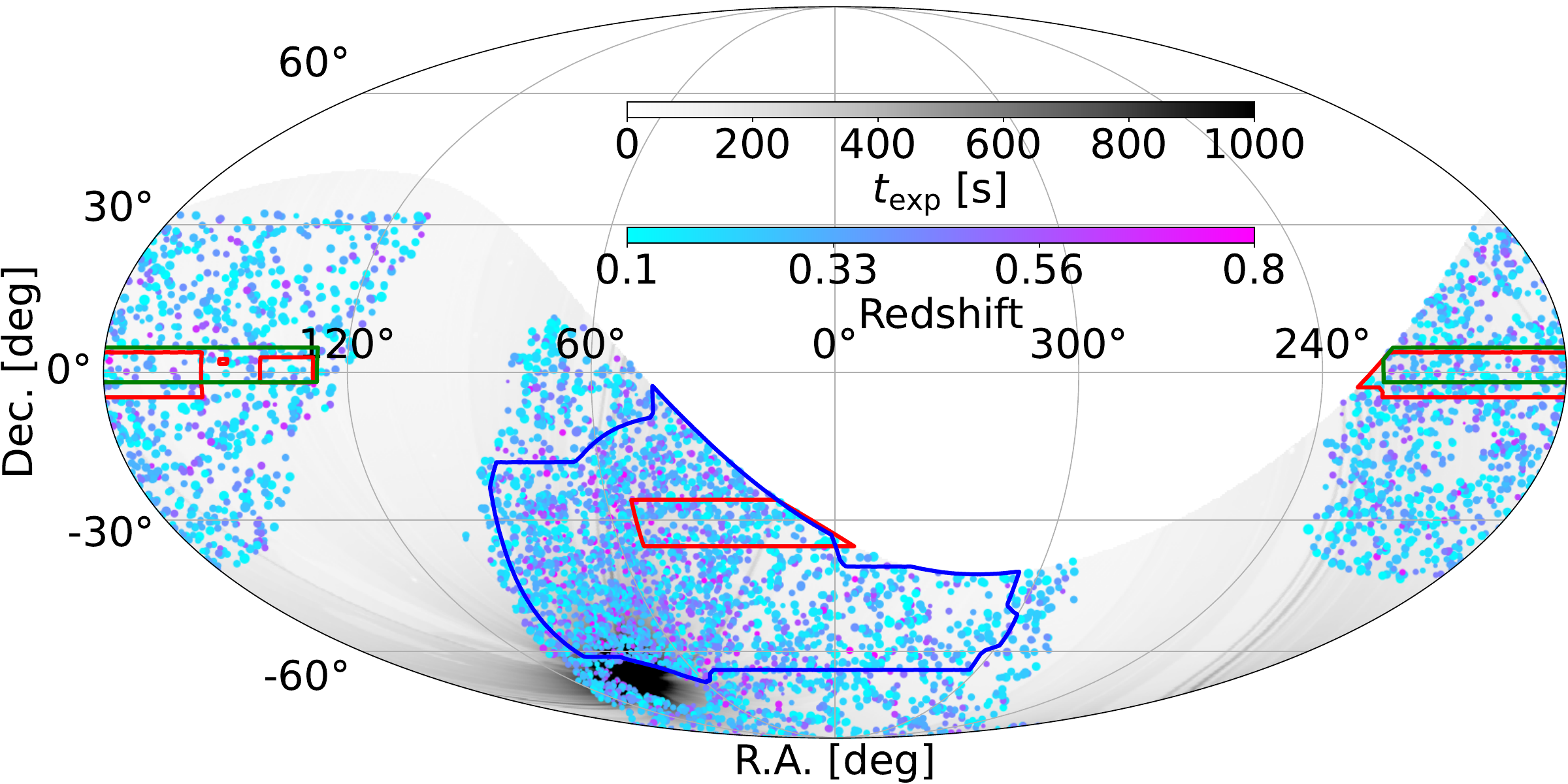} 
\caption{Sky distribution of the 5259 confirmed clusters of galaxies in the cosmology subsample, detected in the western Galatic hemisphere of the \erass All-Sky Survey in equatorial coordinates. The common footprint between the Legacy Survey-South and \erass reaches up to 12791~deg$^2$. The sample covers a redshift range of 0.1 to 0.8. The redshifts of the sources are color-coded, and the sizes of the marks are 
proportional to the observed \erosita\ X-ray count rate. The overlaid image in the background shows the exposure map of the \erass Survey in the 0.2-2.3 keV energy band. The regions plotted in solid lines show the footprints of optical surveys employed in the week lensing mass calibration DES in blue, KiDS in red, and HSC~Y3 in green, covering 4968~deg$^2$ in total.} 
\label{fig:skydist}
\end{figure*}

This work presents the constraints on cosmological parameters from the clusters of galaxies detected in the western Galatic hemisphere in the first \erosita\ All-Sky Survey.
As one of the two X-ray telescopes on board the Spectrum Roentgen Gamma Mission ({\it SRG}), \erosita,  was launched on July 13, 2019. The first All-Sky Survey (\erass) was successfully executed between December 2019 and June 2020 \citep{Predehl2021, Sunyaev2021} and has resulted in the detection of nearly 930~thousand X-ray sources in the western Galactic hemisphere (179.9442~deg~$< l <$~359.9442~deg), where the data and publication rights belong to the \erosita\ German Consortium \citep{Predehl2021, Merloni2024}. In this first All-Sky Survey, we detect a total of 12,247 optically confirmed galaxy groups and clusters spanning the redshift range $0.003 < z < 1.32$ with a sample purity level of 86\% \citep{Bulbul2024, Kluge2024}. In this work, we use a subsample of this primary \erass cluster catalog, namely, the cosmology subsample constructed with a strict X-ray selection criterion to increase the purity of the sample to about 94\%, as described in detail in \citet{Bulbul2024}. In the cosmology subsample, we optically confirm 5,259 securely detected clusters in the 12,791~deg$^{2}$ common footprint of \erass and the DESI Legacy Survey DR10-South in the photometric redshift range of $0.1<z_{\lambda}<0.8$. This catalog represents the largest ICM-selected cluster sample utilized for cosmological studies. 

Besides the statistical power ensured by the extensive \erass cluster catalog, our analysis offers significant improvements over previous works. 
We employ the observed X-ray count rate, which has the advantage of being coupled with the detection and selection process in the mass calibration phase. The selection effects are modeled using the end-to-end simulations of the \erosita sky generated based on realistic models of the sources and their distribution in the sky \citep{Comparat2020, Seppi2022}. We incorporate the selection process into a forward model to represent the entire population of detected clusters in a Bayesian framework. Finally, the residual 6\% contamination in our sample is statistically accounted for using a mixture model \citep[see][]{Bulbul2024, Kluge2024}. For the mass calibration, we utilize a scaling relation between X-ray count rate and cluster mass obtained from the three weak lensing surveys:  Dark Energy Survey Year Three (DES~Y3), the Hyper Supreme-Cam Subaru Strategic Program Year Three (HSC Y3) Survey, and the Kilo Degree Survey (KiDS) for a total of 2588 clusters in an area of $4968$~deg$^{2}$ overlapping with \erass \citep{Chiu2023a, Kleinebreil2024}. 
The biases due to miscentering, baryonic effects, and correlated large-scale structure along the line of sight are fully accounted for in our weak lensing analysis \citep[see ][ for detailed analyses and comparisons]{Grandis2024, Kleinebreil2024}.

This paper is organized as follows. In Sect.~\ref{sec:data}, we summarize the datasets in X-ray (\erosita), optical (DESI Legacy Survey), and weak lensing (DES, KiDS, and HSC) utilized in our analysis. The selection function and Bayesian likelihood modeling method are presented in Sect.~\ref{sec:selection} and Sect.~\ref{sec:likelihood}. The validation process of the cosmology pipeline and the blinding strategy are described in Sect.~\ref{sec:validation_and_blinding}. In Sects.~\ref{sec:scaling_relations} and ~\ref{sec:results}, we present our results on scaling relations and various models for cosmology. We discuss the systematics in our results and provide an outlook for future \erosita\ surveys in Sect.~\ref{sec:discussion}. We provide our summary in Sect.~\ref{sec:summary_conclusions}. Our convention is as follows: the overdensity radius of $R_{500}$ is defined as the radius within which the mean cluster density is 500 times the cosmic critical density at the cluster's redshift. The mass $M_{500}$ is the cluster mass within a sphere of $R_{500}$. With $\log$, we  are referring to the natural logarithm, and when using base 10, we explicitly indicate it with the subscript $\log_{10}$. 
Throughout this paper, we use the terms consistency, tension, and discrepancy multiple times: we define here that by "consistent" we mean less than 3$\sigma$ difference, by "in tension" we mean between 3$\sigma$ and 5$\sigma$ in difference, and by "discrepant," we mean above 5$\sigma$ in difference.

\section{Galaxy cluster survey data }
\label{sec:data}

\subsection{\erass Cluster Catalog}
\label{sec:catalog}

The cluster catalog utilized in this work is based on the soft X-ray band (0.2$-$2.3~keV) catalog of the first \erosita All-Sky Survey in the western Galactic hemisphere presented in \citet{Merloni2024}. The details of the detection and cleaning processes and compilation of the cluster catalog are described in detail in \citet{Bulbul2024}. We restrict ourselves to the cosmology subsample in this work, a subsample of the primary galaxy groups and clusters catalog. To compile the cosmology subsample, \citet{Bulbul2024} applied a conservative extent likelihood cut of \extlike$>6$ to the primary sample, resulting in 11,141 cluster candidates detected as extended in the clean \erass cluster catalog. 

The confirmation, identification, and redshift measurements of the X-ray-selected cluster candidates are typically performed using the publicly available DESI Legacy Survey Data Release 10 data (LS DR10). LS DR10 is a compilation of three surveys that cover most of the extragalactic sky \citep{Dey2019, Sevilla-Noarbe2021}. The cosmology subsample is compiled using only the southern area below Dec.$\lesssim32.5\degr$ (LS DR10-South hereafter) to maximize the homogeneity, as the observations have been taken with the same telescope and set of filters. This ensures that the subset of the \erass clusters has homogeneous systematics on the photometric redshifts ($\hat{z}_\lambda$) and richness ($\hat{\lambda}$) measurements. The common footprint between the \erosita survey in the western Galactic hemisphere and the southern LS DR10-South footprint is 12,791~deg$^2$, and contains 8,129 cluster candidates \citep{Kluge2024}. 

We performed the confirmation and identification with the red sequence-based cluster finder algorithm \eromapper\ \citep{IderChitham2020, Kluge2024}. For the catalog used in this work, we combined the $g$, $r$, and $z$ filter bands to ensure the homogeneity of the richness, redshifts, and contamination. 
In the cosmology catalog used in this work, we adopted photometric redshifts for the detected cluster to ensure homogeneity and uniform treatment for all the optically measured properties \citep[see][for further details]{Bulbul2024}.
Of the 7,077 candidate clusters in the LS DR10-South footprint, we were able to identify the optical counterparts of 6,562 securely detected clusters.

We note that the detection of faint galaxies below the limiting luminosity $L>0.2$~$L_{*}$ is highly uncertain, and the measured richness artificially increases because of Eddington bias. To obtain reliable photometric redshift measurements with well-controlled biases and uncertainties, we applied a further flag of {\texttt{IN$\_$ZVLIM==True}} \citep[see][ for further details]{Kluge2024}. Additionally, we limited our redshift range to $0.1\,<\,z\,<\,0.8$, where the photometric redshifts are the most reliable \citep[see][for details]{Kluge2024}. 
As a final cut, we removed all the clusters with measured richness smaller than 3 to decrease the contamination as these would be cases with no optical cluster counterpart \citep[see][Sect. 3.2.3]{Kluge2024}. This is because they would have a significantly larger chance of being contaminants and, thus, be likely to have the wrong photometric redshifts assigned.
The application of richness measurements below 20 has been shown to be problematic in cluster number counts experiments with optical selection only \citep{Abbott2020}. However in our case, richness is not used as a direct proxy for cosmology; rather, it is mostly used as a clean-up quantity that is very helpful in statistically discriminating between clusters and contaminations in our sample (see Sect.~\ref{sec:mixture}). Furthermore, richness for an ICM-selected sample (as in our case) is well behaved with small scatter \citep{Saro2015, Bleem2020, Grandis2021b, Giles2022}. The final cosmology subsample comprises 5,259 securely confirmed galaxy clusters in the 12,791~deg$^{2}$ LS DR10-South area. The projected spatial distribution of the clusters in this sample is shown in Fig.~\ref{fig:skydist}, overlaid on the exposure map of the \erass survey. The median redshift in the catalog is 0.29. The redshift and richness histograms of all confirmed clusters are presented in the leftmost panel of Fig.~\ref{fig:hist}. The mixture model, extensively described in Sect.~\ref{sec:mixture}, implies that the purity of this sample is at the level of 94\%, ideal for cosmological studies with cluster counts \citep[see][for further details]{Bulbul2024, Kluge2024}.

\subsection{Selection of the X-ray observable}

The choice of an adequate X-ray observable employed in the internal cluster mass calibration process is vital in cluster cosmology studies. Ideally, the X-ray observable employed in the selection process should be adequately included in the selection model and should show a correlation with cluster mass. Additionally, if the X-ray observable has a negligible variation over the sky and a minimal dependence on cosmology  (which are not strictly necessary), the computation cost of the forward hierarchical model significantly decreases. In this regard, we examine extensively several observables, for instance: the extent likelihood, \extlike, detection likelihood, \detlike, and count-rate. In addition, we included commonly used observables in the literature, such as luminosity, gas mass, and the mass proxy $Y_{X}$ in the observer's frame, along with their correlation with mass using the \erosita's digital twin \citep{Seppi2022}. Among these observables, we find that the X-ray count rate satisfies all the conditions mentioned above, especially with the clear advantage that it has minimal dependence on the underlying cosmological model. Furthermore, the sky variation can be further decreased, and, similarly to how we have divided the counts by exposure map to correct for exposure time variation over the sky, we correct the count rate for line of sight absorption due to hydrogen in the so-called \nh-corrected count rate.

The raw count rate that is the output of the detection chain presented in \citet{Merloni2024} is only a rough estimate of the true clusters' count rate and shows a large scatter in correlations with cluster mass. We improve these pipeline measurements through a rigorous X-ray analysis correcting for the Galactic absorption and a more accurate ICM and background modeling to reduce the scatter with the total mass. The details of the X-ray processing are provided in \citet{Bulbul2024}; we briefly summarize the process here. X-ray image fitting is performed using the MultiBand Projector in 2D (\mbproj) tool \citep{Sanders2018}, which forward-models background-included X-ray images of galaxy clusters to fit cluster and background emission in a total of seven bands between 0.3~keV to 7~keV. We obtain the corrected count rate from the best-fit cluster physical model, including a correction for Galactic hydrogen absorption at photometric redshifts $z_{\lambda}$. The measured count rate in the 0.2$-$2.3~keV band is used in the further weak lensing mass calibration process and is also shown in the rightmost
panel  of Fig.~\ref{fig:hist} and has been released in the cluster catalog \citep{Bulbul2024}.

\begin{figure*}
\includegraphics[width=\textwidth, clip]{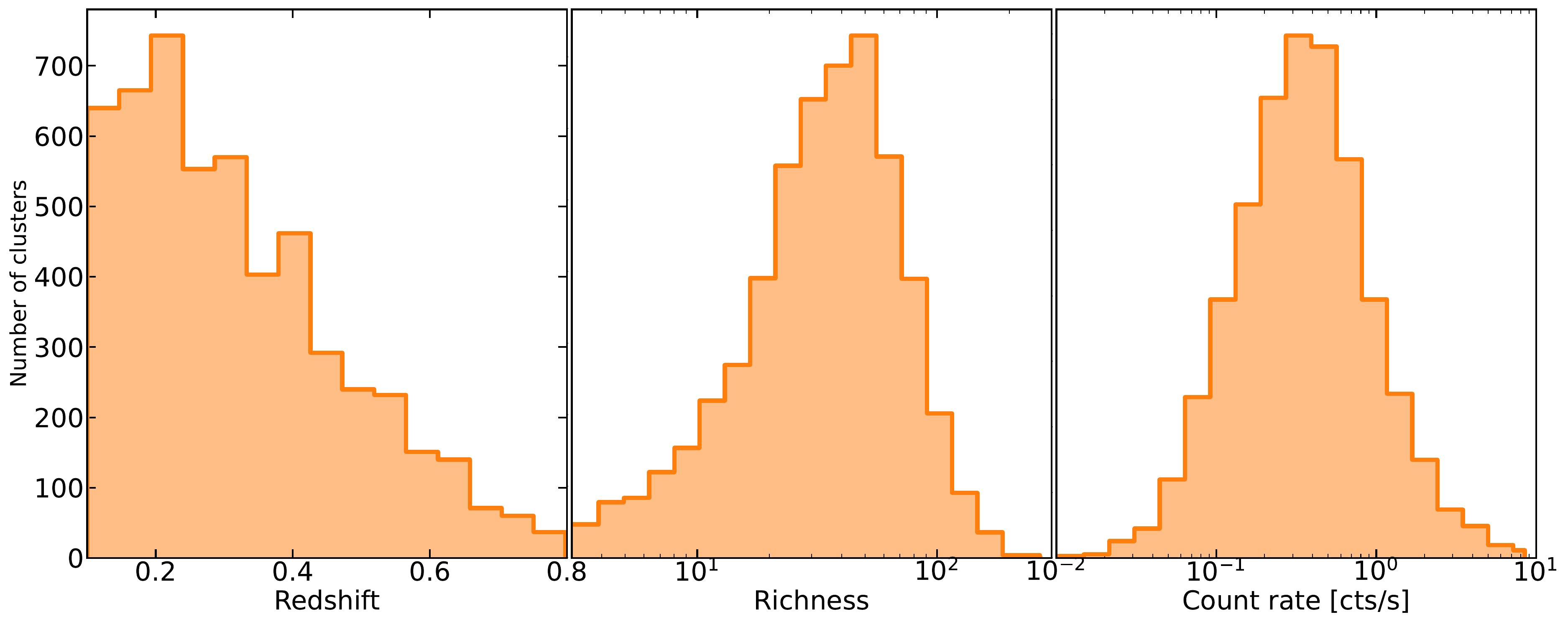}
\caption{Photometric redshift, optical richness, and X-ray count rate histogram of the 5259 confirmed \erass clusters in the cosmology subsample (top). The median redshift of the \erass cosmology catalog is 0.29, the median richness is 35, and the median count rate is 0.33 cts/s. }
\label{fig:hist}
\end{figure*}

\subsection{Weak-lensing follow-up observations and analysis}

In cosmology studies with galaxy cluster abundances, the cluster mass function should be determined with minimal bias. The X-ray observations alone do not provide direct cluster mass measurements, and the usage of hydrostatic mass estimates from X-rays is known to produce significant biases in cosmology results \citep{PlanckCollaboration2013}. Therefore, we employed weak-lensing follow-up observations of the \erosita-selected clusters to obtain reliable mass measurements with minimal bias in our mass calibration and cosmology analysis. This work uses dedicated analysis of the weak lensing signal around the clusters in three major optical surveys, HSC, DES, and KiDS, which have overlapping footprints with \erass in the western Galactic hemisphere. The common footprints of these surveys, covering a total area of 4968~deg$^2$, are shown in Fig.~\ref{fig:skydist}. In particular, \erass has a 4060~deg$^2$ common area with DES, 679~deg$^2$ with HSC, and 1108~deg$^2$ with KiDS\footnote{The reported areas include the entire footprint, without taking into account any masking or regions where shape catalog was not available.}. The detailed analysis methods and the mass calibration with these surveys are covered in our companion papers by \citet{Grandis2024} and \citet{Kleinebreil2024}, and are similar to \citet{Chiu2022} for HSC. A thorough comparison of the results from these independent surveys is presented in \citet{Kleinebreil2024}. In the following sections, we summarize the shear extraction analysis and mass calibration strategy used in these three surveys.

\subsubsection{Dark Energy Survey}

\label{sec:DES}

Here, summarize here the measurements and calibration strategy presented in \citet{Grandis2024}, aimed at extracting the tangential shear profile around \erass clusters in the DES footprint. In this analysis, we utilize observations from the first three years of observations (DES~Y3), covering the full DES survey footprint. The DES~Y3\footnote{DES~Y3 catalogs are publicly available at \href{https://des.ncsa.illinois.edu/releases/y3a2/Y3key-catalogs}{https://des.ncsa.illinois.edu/releases/y3a2/Y3key-catalogs}} shape catalog \citep{Gatti2022} is built from the $r, i, z$-bands using the \textsc{Metacalibration}
pipeline \citep{Huff2017, Sheldon2017}. Our analysis uses the same selection of lensing source galaxies in tomographic bins as the DES~Y3 3x2~pt analysis \citep{Abbott2021}. This selection is defined and calibrated in \citet{Myles2021} and \citet{Gatti2022}. We perform a background selection for each \erass cluster and group by weighting the respective tomographic bins and excluding tomographic source redshift bins whose median redshift is smaller than the cluster redshift. In the calibration steps, we provide
1) a fitting formula for the effective shape noise as a function of cluster redshift,
2) a measure of the contamination of our background selection by cluster sources, validated with three independent measurement techniques,
3) a statistical test that the cross shear is consistent with zero,
4) tangential shear profiles around individual clusters binned in angular annuli, with their respective source redshift distributions and shape noise induced uncertainty, and 
5) a percent-level accurate calibration of the shape and photo-$z$ uncertainties propagated from the calibrations as mentioned above.

Considering the overlap between the DES~Y3 and the \erass footprints, we produced tangential shear data for $2201$ \erass galaxy clusters, with a total signal-to-noise ratio (S/N) of 65 in the tangential shear profile.

\subsubsection{The Hyper Suprime-Cam (HSC) survey}
\label{sec:HSC}

We extract the weak-lensing measurements of \erass clusters using the data from the Hyper Suprime-Cam Subaru Strategic Program \citep[HSC SSP,][]{Aihara2018}, a wide and deep multi-band ($g$, $r$, $i$, $z$, and $Y$) survey in the optical. We make use of the latest three-year weak-lensing data (S19A) \citep{Li2022} with a total area coverage of $\approx500$~deg$^2$, by which $96$ \erass clusters from the western Galactic half of the \erosita sky are covered. As detailed in \citet{Li2022}, the shape measurements in the HSC weak-lensing data are carried out in the $i$-band imaging with a limiting magnitude of $i\leq24.5$~mag, and have been extensively calibrated through image simulations \citep{Mandelbaum2018b}. This results in a shape catalog with a density of $\approx20~\mathrm{galaxies}/\mathrm{arcmin}^2$, which has been used to deliver robust constraints on cosmology using cosmic shears \citep{Dalal2023,Li2023,Sugiyama2023,More2023,Miyatake2023}.
For the weak-lensing analysis of \erass clusters, we follow the same procedure as presented in \citet{Chiu2022}, which we refer to for more details.

The weak-lensing measurements were extracted based on individual clusters.
For each \erass cluster in the HSC S19A footprint, lensing sources were selected on the basis that they were securely located at the background of the cluster. This was done by requiring $\int_{z_{\mathrm{cl}} + 0.2}^{\infty} P\left(z\right)\mathrm{d}z > 0.98$, where $z_{\mathrm{cl}}$ is the cluster redshift and $P\left(z\right)$ is the photometric redshift distribution of each source. We used the  \texttt{DEmP} code \citep{Hsieh2014} to estimate the photometric redshift \citep{Tanaka2018,Nishizawa2020}. The stacked photometric redshift distribution of the sources selected in non-cluster regions was used to estimate the lensing efficiency, which is needed to obtain the weak-lensing mass from the modeling of the shear. With the selected source sample, the tangential shear profile $g_{t}\left(\theta\right)$ as a function of the angular radius $\theta$ around the cluster center was calculated by following Sect.~3.4 in \citet{Chiu2022}.
The best-fit X-ray peak ($\mathtt{RA\_XFIT}$ and $\mathtt{DEC\_XFIT}$), the output of the \mbproj\ provided in the cosmology catalog of \citep{Bulbul2024}, was used as the cluster center to extract the shear profiles around.
The radial binning was defined as ten logarithmic bins in the angular space, corresponding to $0.2~\mathrm{Mpc}/h$ and $3.5~\mathrm{Mpc}/h$ in the physical coordinate in a fiducial flat \lcdm cosmology with $h \equiv \frac{ H_{0} }{ 100~\mathrm{km}/\mathrm{s}/\mathrm{Mpc} } = 0.7$ and $\Omega_{\mathrm{m}} = 0.3$. We discarded the innermost three bins to mitigate the systematics in the weak-lensing modeling at the cluster core ($\lesssim0.5~\mathrm{Mpc}/h$). The uncertainty of the shear profile arising from the shape noise and uncorrelated large-scale structures is accounted for in the weak-lensing covariance matrix by following Sect.~3.5 in \cite{Chiu2022}.

The contamination of cluster members to the shear signal is quantified by a $P\left(z\right)$-decomposition method \citep{Gruen2014,Melchior2017,Varga2019}.
We find no sign of cluster member contamination in the stacked shear profiles in different redshift and richness bins. Conservatively, and as in the previous analysis of the eROSITA Final Equatorial Depth Survey (eFEDS) clusters \citep{Chiu2022}, we set the cluster member contamination to be zero across the radial bins with a $2\sigma$ upper limit of $6 \%$ at the innermost bin ($R \approx 0.2~\mathrm{Mpc}/h$). It then decays outwards following a projected Navarro-Frenk-White (NFW) model \citep{Navarro1996}.
With a calibration against the COSMOS 30-band photo-$z$ sample and a spectroscopic sample, the bias in the photometric redshift of the selected sources can be quantified as $\lesssim 6 \%$ for the cluster redshift $z_{\mathrm{cl}}\lesssim1.2$ and accounted for in the weak-lensing modeling.
The multiplicative bias, $m$, in the shape measurements as a function of the source redshift is quantified with a systematic uncertainty, $\delta m$, at $1.7 \%$ level \citep{Mandelbaum2018b}.

As a result, the shear profile $g_{t}\left(\theta\right)$, the lensing covariance matrix that serves the measurement uncertainty and the photometric redshift distribution of the selected source sample are obtained as the HSC weak-lensing data products for 96 \erass clusters, with a total S/N of 40. The systematic uncertainties, including the cluster member contamination, the photo-$z$ bias, and the multiplicative bias associated with these weak-lensing measurements, are quantified and taken into account in deriving the weak-lensing mass bias $b_{\mathrm{WL}}(M,z)$ \citep[see][]{Kleinebreil2024}.

\subsubsection{Kilo Degree Survey}

We used the gold sample of weak lensing and photometric redshift measurements from the fourth data release of the Kilo-Degree Survey\footnote{KiDS catalogs are publicly available at \href{https://kids.strw.leidenuniv.nl/DR4/KiDS-1000\_shearcatalogue.php}{https://kids.strw.leidenuniv.nl/DR4/KiDS-1000\_shearcatalogue.php}} \citep{Kuijken2019, Wright2020, Hildebrandt2021, Giblin2021}, hereafter referred to as KiDS-1000.

We extracted individual reduced tangential shear profiles for a total of 236 \erass galaxy clusters in both the KiDS-North field (101 clusters) and the KiDS-South field (136 clusters), as both have overlap with the \erass footprint with a total S/N of 19. A detailed presentation of the method is given in \citet{Kleinebreil2024}. Our analysis was run separately in the tomographic redshift bins of KiDS-1000, which provide spectroscopically calibrated redshift distributions for five photometric galaxy redshift bins. We also used the enhanced redshift calibration presented in \citet{vandenBusch2022}, which provides the basis for calibrating the shape measurement and photo-z systematic uncertainties.

Our analysis includes a model for the cluster member contamination that is based on radial background galaxy number density profiles \citep[e.g.][]{Applegate2014}. The model accounts for the effect of blending background galaxies with cluster galaxies on the number density profiles. For this, we simulated KiDS-like galaxy images using \uppercase{galsim} \citep{Rowe2015}, inject them into the actual $r$-band detection images of KiDS, and re-run the source detection to extract radial detection probability profiles.

We point out the importance of KiDS because it provides valuable overlap with HSC and DES, thus allowing us to perform several consistency checks and thorough comparisons between three surveys employed in the mass calibration in this work \citep{Kleinebreil2024}. In particular, there are 45 clusters in the overlap region between HSC and KiDS and 125 clusters in the overlap region between KiDS and DES. We also remark that there is no overlap between DES and HSC in the sky area relevant to the cluster cosmology catalog; therefore, these consistency checks are vital to show the robustness of our analysis.

\section{The X-ray selection function}
\label{sec:selection}

Determining the mass distribution and the consequent cosmological
analyses with cluster surveys heavily depend on the accurate knowledge of the survey's incompleteness, contamination, and selection effects. To characterize the survey selection function, we use the high fidelity end-to-end simulations of \erass from the \erosita's digital twin \citep{Seppi2022}. In this section, we provide brief descriptions of the numerical simulations and source modeling employed in creating the selection function since the detailed modeling and testing of it are already published in \citet{Comparat2019, Comparat2020}, \citet{Liu2022}, \citet{Seppi2022}, and \citet{Clerc2024}.

\subsection{Simulations}
\label{sec:sims}

We used all the snapshots of UNIT1i dark matter-only simulation with redshift $z<6.1$ to generate a full sky light cone \citep{Chuang2019}. 
Its box size (1~$h^{-1}$Gpc) and resolution (4096$^3$ particles) enable reliable prediction of the large-scale structure and its haloes down to 10$^{11}$ M$_\odot$. 
This resolution is sufficient to model the X-ray active galactic nucleus (AGN) population at the depth of \erass \citep{Comparat2019}. 
The box size is large enough to contain a fair census of massive clusters and to suppress finite volume effects \citep{Comparat2017, Klypin2019}.
The light cone is populated using models of the X-ray emission from clusters and AGNs \citep{Comparat2019, Comparat2020, Liu2022, Seppi2022}.

In the digital twin, the AGNs are simulated with an abundance matching relation between X-ray luminosity and stellar mass \citep{Comparat2019}. By construction, this reproduces the AGN luminosity function as modeled by \citet{Aird2015}. We find that the AGN model performs well as it reproduces the general statistics of the AGN population: the source number with respect to the detection flux distribution (log~N$-$log~S), luminosity function, and two-point correlation function \citep{Liu2022, Comparat2023}.

The model for clusters of galaxies is based on the X-ray observations in the literature \citep[e.g.,][]{Reiprich2002, Pierre2016, Adami2018, Sanders2018, Eckert2019}. 
A Gaussian process randomly draws temperature, mass, redshift, and emissivity profiles from a combined covariance matrix. The resulting model from \citet{Comparat2020}  reproduces the scaling relations between mass, temperature, and luminosity \citep[e.g.,][]{Giles2016} at the cluster scale. 
However, since the model relies on observed clusters with a high S/N it is not trained at the low-mass group's scale ($M_{500c}<5\times 10^{13}$M$_\odot$), and it ends up over-predicting the X-ray luminosity of these objects. To overcome this effect, the fluxes are scaled to match the expected scaling relation between luminosity and stellar mass \citep{Anderson2015} in the low-mass regime \citep[see][ for details]{Seppi2022}. 
The profiles are assigned to dark matter halos by a near-neighbor process, where the central emissivity is related to the dynamical state of the dark matter halo \citep[see][for a detailed discussion on dark matter halo property and dynamical state]{Seppi2021}. In total, we simulated 1,116,758 clusters and groups.

\subsection{Event generation and catalog inference}

To mimic the \erosita survey, the next step is to generate X-ray events with the \texttt{SIXTE} simulator \citep{Dauser2019}. \texttt{SIXTE} uses the ancillary response file, the redistribution matrix file, PSF, background, and the \erass attitude file of the spacecraft in the most up-to-date calibration database in a self-consistent manner with the observations. This method allows the simulations to accurately account for the instrumental response and exposure variations across the sky due to the scanning strategy of \textit{SRG}. The output is a collection of event lists with a realistic angular position, energy, and arrival time. These event lists are processed through the \esass software \citep{Brunner2022} as described in \citet{Merloni2024}. We matched the resulting source catalog to the input catalog from simulations by tracing the origin of each photon. The sources are grouped into five classes, described as follows:
1) \textit{PNT} is a primary counterpart of a simulated point source. The detected source is uniquely assigned to an AGN or star.
2) \textit{EXT} is a primary counterpart of a simulated extended source. The detected source is uniquely assigned to a cluster.
3) \textit{PNT2} is a secondary counterpart of a simulated point source. The detected source is assigned to an AGN or star, but the primary detection containing a large number of events is already classified as \textit{PNT} or \textit{EXT}. This is a split source corresponding to an AGN or star.
4) \textit{EXT2} is a secondary counterpart of a simulated extended source. The detected source is assigned to a cluster, but the primary detection containing a large number of events is already classified as \textit{PNT} or \textit{EXT}. This is a split source corresponding to a cluster.
5) \textit{BKG} is a false detection due to random fluctuations in the background. The entry in the source catalog is not associated with any simulated source.

For AGNs and stars, we created a single realization of the events. The sources and the realization scheme are described in detail in \citet{Seppi2022}. We create 100 realizations for the clusters initiated with a different random seed. 
The remaining 99 other realizations are used to determine accurately the selection function \citep[see][ for further details]{Clerc2024}. 
Compared to the original simulation \citep[see Sect. 3.2 in][]{Seppi2022}, the generation of cluster events was successful in the whole western Galactic half of the All-Sky survey.

\subsection{Assigning a detection probability}
\label{sec:detection_probability}

We build a detection probability model based on the simulated \erosita survey as described in the previous section. We select Gaussian process (GP) classifiers, which satisfy the requirements of short computational response, reduced prior assumptions, and can issue meaningful uncertainties. Given a set of five selected labels (features) from the simulated catalogs, we train a real-valued latent function over the corresponding five-dimensional parameter space, under the assumption of a GP prior \citep[e.g.,][]{Rasmussen2006}. We map its values to the interval $[0, 1]$ through a "probit" bijection (link function). A Bernoulli likelihood describes the distribution of binary detection flags associated with the simulated clusters, given the values of the mapped functional. Our GP prior assumes a pairwise covariance function of the form $k(r) = \sigma^2 \exp(-r^2/2)$, with $\sigma$ being a free hyperparameter and $r$ as the Euclidian distance between pairs of points, normalized along each dimension by a free hyperparameter (lengthscale). This kernel is the widely-used stationary squared exponential \citep{Rasmussen2006}~; lengthscales set the size of "wiggles" in the model while $\sigma$ controls the average deviation of the latent function from its mean (that is supposed uniform). Hyperparameters are optimized to minimize the marginal likelihood in the nominal implementation of GP classification. The quantity of interest is the posterior of the latent function given the set of training points. For any new set of feature values, the posterior predicts the distribution of the corresponding values of the latent function~; a probabilistic prediction can be obtained by integrating this distribution against the link function.

In practice, we select two-thirds of the simulated clusters in the \erass sky to train a classifier with features being the true redshift, $z$, the Galactic column density (\nh)-corrected count rate, $C_R$\footnote{We note that by correcting the observed counts using exposure map to get count rate and correcting for the absorption due to hydrogen we effectively make the X-ray observable sky independent.}, 
and the sky position, $\mathcal{H}_i$ (this indicates a combination of three sky-dependent quantities, uniquely determined by its sky coordinates: the local background surface brightness, the local \nh, and the exposure time)
A sky mask is applied to reproduce the sample selection, and we only considered objects simulated at 
$0.05<z<0.85$, slightly larger than the cosmology subsample to take into account leaking in and out of detected clusters due to redshift uncertainty.
After normalizing the feature values into an adequate parameter space, we resort to using stochastic variational Gaussian processes \citep[SVGP][]{Hensman2015} implementation in the GPy library \citep{gpy2014} by taking advantage of its numerical scalability properties, that is well-suited to our large training set (of size $\simeq 4.8 \times 10^5$). The sparse GP classification scheme \citep{Titsias2009} embedded in the algorithm relies on a small set of inducing points (30 in our case). These inducing points "summarize" the whole training dataset and are optimized with other hyperparameters. SVGP implements stochastic optimization on several mini-batches and variational inference \citep{Hensman2013} to approximate the marginal likelihood, further increasing efficiency while preserving good convergence properties. We found that minibatches of size $2^{16}$ and 12000 iterations provide good results within a tractable time. The optimization of a set of 651 hyperparameters\footnote{With $\sigma$, the lengthscales along the five dimensions, $30 \times 5$ coordinates for the inducing points, $30$ numbers for the mean values at the inducing points, and $30\times31/2$ values for the Cholesky decomposition of the covariance matrix between those values.} relies on the {\sc Adadelta} method \citep{Zeiler2012} as implemented in the {\tt climin} library.

\subsection{Selection models for validation and blinding}
\label{sec:selfunc_blinding}

For validation and blinding purposes (see details in Sect.~\ref{sec:validation_and_blinding}), six models were computed, differing only by the exact definition of an object that qualifies as "detected" in the simulation. We modified the lower thresholds in selection parameters (\extlike, \ext), choosing (3.6, 16.6), (3.9, 15.4), (4.4, 15.1), (4.8, 13.7), (5.5, 11.6), and (6, 0). By design, the six models predict different populations of detected clusters, yet we have tuned the thresholds to ensure the total number of objects is similar. Among these models, only one corresponds to the true set of selection cuts, which is described in Sect.~\ref{sec:data} (\extlike~$>6$, \ext~$>0$). Each model is tested against the remaining third of the simulation sets (each of size $\simeq 2.5 \times 10^5$). Figure~\ref{fig:checking_selection_model} shows the outcome of testing the true selection model against a concatenation of 10 test samples. The figure highlights the satisfactory agreement between the model-predicted detection probability and the actual detection rate per bin to reflect the internal consistency of our selection model. In addition to this test, the total number of predicted detections is checked to be consistent with the number of actual detections in the test samples. We repeat this series of tests using a collection of 10 independent Poissonian realizations of the \erass All-Sky survey to ensure the stability of the results against photon shot noise.

\begin{figure}
 \centering
 \includegraphics[width=\linewidth]{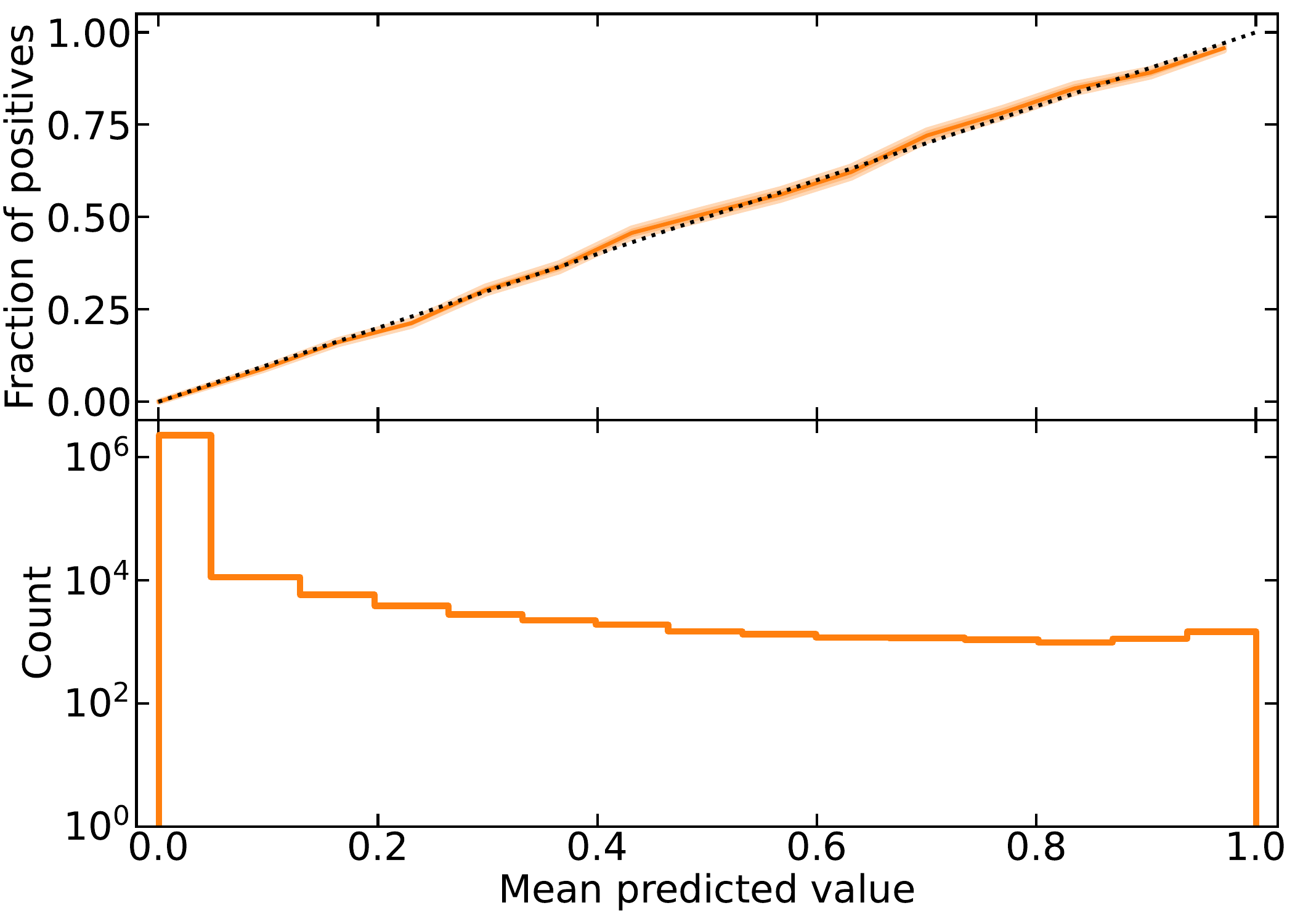}
 \caption{Reliability of the selection function model, as estimated from a test sample comprising 2,566,057 simulated galaxy clusters, none are used in the training phase. The x-axis represents the detection probability predicted by the selection function model. \emph{Top panel:}  y-axis representing the fraction of detected objects among all simulated objects in each bin. The shaded area corresponds to the 1-$\sigma$ binomial uncertainty. The dotted line represents the 1:1 relation. \emph{Bottom panel:} Histogram of clusters in the test sample per bin of predicted detection probability.} 
 \label{fig:checking_selection_model}
\end{figure}

\begin{figure}
 \centering
 \includegraphics[width=\linewidth]{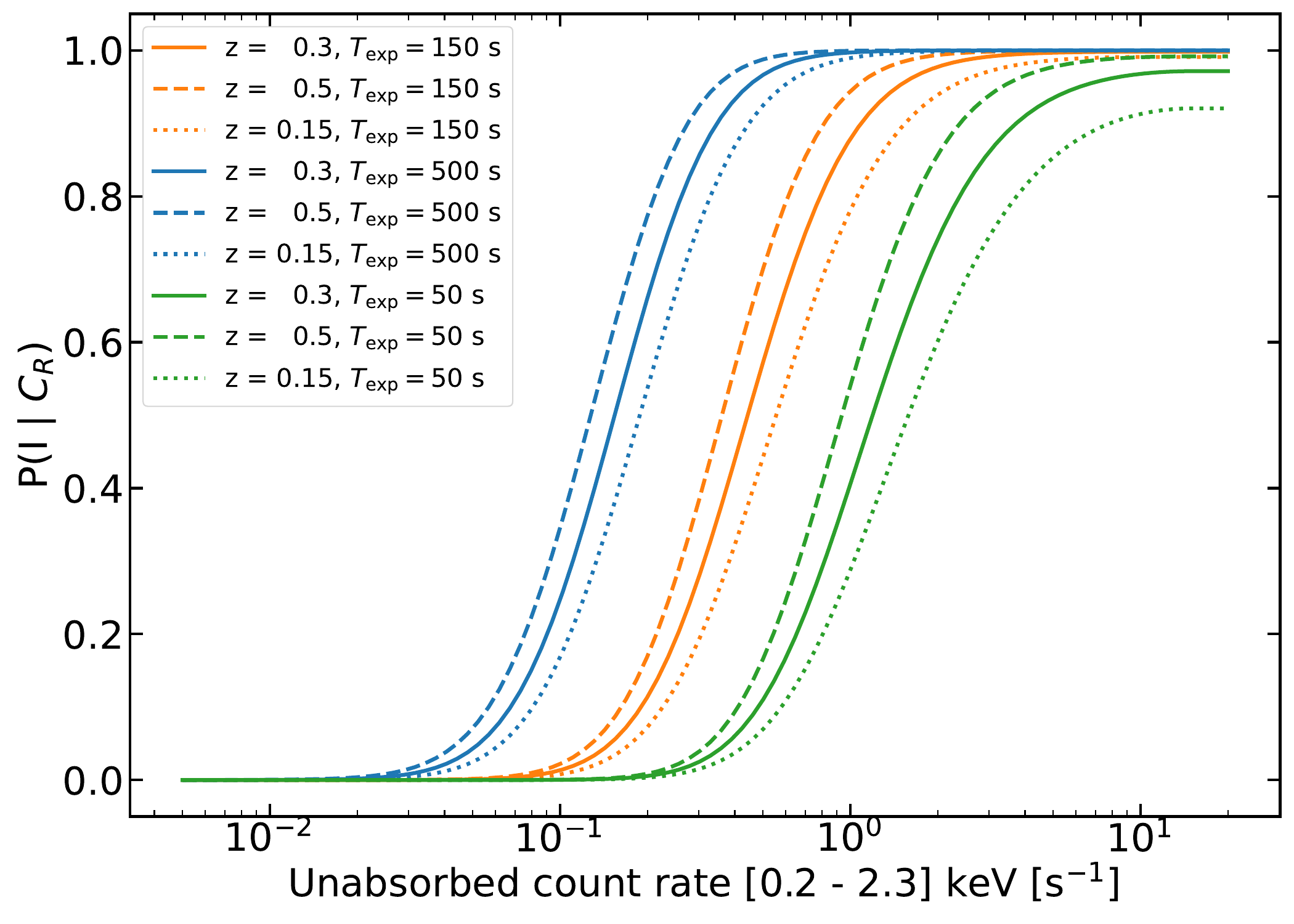}
 \caption{Representation of a model $P(I | z, C_R, \mathcal{H}_i)$ for fixed values of cluster redshift, z = [0.15, 0.3, 0.5], local galactic absorption \nh = $3.5 \times 10^{20} \mathrm{cm}^{-2}$, a nominal background level $5.2 \times 10^{-15} \mathrm{erg/s/cm^2/arcmin^2}$, three values of exposure times [50, 150, 500] s, as a function of count-rate. 
 }
 \label{fig:sel_func_summary}
\end{figure}

While the most valuable outcome of the models is the expectation values for the probabilistic predictions $P(I | z, C_R, \mathcal{H}_i)$, the probability of detecting a cluster
\citep[represented by $I$, standing for inclusion or detection, as in ][]{Mantz2010}
at a given sky position ($\mathcal{H}_i$), a given redshift ($z$), and a specific intrinsic count rate ($C_R$), the very nature of GP enables deriving confidence levels around the expectation, conditioned to the training set and the GP prior; in general, the width of confidence levels correlates with the sparsity of the training sample, that depends on location in the feature parameter space. This assigned probability is fully integrated into the likelihood for modeling the selection function. 
In Figure~\ref{fig:sel_func_summary}, we show a summary plot showing example outputs of the selection function at different exposure times and redshifts as a function of count rate.

\subsection{Optical selection function}

As described in Sect.~\ref{sec:catalog}, the cosmology cluster catalog is compiled after optical measurements of richness and redshift measured through \eromapper run on the initial cluster catalog as described in \citet{Kluge2024}.
Effectively, these are redshift and richness cuts, $0.1 < \hat{z} < 0.8$ and $\hat{\lambda} > 3$, which are applied to the primary X-ray catalog presented in \citet{Bulbul2024}. This means our selection is a two-step process, where X-ray selection is applied first and optical selection later. 

This implies that the redshift and richness cuts do not affect the X-ray selection function as these cuts are applied after the first initial X-ray selection. The modeling of this optical selection through the optical selection function is given by the Heaviside step function, which (in practice) affects only the integration limits in likelihood terms where richness or redshift are involved, as discussed in \citet{Grandis2020}. 

We stress that the role of the optical selection function is to model the contamination in the cluster sample. The cluster detection and completeness are purely governed by the X-ray detection pipeline and modeled by the X-ray selection function through the eROSITA's digital twin \citep{Seppi2022}. The optical identification process does not remove a significant number of clusters in the sample owing to the low richness limit ($\hat{\lambda} > 3$) we employ in this process. To demonstrate this, \eromapper\ was run on SZ-based cluster samples, which are known to be complete. In the companion paper \citep{Kluge2024} Sect.~5, we show that more than 97\% (optical completeness) of SZ-detected clusters are identified by \eromapper. The remaining few percent are ambiguous cases where clusters overlap along the line of sight (and therefore also the X-ray signals overlap), a known effect for cluster samples. This test demonstrates that our optical confirmation process does not yield to optical incompleteness and the method presented in this section successfully models both purity and completeness of the sample. 
The small impact of optical selection is further illustrated in Appendix~\ref{app:Optical and X-ray completeness} (particularly in Fig.~\ref{fig:completeness}).

\section{Likelihood}

\label{sec:likelihood}
This section derives the global likelihood function used to fit the data, sample the contamination, and the scaling relations to produce the cosmological inference. In the following, a variable with a ``hat'' indicates an observed quantity, while without it indicates an intrinsic quantity, for instance $\hat{C}_R$ indicates the observed \nh-corrected count rate, while $C_R$ indicates the intrinsic \nh-corrected count rate.

\subsection{Global likelihood function}

The likelihood is the probability of the data set modeled as a function of the parameters of the statistical model. Similarly to the previous studies in the literature \citep[e.g.,][]{Mantz2010, Mantz2015, Bocquet2019, Zubeldia2019, Chiu2023a, Bocquet2024}, we started from a Poisson-differential log-likelihood for our set of observables, X-ray \nh corrected count rate. $\hat{C}_R$, redshift, $\hat{z}$, sky position, $\hat{\mathcal{H}}_i$, richness, $\hat{\lambda}$, and weak-lensing tangential shear profile, $\hat{g}_t$;

\begin{align}
\log \mathcal{L} &= \sum_j \log 
\left(
\frac{dN_{tot}}{d\hat{C}_R d\hat{z} d\hat{\lambda} d\hat{g}_t d\hat{\mathcal{H}}_i} 
P(I | \hat{C}_R, \hat{z}, \hat{\lambda} ,\hat{\mathcal{H}}_i, \hat{g}_t) 
\right)
\nonumber \\
& - \idotsint \bigg[
\frac{dN_{tot}}{d\hat{C}_R d\hat{z} d\hat{\lambda} 
 d\hat{g}_t d\hat{\mathcal{H}}_i} 
P(I | \hat{C}_R, \hat{z}, \hat{\lambda}, \hat{\mathcal{H}}_i, \hat{g}_t) \nonumber \\
& \hspace{2cm} d\hat{C}_R d\hat{z} d\hat{\lambda} 
 d\hat{g}_t d\hat{\mathcal{H}}_i \bigg],
 \label{eq:Poisson_log_like}
\end{align}

\noindent where the first term runs over all detected clusters, while the second one is a computation representing the total number of objects detected in the \erass survey.

We highlight that (as tends to happen in the process of detecting and including clusters in our dataset) our selection function does not depend on the measured tangential shear profile. Further, it is crucial that we do not discard the measured shear profiles whose signal is consistent with noise from the fitting process to avoid having an effect on the selection function. 
Furthermore, since the likelihood mimics the detection and measuring process, the selection function can be separated into the X-ray selection followed by the optical selection, as follows:

\begin{equation}
P(I | \hat{C}_R, \hat{z}, \hat{\lambda}, \hat{\mathcal{H}}_i, \hat{g}_t) = 
P(I | \hat{C}_R, \hat{z}, \hat{\mathcal{H}}_i)
\,
\Theta(\hat{\lambda} > 3),
\end{equation}

\noindent where $\Theta(\hat{\lambda} > 3)$ is the Heaviside step function, equals to 1 where $\hat{\lambda} > 3$, and equal to 0 where $\hat{\lambda} < 3$. This cut is necessary to reproduce the richness cut applied by the optical confirmation process with \eromapper code runs. For more information, we refer to Sect.~\ref{sec:catalog} and \citet{Kluge2024}.

We, therefore, isolate richness and tangential shear profile likelihood terms. In practice, the first term in Equation~\ref{eq:Poisson_log_like} can be simplified into:

\begin{equation}
\frac{dN_{tot}}{d\hat{C}_R d\hat{z} d\hat{\lambda} d\hat{g}_t d\hat{\mathcal{H}}_i} = 
\frac{dN_{tot}}{d\hat{C}_R d\hat{z} d\hat{\mathcal{H}}_i} 
\,
P(\hat{\lambda}, \hat{g}_t | \hat{C}_R, \hat{z}, \hat{\mathcal{H}}_i).
\end{equation}

Given these considerations, it is therefore consequent that the measurement of tangential shear profiles does not directly affect the total number of detected objects. This means that in the likelihood term that computes the total number of detected objects, the second of Equation~\eqref{eq:Poisson_log_like}, the dependence on tangential shear disappears, and Equation~\eqref{eq:Poisson_log_like} therefore becomes;

\begin{align}
\log \mathcal{L} &= \sum_j \log 
\left(
\frac{dN_{tot}}{d\hat{C}_R d\hat{z} d\hat{\mathcal{H}}_i} 
P(I | \hat{C}_R, \hat{z} ,\hat{\mathcal{H}}_i) \Theta(\hat{\lambda} > 3) 
\right)
\nonumber \\
& + \sum_j \log 
P(\hat{\lambda}, \hat{g}_t | \hat{C}_R, \hat{z}, \hat{\mathcal{H}}_i, I) \nonumber \\
& - \idotsint \bigg[
\frac{dN_{tot}}{d\hat{C}_R d\hat{z} d\hat{\lambda} 
 d\hat{\mathcal{H}}_i} 
P(I | \hat{C}_R, \hat{z}, \hat{\mathcal{H}}_i) 
\Theta(\hat{\lambda} > 3)
\nonumber \\
& \hspace{2cm} d\hat{C}_R d\hat{z} d\hat{\lambda} d\hat{\mathcal{H}}_i \bigg]
\label{eq:Poisson_log_like_step2}
\end{align}

Additionally, although optical richness and weak lensing shear profiles are correlated signals, we can safely assume that the correlation between the two quantities is a subdominant factor. The richness used in this work involves a level of X-ray information which makes the richness better behaved than if it was measured from a purely optical survey without the X-ray or SZe information \citep{Saro2015, Bleem2020, Grandis2021b, Giles2022}. In fact, the extra prior regarding the location of the center of the halo and the presence of an X-ray emitting halo allow for more accurate richness measurement. Additionally, a significant reduction in the number of halos with low mass and spurious sources not associated with X-ray emitting halos that pollute the mass-observable space because of line-of-sight projection effects is elevated in the ICM-selected samples. As a consequence, the intrinsic correlation between weak lensing mass and richness is expected to be small as it is absorbed in the other two correlations we consider in this work, specifically the correlation between WL mass and count rate and the correlation between richness and count rate. Therefore, the likelihood of observing the measured tangential shear profile and the measured richness are uncorrelated and can be simplified as:

\begin{equation}
P(\hat{\lambda}, \hat{g}_t | \hat{C}_R, \hat{z}, \hat{\mathcal{H}}_i, I) = 
P(\hat{\lambda} | \hat{C}_R, \hat{z}, \hat{\mathcal{H}}_i, I)
\,
P(\hat{g}_t | \hat{C}_R, \hat{z}, \hat{\mathcal{H}}_i, I).
\end{equation}
Thus, we can write our final log-likelihood as:
\begin{align}
\log \mathcal{L} &= \sum_j \log 
\left(
\frac{dN_{tot}}{d\hat{C}_R d\hat{z} d\hat{\mathcal{H}}_i} 
P(I | \hat{C}_R, \hat{z} ,\hat{\mathcal{H}}_i)
\right)
\nonumber \\
& - \idotsint \bigg[
\frac{dN_{tot}}{d\hat{C}_R d\hat{z} d\hat{\lambda} 
 d\hat{\mathcal{H}}_i} 
P(I | \hat{C}_R, \hat{z}, \hat{\mathcal{H}}_i) 
\Theta(\hat{\lambda} > 3)
\nonumber \\
& \hspace{2cm} d\hat{C}_R d\hat{z} d\hat{\lambda} d\hat{\mathcal{H}}_i \bigg] \nonumber \\
& + \sum_j \log 
P(\hat{\lambda} | \hat{C}_R, \hat{z}, \hat{\mathcal{H}}_i, I) \nonumber \\
& + \sum_j \log 
P(\hat{g}_t | \hat{C}_R, \hat{z}, \hat{\mathcal{H}}_i, I).
\label{eq:Poisson_log_like_final}
\end{align}

The first term in Equation~\eqref{eq:Poisson_log_like_final} represents the number count likelihood coming from X-ray information, and the second term is the expected total number of detected objects, which depends on the selection function. In contrast, the latter two terms describe the information we obtain from our follow-up observables: optical richness and tangential shear.

In the following subsections, we expand on the terms in the likelihood: first, we show how our mixture model allows us to distinguish the clusters from the contamination, for instance, AGNs and background fluctuations, in the sample. We then examine the X-ray likelihood term, the joint X-ray/optical cluster total number of predicted clusters, the WL mass calibration, and the richness mass calibration.
The integrals are computed to obtain the likelihood value for a set of parameters in a range where the integrands differ from zero. Specifically, redshift from 0 to 0.9, mass from $5 \times 10^{12} M_\odot$ to $5 \times 10^{15} M_\odot$, richness from 1 to 600, and count rate from $10^{-3}$~cts/s to 100~cts/s.

\subsection{Mixture method for the contamination modeling}
\label{sec:mixture}

Statistical studies previously performed with limited samples of clusters often had a negligible fraction of contaminants. They did not require extensive modeling as the fraction of expected contaminants was sub-dominant compared to the Poisson noise \citep{Planck2016, Garrel2022}. The effects of contamination become relevant only when the contamination fraction exceeds the Poisson noise of the data. It has recently become a routine to include the contamination modeling in cluster abundance studies \citep[see the implementation in the SPT-SZ sample][]{Bocquet2019}\footnote{In \citep{Bocquet2019}, the authors compute the corresponding likelihood only for a fraction of systems that are not optically confirmed. This is different than our implementation presented here}.
As large cluster samples from wide-area surveys become available, contamination modeling becomes crucial in statistical cosmological studies with cluster abundances.

In \erass\ cosmology subsample, after the \extlike~$>6$ cut is applied, the expected contamination level is estimated to be 20\% from our realistic simulations \citep{Seppi2022}. The optical confirmation process is expected to reduce the contamination to $\sim$6\% after removing the unconfirmed sources in the sample \citep{Kluge2024, Bulbul2024}. The Poissonian noise of our data ($\sqrt{N} / {N}$) is at the 1.4\% level; therefore, in our analysis, contamination modeling is crucial to derive unbiased results from statistical fits. To this end, we develop a mixture model, which is a probabilistic model that represents multiple sub-populations of sources within a given sample without requiring exact knowledge of the nature of each object.

Informed by the prior work on the X-ray selection on simulation \citep{Seppi2022}, we model our cluster sample as the mixture of three populations: clusters, AGNs, and random sources (RSs) from background fluctuations. Motivated by the fact that for clusters, both richness and count rate correlate with halo mass at a given redshift, we model the cluster population with multivariate observable mass scaling relations (see Sect.~\ref{sec:cluster_number_count_likelihood}). For the other two populations, the X-ray emission is modeled as uncorrelated with the richness--redshift distribution, with the count rate distribution as a function of sky position coming from our X-ray simulations and the richness--redshift distribution coming from optical follow-up runs of the point source catalog and random position for the AGN and RS populations, respectively. This serves two related purposes. Firstly, we can fit for the contamination of our sample on the fly. Secondly, we ensure that for the cluster population, the optical signal and the derived redshift are linked to the same host halo and its X-ray signature, preventing chance superpositions. In the context of X-ray-selected cluster samples, erroneously associating a high redshift to a lower count rate X-ray signature would mislead us into interpreting that object as a high mass, high redshift cluster, biasing the cosmological inference. As a consequence, we need to model the richness--mass relation of the cluster population (see Sect.~\ref{sec:optical_likelihood}).

In practice, we can a-posteriori assign a probability to each source belonging to one or the other sub-population based on its measured properties \citep{Kluge2024}. 
In detail, our realistic simulations suggest that the three classes of X-ray sources comprise the extended source catalog: clusters, AGNs, and RSs from background fluctuations. 
Clusters are the population of interest and form the majority of the cleaned extended source sample ($\sim$90-95\%). The global population modeling can be represented as a sum of cluster distribution, which is cosmology dependent, and the distribution of RSs and AGNs, which are assumed to have known probability density functions but unknown overall numbers, thus allowing for a simultaneous fit for cosmology as well as the overall number of contaminants in the likelihood Equation~\eqref{eq:Poisson_log_like_final}.

The complete derivation of the mixture model and its effects on each likelihood term in Equation~\eqref{eq:Poisson_log_like_final} is shown in Appendix~\ref{app:mixture}.
In short, the probability density functions for every observables $\hat{\mathcal{O}}$ is written as:

\begin{equation}
 P(\hat{\mathcal{O}}) = \sum_{\rm i \in \{ C, \ AGN, \ RS \}} f_{\rm i} \, P(\hat{\mathcal{O}} | {\rm i}),
\end{equation}

\noindent where this equation holds for all our observables $\hat{\mathcal{O}}$ as $\hat{C}_R$, $\hat{z}$, $\hat{\lambda}$, and $\hat{g}_t$. $f_{\rm i}$ are the fraction of each population in the \erass sample and sum up to unity. Therefore, the mixture model adds two extra parameters to the fit, which respectively describe the amount of AGNs and RSs in our cluster cosmology sample, while the constraint of unity summed fraction fixes the number of clusters.

\subsection{Cluster number counts}

\label{sec:cluster_number_count_likelihood}

This subsection focuses on the differential cluster number counts term, $\frac{dN_C}{d\hat{C}_R d\hat{z} d\hat{\mathcal{H}}_i} P(I | \hat{C}_R, \hat{z}, \hat{\mathcal{H}}_i)$, which is a significant term in Equation~\eqref{eq:Poisson_log_like_final}, being involved in both terms of the differential Poissonian likelihood required to compute the number counts likelihood.
Note that the corresponding terms for AGNs and RSs from the mixture model are directly recoverable from the mixture model; see Equation~\eqref{mix:dN}. 

We consider this term as a marginalized likelihood over true count rate, true mass, and true redshift.\footnote{Technically also over the true cluster position; however, we assume that cluster positions are measured precisely, e.g., $P(\hat{\mathcal{H}}_i | \mathcal{H}_i) = \delta (\hat{\mathcal{H}}_i - \mathcal{H}_i)$, where $\delta$ represents the Dirac's delta function. Indeed, the relevant observing conditions do not vary within the astrometric uncertainty on the cluster position.}

\begin{align}
\frac{dN_C}{d\hat{C}_R d\hat{z} d\hat{\mathcal{H}}_i} & P(I | \hat{C}_R, \hat{z}, \hat{\mathcal{H}}_i) =
P(\hat{\mathcal{H}}_i)
 \iiint 
P(\hat{z} | z) 
P(I | z, C_R, \hat{\mathcal{H}}_i) \nonumber \\
&
\quad
\times
P(\hat{C}_R | C_R )
 \frac{dN}{dM dV} \frac{dV}{dz}
 P(C_R | M, z) 
 dM 
 dC_R
 dz
\label{eq:X-ray_dN_dobs}
\end{align}

\noindent where $P(\hat{\mathcal{H}}_i)$ is the uniform probability over the sphere that a cluster exists at a specific sky position $\hat{\mathcal{H}}_i$, $P(\hat{z} | z)$ expresses the uncertainty on our redshift measurements, provided by \eromapper \citep{Kluge2024},
$P(I | z, C_R, \hat{\mathcal{H}}_i)$ is the X-ray selection function (see Sect.~\ref{sec:selection} and for more details \citet{Clerc2024}), $P(\hat{C}_R | C_R )$ is the uncertainty on the X-ray count rate, measured by \mbproj, 
$\frac{dN}{dM dV}$ is the halo mass function\footnote{
$\frac{dN}{dM dV} = f(\sigma) \frac{\rho}{M} \frac{d\log(\sigma^{-1})}{d\log M}$ where $f(\sigma)$ is a function of $\sigma(R, z)$, which is defined as 
$\sigma^2(R,z) = \frac{1}{2 \pi^2} \int k^2 P(k, z) |W(kR)|^2 dk$}, 
which directly depends on cosmology,
$\frac{dV}{dz}$ is the fractional comoving volume in an infinitesimal redshift shell \citep[see Equation~(28) in ][]{Hogg1999}%
\footnote{We point out that $\frac{dN}{dM dV}$ is the halo mass function as a function of comoving volume, while $\frac{dN}{dM dV} \frac{dV}{dz}$ is the halo mass function as a function of redshift. Cosmology calculators, like \texttt{pyccl}, give the user the former, and the latter is left to calculate for the user.},
$P(C_R | M, z)$ is the log-normal distribution around the X-ray scaling relation and can be described as:

\begin{equation}
P(C_R | M, z) = \mathcal{LN} (\left\langle C_R | M, z \right\rangle, \sigma_X)
\label{eq:CR_given_M_z}
.\end{equation}

\noindent We refer to Sect.~\ref{sec:scaling_relations} for details on the parameterized scaling relation.
Furthermore, the total number of objects detected in \erass, depends on the calculation of the number of clusters and on the fraction of contaminants, as shown in Equation~\eqref{mix:N}. The total number of detected clusters is simply the integration over the observable space of the differential Poisson term, and can be computed via the equation:

\begin{equation}
N_{\rm C} = \iiiint \frac{dN_{\rm C}}{d\hat{C}_R d\hat{z} d\hat{\lambda} 
 d\hat{\mathcal{H}}_i} 
P(I | \hat{C}_R, \hat{z}, \hat{\mathcal{H}}_i) 
\Theta(\hat{\lambda} > 3) d\hat{C}_R d\hat{z} d\hat{\lambda} d\hat{\mathcal{H}}_i
\label{eq:ntot0}
\end{equation}

\noindent where several simplifications are applied.
First, $P(\hat{z} | z)$, the probability of observing a cluster with measured redshift $\hat{z}$ given the true redshift $z$, is modeled as a mixture of two normal distributions whose parameters are calibrated by \eromapper, expressed as:

\begin{align}
P(\hat{z} | z) &= 
(1-c_z) \, \mathcal{N} (b_z \, z, \sigma_z (1+z)) \nonumber \\ 
& \quad + c_z \, \mathcal{N} (b_z \, z + c_{\mathrm{shift}, z}, \sigma_z (1+z)),
\label{eq:zhat_given_z}
\end{align}

\noindent where $c_z$ is the fraction of objects for which we measure a redshift shifted by some constant $c_{\mathrm{shift}, z}$, $b_z$ is a systematic bias in our redshift estimate, and $\sigma_z$ is the relative error on the measured redshift. 
We can, therefore, express $g(z, \mathcal{H}_i) = \int P(\hat{z} | z) w(\hat{z}, \mathcal{H}_i) d\hat{z}$ as an analytic expression for each sky position, with $w(\hat{z}, \mathcal{H}_i)$ being the filter function that is redshift and sky position-dependent, as follows:

\begin{align}
w(\hat{z}, \mathcal{H}_i) = 
\begin{cases}
 1, & \text{if}\ (\hat{z} > 0.1) \ \& \ (\hat{z} < \text{min}(0.8, \, \hat{z}_{\rm max}(\hat{\mathcal{H}}_i)), \\
 0, & \text{otherwise,}
\end{cases}
\end{align}

\noindent where $\hat{z}_{\rm max}(\hat{\mathcal{H}}_i)$ is the maximum measurable redshift at sky position $\hat{\mathcal{H}}_i$, given by the ZVLIM map \citep[see][ for further details]{Kluge2024}.

Then, $P(\hat{\lambda} | \lambda)$, the probability of observing a cluster with measured richness $\hat{\lambda}$ given the true richness $\lambda$, is modeled as 
in previous cosmological analyses with cluster counts \citep{Abbott2020, Chiu2023a, Bocquet2024} as a Poisson distribution in its Gaussian limit. A Poisson distribution is a reasonable theoretical assumption as richness effectively counts galaxy members, and partly demonstrated in \citet{Costanzi2019r}. Therefore, we have:

\begin{equation}
P(\hat{\lambda} | \lambda ) = \mathcal{N}(\lambda, \sqrt{\lambda}).
\label{eq:lambdahat_given_lambda}
\end{equation}

Thus, we can express the optical completeness term $f(\lambda) = \int P(\hat{\lambda} | \lambda ) \Theta(\hat{\lambda} > 3) d\hat{\lambda}$ term analytically.%
\footnote{In short, for redshift and richness, for which the distributions are normal distributions integrated in some bounded range, the integral is expressed as an error function.}
The effects of the optical cut, $\hat{\lambda} > 3$, on the overall sample completeness is very small, decreasing the overall number of clusters computed from Equation~\eqref{eq:ntot0}, and thus the completeness by 1.1\%. Furthermore, photometric redshifts from optical structures with 3 or fewer member galaxies are arguably quite unreliable.

Finally, since $\hat{C_R}$ appears only in the term $P(\hat{C_R} | C_R)$, and since probabilities are normalized, the integral over measured count rate is unity and does not affect the calculation of the total number of objects. Therefore, the total number of clusters is expressed as:

\begin{align}
N_{C} = \iiiint &
P(I | z, C_R, \hat{\mathcal{H}}_i) 
g(z, \hat{\mathcal{H}}_i)
f(\lambda)
P(C_R, \lambda | M, z) 
P(\hat{\mathcal{H}}_i) \nonumber \\
 & \times \frac{dN}{dM dV} \frac{dV}{dz} 
 dM 
 dC_R 
 dz 
 d\lambda
 d\hat{\mathcal{H}}_i ,
 \label{eq:NC_tot}
\end{align}

To conclude, we now have expressed explicitly all the ingredients required in order to model the cluster number counts likelihood in Equation~\eqref{eq:X-ray_dN_dobs} and to compute the total number of clusters that are detected in Equation~\eqref{eq:NC_tot}.

\subsection{Weak-lensing mass calibration likelihood}\label{sec:wl--mass-cali}

This subsection focuses on the weak lensing likelihood, which is crucial for mass calibration.
We make use of Bayes' theorem to express the weak lensing likelihood as:

\begin{equation}
 P(\hat{g}_t | \hat{C}_R, \hat{z}, \hat{\mathcal{H}}_i, I, C) = 
 \frac{P(\hat{g}_t, \hat{C}_R, \hat{z}, \hat{\mathcal{H}}_i, I, C)}{P(\hat{C}_R, \hat{z}, \hat{\mathcal{H}}_i, I, C)} =
 \frac{N_{\rm WL}}{D_{\rm WL}}
,\end{equation}

\noindent where we have factored out the numerator $N_{\rm WL}$ and the denominator $D_{\rm WL}$; otherwise, the expression will not fit a double column article. Similarly to what we have done in Sect.~\ref{sec:cluster_number_count_likelihood}, we marginalize the expressions $N_{\rm WL}$ and $D_{\rm WL}$, similarly to Equation~\eqref{eq:X-ray_dN_dobs}, as follows:

\begin{align}
N_{\rm WL} = P(\hat{\mathcal{H}}_i) & \iiint 
 P(I | C_R, \hat{z}, \hat{\mathcal{H}}_i)
 P(\hat{C}_R | C_R) 
 P(M, \hat{z})
 \nonumber \\
& \times 
 P(\hat{g}_t | M_{\rm WL}, \hat{z}) 
 P(M_{\rm WL}, C_R | M, \hat{z})
 dM
 dC_R
 dM_{\rm WL}
\end{align}

\noindent and

\begin{align}
D_{\rm WL} =
P(\hat{\mathcal{H}}_i) \iint &
 P(I | C_R, \hat{z}, \hat{\mathcal{H}}_i)
 P(\hat{C}_R | C_R) 
 P(M, \hat{z}) \nonumber \\
 & \times P(C_R | M, \hat{z}) 
 dM
 dC_R.
\end{align}

We note that the sky position probabilities, $P(\hat{\mathcal{H}}_i)$, cancel out in the ratio of $N_{\rm WL}$~/~$D_{\rm WL}$ in the weak lensing mass calibration term. $P(\hat{g}_t | M_{\rm WL}, \hat{z})$ is the likelihood term that compares the measured weak lensing shear with the one predicted from the weak lensing mass \citep{Grandis2024}. Furthermore, in both the numerator and denominator, we have the error on the measured count rate $P(\hat{C}_R | C_R)$, the selection function $P(I | C_R, \hat{z}, \hat{\mathcal{H}}_i)$, the normalized halo mass function, $P(M, \hat{z})$. However, since the normalization factor (needed to compute $P(M, z)$ from $\frac{dN}{dM dV} \frac{dV}{dz}$) is present both at the numerator and denominator, it cancels itself out, and the halo mass function, $\frac{dN}{dM dV} \frac{dV}{d\hat{z}}$, can be used directly in the previous equations. 
All the details regarding the expressions for these terms are described just below Equation~\eqref{eq:X-ray_dN_dobs}.

Finally,
$P(M_{\rm WL}, C_R | M, \hat{z})$,
the scaling relation term between mass and both count rate and weak lensing mass that takes into account the intrinsic correlation between these quantites, is a bi-variate log-normal distribution $\mathcal{BN}(\bar{\mu}, \bar{\Sigma}),$ where

\begin{equation}
\bar{\mu} = [\left\langle \log C_R | M, z\right\rangle, \left\langle \log M_{\rm WL} | M, z\right\rangle],
\label{eq:Mwl_given_M_z}
\end{equation}
\begin{equation}
\bar{\Sigma} =
 \begin{bmatrix}
 \sigma_{C_R}^2 & \rho_{M_{\rm WL}, C_R} \sigma_{C_R} \sigma_{M_{\rm WL}} \\ 
 \rho_{M_{\rm WL}, C_R} \sigma_{C_R} \sigma_{M_{\rm WL}} & \sigma_{M_{\rm WL}}^2 
 \end{bmatrix},
\end{equation}

\noindent where $\left\langle\log C_R | M,z\right\rangle$ and $\left\langle\log M_{\rm WL} | M, z\right\rangle$ are the scaling relations between true cluster mass and count rate and weak lensing mass, respectively, see Sect.~\ref{sec:scaling_relations} for further details, and $\rho_{M_{\rm WL}, C_R}$ is the intrinsic correlation between weak lensing mass and count rate.

\subsection{Optical likelihood}\label{sec:optical_likelihood}

Similarly to the previous subsection, we apply Bayes's theorem to express the optical likelihood, meaning the probability of measuring the observed richness for a cluster that has an observed count rate and redshift at a specific sky position, as:

\begin{equation}
 P(\hat{\lambda} | \hat{C}_R, \hat{z}, \hat{\mathcal{H}}_i, I, C) = 
 \frac{P(\hat{\lambda}, \hat{C}_R, \hat{z}, \hat{\mathcal{H}}_i, I, C)}{P(\hat{C}_R, \hat{z}, \hat{\mathcal{H}}_i, I, C)} = \frac{N_{\rm opt}}{D_{\rm opt}}
,\end{equation}

\noindent where (as before) we have factored out the numerator $N_{\rm opt}$ and the denominator $D_{\rm opt}$.
As before, we expand the numerator and denominator as:

\begin{align}
N_{\rm opt} = P(\hat{\mathcal{H}}_i) & \iiint P(I | C_R, \hat{z}, \hat{\mathcal{H}}_i) 
P(\hat{C}_R | C_R)
P(M, \hat{z}) \nonumber \\
& \times P(\hat{\lambda} | \lambda) 
\Theta(\hat{\lambda} > 3) 
P(\lambda, C_R | M, \hat{z}) 
 dM
 dC_R
 d\lambda,
\end{align}
\begin{align}
D_{\rm opt} =
P(\hat{\mathcal{H}}_i) \iint &
 P(I | C_R, \hat{z}, \hat{\mathcal{H}}_i)
 P(\hat{C}_R | C_R)
 P(M, \hat{z}) \nonumber \\
 & \times P(C_R | M, \hat{z}) 
 dM
 dC_R.
\end{align}

\noindent Here, in particular, $P(\hat{\lambda} | \lambda)$ is the likelihood term that evaluates the likelihood of observing a richness given the true richness, given by the measurement uncertainty on the richness given by eROMapper. Similarly to the weak lensing likelihood, we have that $P(I | C_R, \hat{z}, \hat{\mathcal{H}}_i)$ is the X-ray selection function, $\Theta(\hat{\lambda} > 3)$ is the optical selection function\footnote{Even though the optical selection function has no effect since this term is computed only for clusters that are detected, we still leave it inside the equation to recall that the optical uncertainty term $P(\hat{\lambda} | \lambda)$ is not normalized from $-\infty$ to $+\infty$, but from 3 to $\infty$, where 3 is our implicit optical cut.}, $P(M, \hat{z}) $ is the normalized mass function, $P(\hat{C}_R | C_R)$ is the error on the measurement of the count rate, $P(\lambda, C_R | M, \hat{z}) $ is a bivariate log-normal distribution $\mathcal{MN}(\bar{\mu}, 
\bar{\Sigma})$ where

\begin{equation}
 \bar{\mu} = [\left\langle \log C_R | M, \hat{z} \right\rangle , \left\langle \log \lambda | M, z \right\rangle]
 \label{eq:rich_given_M_z} 
,\end{equation}
\begin{equation}
\bar{\Sigma} =
 \begin{bmatrix}
 \sigma_{C_R}^2 & \rho_{\lambda, C_R} \sigma_{C_R} \sigma_{\lambda} \\ 
 \rho_{\lambda, C_R} \sigma_{C_R} \sigma_{\lambda} & \sigma_{\lambda}^2 + \sigma_P^2
 \end{bmatrix}
,\end{equation}

\noindent where $\left\langle \log C_R | M,z\right\rangle$ and $\left\langle \log \lambda | M, z\right\rangle$ are the scaling relations between true mass, count rate, and optical richness, details are given in Sect.~\ref{sec:scaling_relations}, while $\rho_{\lambda, C_R}$ is the intrinsic correlation between richness and count rate and where $\sigma_{\lambda}$ is richness intrinsic scatter, and $\sigma_P$ is a Poisson term, defined as in \citet{Costanzi2019r} by $\frac{\left\langle \lambda | M, z \right\rangle - 1}{\left\langle \lambda | M, z \right\rangle^2}$.
We point out that the denominator for the optical, $D_{\rm opt}$, and weak lensing, $D_{\rm WL}$, likelihoods look identical; however, we recall that while the optical term runs over all the objects in the sample, the weak-lensing term runs over only the objects that have tangential shear measurements. 

\subsection{Parameters and priors}
In Table~\ref{tab:parameters_priors}, we list all the parameters fitted in our Bayesian statistical model. We include their units, descriptions, and prior functions used.
We note that our analysis is not sensitive to some cosmological parameters. For this reason, to avoid the chains reaching unphysical values but still allowing marginalization over these parameters, we limit their range using realistic priors from previous cosmological experiments. Specifically, we limit $\Omega_\mathrm{b}$ in a range allowed by the Big Bang Nucleosynthesis \citep{Beringer2012, Cooke2016, Cooke2018,Mossa2020,Pitrou2021}, $n_s$ in a range round \planck CMB results \citep{Planck2020}. For $H_0$, against which we are not sensitive, we use the \planck CMB \lcdm results with a tight Gaussian prior set around them \citep{Planck2020}. Given the so-called $H_0$ tension, we have checked that using a prior from distance ladder method \citep[$H_0 \sim \mathcal{N}(73.04, 1.04) \, \mathrm{km ~ s^{-1} ~ Mpc^{-1}}$,][]{Riess2022} or a flat prior ($H_0 \sim \mathcal{U}(60, 80) \, \mathrm{km ~ s^{-1} ~ Mpc^{-1}}$) does not affect our results and conclusions.

\begin{table*}[h]
 \caption{Priors used on the free parameters in the fitting, divided into several parameter types for clarity.}
 \label{tab:parameters_priors}
 \centering
 \begin{tabular}{llll}
 \hline
 \hline
 Parameter & Units & Description & Prior \\
 \hline
 \multicolumn{4}{l}{$\bullet$ Cosmology}\\
 $\Omega_{\mathrm{m}}$ & - & Mean matter density at present time & $\mathcal{U}(0.05, 0.95)$ \\
 $\log_{10} A_s$ & - & Amplitude of the primordial power spectrum & $\mathcal{U}(-10, -8)$ \\
 $H_0$ & $\frac{\frac{\rm km}{\rm s}}{\rm Mpc}$ & Hubble expansion rate at present time & $\mathcal{N}(67.77, 0.6)$ \\
 $\Omega_{\mathrm{b}}$ & - & Mean baryon density at present time & $\mathcal{U}(0.046, 0.052)$ \\
 $n_s$ & - & Spectra index of the primordial power spectrum & $\mathcal{U}(0.92, 1.0)$ \\
 $w_0$ & - & Dark energy equation of state. Fixed to -1 in \lcdm & $\mathcal{U}(-2.5, -0.33)$ \\
 $\sum m_\nu$ & eV & Summed neutrino masses. Fixed to 0~eV in \lcdm & $\mathcal{U}(0, 1)$ \\
 \hline
 \multicolumn{4}{l}{$\bullet$ X-ray scaling relation}\\
 $A_X$ & - & Normalization of the $M - C_R$ scaling relation & $\mathcal{U}(0.01, 3)$ \\
 $B_X$ & - & Mass slope of the $M - C_R$ scaling relation & $\mathcal{U}(0.1, 5)$ \\
 $D_X$ & - & Luminosity distance evolution of the $M - C_R$ scaling relation & Fixed to -2 \\
 $E_X$ & - & Scale factor evolution of the $M - C_R$ scaling relation & Fixed to 2 \\
 $F_X$ & - & Redshift evolution of the mass slope of the $M - C_R$ scaling relation & $\mathcal{U}(-5, 5)$ \\
 $G_X$ & - & Redshift evolution of the normalization of the $M - C_R$ scaling relation & $\mathcal{U}(-5, 5)$ \\
 $\sigma_{X}$ & - & Intrinsic scatter of the $M - C_R$ scaling relation & $\mathcal{U}(0.05, 2)$ \\
 \hline
 \multicolumn{4}{l}{$\bullet$ Weak lensing mass calibration}\\
 $A_{\rm WL}$ & - & Scatter in the weak lensing bias from the first principal component & $\mathcal{N}(0, 1)$ \\
 $B_{\rm WL}$ & - & Scatter in the weak lensing bias from the second principal component & $\mathcal{N}(0, 1)$ \\
 $C_{\rm WL}$ & - & Standardize mass slope of the weak lensing bias & $\mathcal{N}(0, 1)$ \\
 $D_{\rm WL}$ & - & Redshift dependent intrinsic scatter in the weak lensing bias & $\mathcal{N}(0, 1)$ \\
 $\rho_{M_{\rm WL}, C_R}$ & - & Intrinsic correlation between weak lensing mass and count rate & $\mathcal{U}(-0.9, 0.9)$ \\ 
 \hline
 \multicolumn{4}{l}{$\bullet$ Richness mass calibration}\\
 $\log A_{\lambda}$ & - & Normalization of the $M - \lambda$ scaling relation & $\mathcal{U}(1.4, 6)$ \\
 $B_{\lambda}$ & - & Mass slope of the $M - \lambda$ scaling relation & $\mathcal{U}(0, 2)$ \\
 $C_{\lambda}$ & - & Redshift evolution of the normalization of the $M - \lambda$ scaling relation & $\mathcal{U}(-2, 2)$ \\
 $D_{\lambda}$ & - & Redshift evolution of the mass slope of the $M - \lambda$ scaling relation & $\mathcal{U}(-2, 2)$ \\
 $\sigma_{\lambda}$ & - & Intrinsic scatter of the $M - \lambda$ scaling relation & $\mathcal{U}(0.05, 2)$ \\
 $\rho_{\lambda, C_R}$ & - & Intrinsic correlation between richness and count rate & $\mathcal{U}(-0.9, 0.9)$ \\
 \hline
 \multicolumn{4}{l}{$\bullet$ Contamination modeling}\\
 $f_{\rm AGN}$ & - & Fraction of AGN contaminants in the extended source sample & $\mathcal{U}(0, 0.1)$ \\
 $f_{\rm RS}$ & - & Fraction of RS contaminants in the extended source sample & $\mathcal{U}(0, 0.15)$ \\
 \hline
 \multicolumn{4}{l}{$\bullet$ Redshift uncertainty}\\
 $\sigma_{z}$ & - & Relative error on the measured redshift & $\mathcal{TN}(0.0050, 0.0011, 0, 1)$ \\
 $b_{z}$ & - & Systematic bias in our redshift estimate & $\mathcal{N}(1.005, 0.037)$ \\
 $c_{z}$ & - & Fraction of objects for which we measure a shifted redshift & $\mathcal{TN}(0.0013, 0.0010, 0, 1)$ \\
 $c_{shift, z}$ & - & Amount of redshift shift for $c_{z}$ fraction of objects & $\mathcal{U}(-0.1, 0.1)$ \\
 \hline
 \hline
 \end{tabular}
 \tablefoot{With $\mathcal{U}({\rm min}, {\rm max})$ we indicate a uniform distribution between "min" and "max." With $\mathcal{N}(\mu, \sigma)$ we indicate a normal distribution centered on $\mu$ and with standard deviation $\sigma$. With $\mathcal{TN}(\mu, \sigma, {\rm min}, {\rm max})$ we indicate a truncated normal distribution centered on $\mu$ and with standard deviation $\sigma$ bounded between "min" and "max."}
\end{table*}

\section{Validation of the cosmology pipeline and the blinding strategy}
\label{sec:validation_and_blinding}

To ensure the results from the run of the cosmology pipeline on the \erosita cosmology subsample are robust against any additional unaccounted systematics, we extensively test our pipeline against the realistic simulations of the \erosita's digital twin \citep{Seppi2022}
and the mock observations that reflect the characteristics of our modeling. We perform our analysis blindly to avoid any unintentional unconscious bias toward the already published results in the literature on cosmology. 
Specifically, we investigate the influence of uncertainties on photometric redshifts, selection function, WL miscentering, other WL-related systematic effects, and contamination modeling on our results. This section describes in detail our testing of the cosmology pipeline, assessment of potential biases, and our blinding strategy.

\subsection{Mocks generation and validation of the pipeline}
\label{sec:validation}

As cosmology studies with cluster counts rely on specific modeling assumptions, such as modeling of the mass bias, selection function, and redshift uncertainty, we tested the robustness of the cosmological framework against the parameters used in this analysis. To verify the cosmology framework we made  the cosmology pipeline and theoretical modeling available to the collaboration members for an independent review of each component.

The validation of the cosmological framework is performed independently. We developed a new pipeline for generating mock catalogs following the physical formalism (described in Sect.~\ref{sec:likelihood} and Sect.~\ref{sec:scaling_relations}), without relying on the \erosita cosmology pipeline. We ensure that the observables generated by these mocks follow the distribution of the \erosita X-ray count rate and optical information from LS DR10-South, and weak lensing follow-up data sets from the DES, KiDS, and HSC surveys. 
Then the cosmology pipeline is tested against these mocks to make sure the pipeline reproduces the input parameters in the assumed theoretical framework.
This approach allows us to independently test the statistical robustness of all the contributions to the total likelihood, shown in Equation~\eqref{eq:Poisson_log_like_final}.

We generated mock cluster catalogs with multi-wavelength information using the dark matter halo mass function package \texttt{hmf} \citep{Murray2013}. In the mocks, we utilized halos with $M_\mathrm{500} > 10^{12} M_\odot$, and distributed them homogeneously across the celestial sphere to reflect the \erosita clusters sample. At the position of each halo, we assigned a local background \citep{Freyberg2020}, absorption from hydrogen \citep{HI4PI2016, Willingale2013}, and exposure time \citep{Predehl2021} by interpolating the corresponding maps consistent with the survey strategy. To compute the expected observables, such as the count rate and richness, we employed the scaling relations described in Sect.~\ref{sec:scaling_relations} with fiducial values from our fits. Observable quantities were then drawn according to the expected statistical distribution. As described in Sect.~\ref{sec:detection_probability}, we computed the detection probability for each halo and select edclusters accordingly based on \erosita's selection function. Finally, we tested the components of the global cosmology likelihood to ensure that systematics trends in the data are recovered well and estimated accurately by the pipeline.

The dark matter surface mass density was calculated by assuming a spherical NFW profile \citep{Navarro1996} for each cluster. We computed the mean expected shear by integrating this value over the background galaxy distribution from the respective survey. The observed shear was deduced by adding errors similar to those observed in the DES shear catalog\citep[see Sect.~\ref{sec:DES} and ][]{Grandis2024}. 
Out of these shear simulations, we selected only those that fall into the optical survey footprint of DES, HSC, and KiDS, while the corresponding shear profiles are assigned to these objects. 
Finally, by sampling their probability distribution, we included the contaminants, RSs and AGNs (see Sect.~\ref{sec:mixture}) and used the contamination fractions from the best fit.

We generated ten different mocks from the same model and used the cosmological pipeline to recover the input parameters, $\sigma_{8}$ and $\Omega_{\mathrm{m}}$. Most of the output parameter values fall well within our 68\% confidence level of the input values. We note that the rest of the parameters also fall well within our 68\% confidence level of the input values; in particular, we also recovered the input number of contamination level accurately in these tests. This validation step ensures that the components of the global likelihood, mass calibration, cluster counts, and contamination modeling work as expected free of additional systematics. As a result of this test, we demonstrate that the pipeline is free of bugs and passes the statistical robustness test successfully. 

To test the effects of the selection on our results, we applied realistic changes to our selection function. We generated six selection functions\footnote{We note that at this point, the "true" selection function is still unknown to us as part of our blinding strategy.} described in Sect.~\ref{sec:selfunc_blinding}. These models differ by the thresholds used to construct the simulation catalog. 
We then generated six mocks, using a different selection function each time, freezing all the other components in the likelihood. We then ran the cosmology pipeline with different selection functions to investigate the impact of the uncertainties on the selection function on our results. We find that the contours shift slightly but always within the 68\% percentiles of the input of the $\sigma_{8}$ and $\Omega_{\mathrm{m}}$ parameters.
 
Recent results by \citet{Sommer2023} have suggested that the underestimated miscentering measurements between ICM centers and halo centroids indicated by the numerical simulations may introduce an additional systematic bias on WL mass measurements by 2–6\%. However, the exact extent of the bias remains to be investigated and depends on the assumed X-ray miscentering in hydrodynamical simulations. Although the uncertainties in miscentering distribution calibrated on hydrodynamical simulations are fully accounted for in our WL shear measurements (see Sect.~\ref{sec:WLSclRel} and \citet{Grandis2024}), we further investigated the effect of an additional miscentering bias and its influence on our measurements by introducing a bias at the levels as suggested by \cite{Sommer2023} in Equation~\eqref{eq:WLbias_given_M_z}. We find that 
a decrease
in the masses as found by \citet{Sommer2023} indeed 
decreases the values of $\Omega_{\mathrm{m}}$ and $\sigma_8$ as anticipated; however, the best-fit values remain statistically consistent within the 68\% contour level of the input values for both parameters.

The redshifts in our catalog were measured using the photometric information from the LS DR10-South $grz$ bands; see Sect.~\ref{sec:catalog} \citep{Kluge2024}. This means that their uncertainties are significantly larger than spectroscopic redshift measurements. This potentially could cause a bias on the cosmological parameters mainly as a result of geometrical effects. This is because $\frac{dN}{dM dV}$ is a weak function of redshift, especially in the redshift interval of interest, but the volume $\frac{dV}{dz}$ and cosmological distances change drastically. We tested this effect for various parameters in Equation~\eqref{eq:zhat_given_z} when generating mocks. 
We find that the parameters that would result in significant bias in $\Omega_{\mathrm{m}}$ and $\sigma_8$ measurements are unrealistic.
For example, a systematic error of 10\% on photo-zs, namely, $\sigma_z \approx 0.1$, affects $\Omega_{\mathrm{m}}$ and $\sigma_8$ by about 3\% and 2\% respectively, and a trend of catastrophic failures on photo-z with occurrence of 10\% results in lower values of $\Omega_{\mathrm{m}}$ by 4\% and higher values of $\sigma_8$ by 2\%. 
The comparisons of the photometric measurements against spectroscopic redshifts in the redshift range 0.1 to 0.8, given in \citet{Kluge2024}, allows us to assess the actual values for the parameters in Equation~\eqref{eq:zhat_given_z}. 
The intrinsic scatter of the photometric redshift with respect to the spectroscopic redshift is $<0.5\%,$ on average, and the fraction of significant catastrophic failures is just 0.13\%; thus, it is much smaller than the extreme parameters required to introduce a significant bias. Therefore, the bias introduced by the use of photometric redshifts on $\Omega_{\mathrm{m}}$ and $\sigma_8$ parameters is negligible.

Given the positive results from the robustness tests and verification of the contamination modeling, assessment of the WL mass, and redshift bias on the cosmological constraints, we do not anticipate any significant bug or unaccounted bias in our pipeline and methodology. Therefore, we conclude that the cosmology pipeline and the methods have been validated successfully.

\subsection{Blinding strategy}
\label{sec:blinding}

Blinding the results prior to publication minimizes preconceived assumptions and prevents us from introducing bias into our analysis. After the validation phase was successfully completed on our simulations and mock data, we froze the cosmology pipeline and apply the following blinding strategy to ensure that our conclusions would remain as objective and unbiased as possible. We generated six different selection function models, as presented in Sect.~\ref{sec:selfunc_blinding}, only one corresponding to the true selection model, namely, with the \extlike thresholds that match the construction of the cosmology subsample. These selection functions were tested on the mocks described in Sect.~\ref{sec:validation} to ensure that they do not display any unphysical behavior that will lead to unintentional blinding. A total of six selection functions were then sent to our external blinding expert, Professor Catherine Heymans. The eROSITA-DE cluster working group received four of these six selection models after the blinding lead shuffled their indices. During this process, the ``true'' selection function model and its index became unknown to the working group and analysis teams. The cosmology analysis for the flat \lcdm model was carried out using the four independent selection functions simultaneously without any prior knowledge of the "true" selection function. This strategy resulted in four independent cosmology analyses and contours. A draft of the paper was prepared on the four distinct cosmology results from the \lcdm analysis and circulated for the working group's review. During this phase, we satisfied the tests described in the validation phase; the cosmology pipeline was extensively tested against any bugs or biases and completed the internal checks on the different components of the analysis (see Sect.~\ref{sec:validation}). We then unblinded the results and completed the rest of the draft of the paper, extending our analysis beyond the \lcdm model without introducing any modifications to the analysis method or results after unblinding.

\section{Scaling relations}
\label{sec:scaling_relations}

This section  describes our formalism for the scaling relations between mass and our observables accounting for the redshift evolution. This step is vital for constructing the mass function from the clusters detected in the \erass survey with X-ray count-rate, optical richness, redshift, and shear profile information. We also compare our findings and the reported best-fit values in the literature from similar studies.

\subsection{X-ray Count Rate - Mass Scaling Relation}

\label{sec:xrayscaling_relations}

\begin{table}[]
\caption{Best-fit parameters of the scaling relations between WL shear and X-ray count-rate and optical richness.}
 \centering
 \begin{tabular}{ l l ll }
 \hline\hline
 Scaling relation & Parameter & Best-fit value \\
 \hline
 $C_{R} - M$ & $A_{X}$ & $0.64^{+0.04}_{-0.06}$\\
 & $B_{X}$ & $1.38^{+0.03}_{-0.04}$\\
 & $F_{X}$ & $-0.33\pm 0.12$\\
 & $G_{X}$ & $0.29^{+0.12}_{-0.13}$\\
 & $\sigma_{X}$ & $0.98^{+0.05}_{-0.04}$\\
 $\lambda - M$ & $A_{\lambda}$ & $3.08\pm 0.05$\\
 & $B_{\lambda}$ & $1.01^{+0.02}_{-0.03}$\\
 & $C_{\lambda}$ & $-0.49^{+0.14}_{-0.15}$\\
 & $D_{\lambda}$ & $0.30\pm 0.11$\\
 & $\sigma_{\lambda}$ & $0.32^{+0.04}_{-0.03}$\\
 \hline\hline
 \end{tabular}
 \tablefoot{The reported uncertainties are the 68th percentiles. The posteriors of the best-fit parameters are shown in Fig.~\ref{fig:LCDM_scaling}.}
 \label{tab:scaling_fit}
\end{table}

\begin{figure*}
 \centering
 \includegraphics[width=0.9\textwidth]{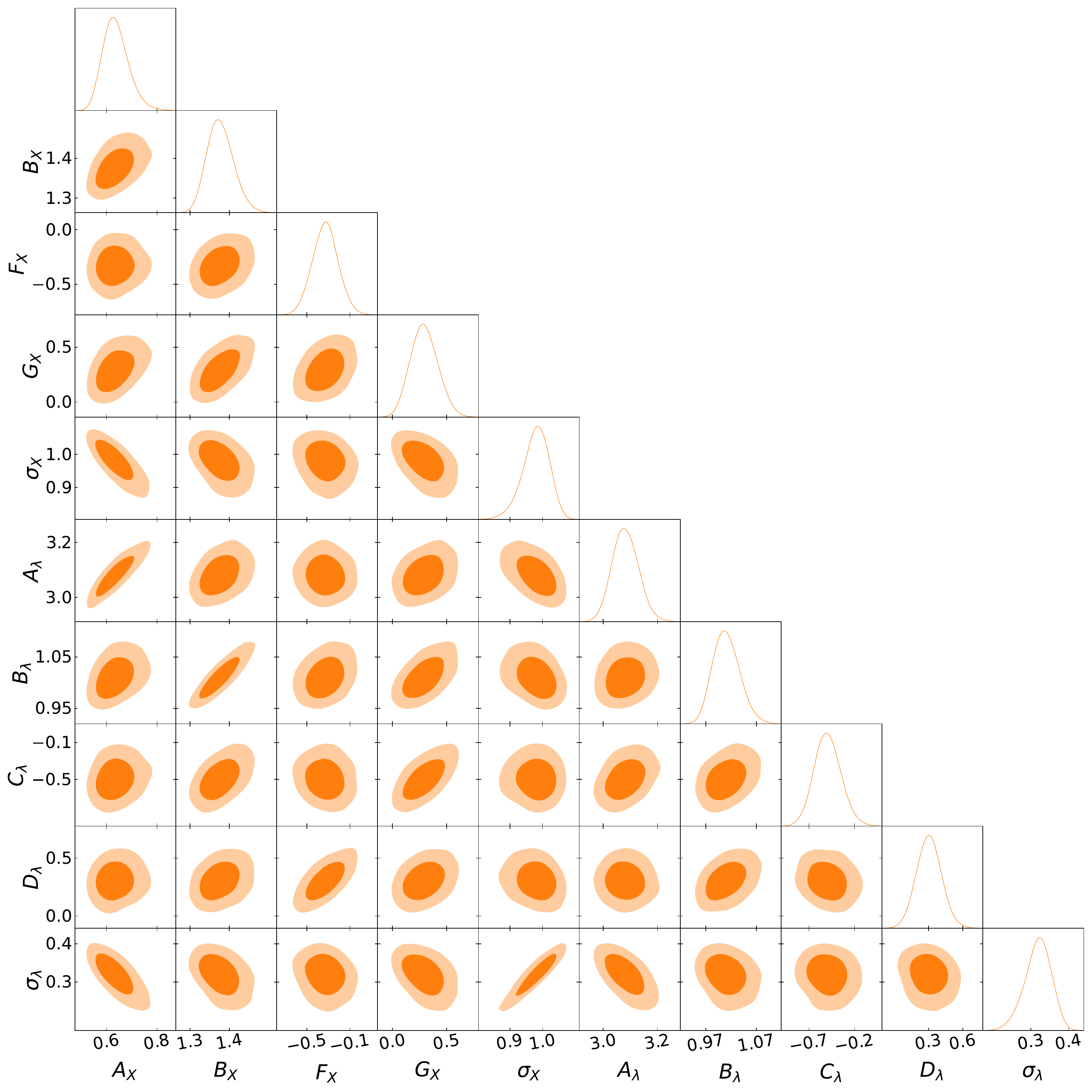}
 \caption{Posteriors on the $C_{R}-M$ and $\lambda-M$ scaling relation parameters. A strong correlation between X-ray and optical scaling relation is visible on the corner plots. The best-fit values and their uncertainties are provided in Table~\ref{tab:scaling_fit}. The solid and shaded orange contour represents the 68\% and 95\% contour levels respectively.}
 \label{fig:LCDM_scaling}
\end{figure*}

The general form of the scaling relation between the X-ray observable $C_R$ and cluster mass, required in Equation~\eqref{eq:CR_given_M_z}, can be described with the following equation:

\begin{equation}
\left\langle \log \frac{C_R}{C_{R,p}} \bigg| M, z \right\rangle = 
\log A_X + 
b_X(z) \, \log \frac{M}{M_p} + e_x(z)
\label{eq:xrayscaling}
.\end{equation}

\noindent We used the $C_{R,p} = 0.1$ cts~/~s as the pivot value for the count rate, $M_p = 2 \times 10 ^{14} M_\odot$, as the pivot value for the mass, and $z_p = 0.35$ as the pivot value for the redshift. The redshift dependent slope of the scaling relation is expressed by the term $b_X(z)$, and the redshift evolution of the scaling relation is expressed by $e_x(z)$.
Explicitly, the former is expressed as:

\begin{equation}
b_X(z) = B_X + F_X \, \log \frac{1+z}{1+z_p}. 
\end{equation}

\noindent Here, $B_X$ is the normalization, 
and $F_X$ allows the slope to be redshift dependent. The self-similar model of \citet{Kaiser1986} predicts the
$F_X$ slope to be zero. In this work, we  set 
the prior for $F_X$ parameter centered around zero to avoid any potential bias in our slopes. 
The redshift evolution term, $e_x(z)$, is given by:

\begin{equation}
e_x(z) = D_X \, \log \frac{d_L(z)}{d_L(z_p)} + E_X \, \log \frac{E(z)}{E(z_p)} + G_X \, \log \frac{1+z}{1+z_p}
\label{eq:e_xz}
,\end{equation}

\noindent for which $E_X = 2$ and $D_X=-2$ are the values adopted by the self-similar model \citep{Kaiser1986} for the case of luminosity; our \nh-corrected count rate has in principle the same dependence except for the K-correction, which is captured by the $G_X$ parameter. The last term, $G_X$, quantifies the deviation from the self-similar model in redshift evolution \citep[e.g., as used in ][]{Bulbul2019}. The intrinsic scatter of our scaling relations is represented by a single value, $\sigma_X$, independent of mass and redshift, 
similar to the studies in the literature on scaling relations from X-ray observations, \citep[for example,][]{Mantz2016, Schellenberger2017I, Lovisari2020, Lovisari2021, Duffy2022, Giles2022}, where the limited number of clusters and mass range does not allow to constrain mass or redshift trend for the scatter. 
This assumption is also adopted in the cosmological analyses utilizing galaxy cluster number counts \citep{Mantz2015, Schellenberger2017, Bocquet2019, Costanzi2021, Bocquet2024}.
Additionally, the X-ray luminosity's scatter as a function of mass has been studied for the mass range similar to the sample used in this work in IllustrisTNG numerical simulations of \citet{Pop2022}; see bottom panel of their Figure~6. The authors show that the scatter shows a subdominant mass trend. As we extend the mass and redshift range of the \erosita\ samples in the deeper surveys, the proper modeling of the intrinsic scatter in cosmology analyses  becomes important.
The best-fit values of the variables in the $C_{R}-M$ scaling relations are provided in Table~\ref{tab:scaling_fit}, and their posteriors are shown in Fig.~\ref{fig:LCDM_scaling}.

The only other works in the literature that use \erosita X-ray count-rate in mass scaling relations are the ones by \citet{Chiu2022, Chiu2023a} on the eFEDS sample. Our best-fit normalization parameter is $A_x = 0.64^{+0.04}_{-0.06}$. The reported normalization parameters are $A_x = 1.24^{+0.22}_{-0.19}$ and $A_x = 1.33^{+0.26}_{-0.20}$ in \citet{Chiu2022, Chiu2023a} in $\sim3\sigma$ tension with our results. In particular, the difference in normalization is due to the observed degeneracy with the scatter, see Fig.~\ref{fig:Ax_scatter} in Appendix~\ref{app:tests}, the selection of the sample \citep{Bahar2022}, and the selection function parameters. 
In fact, in \citet{Chiu2022, Chiu2023a}, the authors have significantly simplified the selection function, particularly in its sky variation. eFEDS covers an area of 140~deg$^2$, albeit much smaller in area compared to \erass, the eFEDS area is not small enough such that fundamental sky properties affecting the X-ray selection function can be safely ignored \citep{Brunner2022}.

We find that the slope of the power-law index of the mass term is $B_x = 1.38^{+0.03}_{-0.04}$. Overall, in terms of mass trends, we find a steeper slope than the self-similar mass trend ($B_{x,ss} = 1$), while many scaling relation studies in the literature report even steeper slopes. We find shallower slope than the reported value of $B_x = 1.58^{+0.17}_{-0.14}$ in \cite{Chiu2022} and that of $B_x = 1.86^{+0.20}_{-0.15}$ in the count rate mass relations provided in \cite{Chiu2023a}. 
The discrepancy is likely due to different treatments of the selection function between these two sets of results. In this work, we accurately model our selection with the one-to-one simulations of the \erosita X-ray sky, while in \cite{Chiu2023a}, an empirical model for the selection is used. In principle, as the X-ray count rate is closely related to the X-ray luminosity, the mass slope of the $L_{X}-M$ scaling relations reported in the literature can be compared to our slope measurements. The slopes reported in the literature range between 1.3 and 1.9. The studies resulting in shallow slopes employ X-ray hydrostatic masses $B_{L_x} = 1.45 \pm 0.07$ in \citet{Maughan2007}, $1.61 \pm 0.14$ in \citet{Vikhlinin2009}, $ 1.62 \pm 0.12$ in \citet{Pratt2009}, $1.34 \pm 0.18$ in \citet{Eckmiller2011}, $1.66 \pm 0.22$ in \citet{Lovisari2015},
and $1.35 \pm 0.07$ in \citet{Schellenberger2017}. Our slopes are consistent with these findings at the $1\sigma$ level. Higher slopes are reported in SZ-selected massive cluster samples covering a wider redshift range \citep[$B_{L_x} = 1.89 \pm 0.18$,][]{Bulbul2019}. When the authors excised the core, they found shallower slopes $1.60^{+0.15}_{-0.16}$, consistent with our finding within uncertainties. Another sample with a higher reported mass slope is the CODEX with the best-fit value of $2.01^{+0.09}_{-0.09}$ \citep{Capasso2020}.

The best-fitting scatter $\sigma_{X}$, $0.98^{+0.05}_{-0.04}$, in our sample is larger than the reported values from $C_{R}-M$ or $L_{X}$-M scaling in the literature, e.g., 0.17 and 0.25 in the X-ray selected REXCESS \citep{Pratt2009}, HIFLUGCS samples \citep{Lovisari2015}, and 0.27 in the SZ selected samples \citep{Bulbul2019}. The log-normal scatter of the soft-band luminosity is constrained to be 0.12 in \citet{Chiu2022} for eFEDS clusters. We also note that the scatter in the C-Eagle cosmological hydrodynamical
simulations of clusters is 0.30, which is also smaller than our measurements \citep{Barnes2017}. This is partially because our mass range and redshift are more extensive than those employed in previous studies. We do not apply any cuts on the cluster morphology and do not excise core emissions. This strategy might lead to a larger spread than previously observed in scaling relations. However, we note that this large scatter has a negligible impact on the accuracy of the cosmological parameters (see Sect.~\ref{sec:results}).

\subsection{Weak-lensing inferred mass-mass scaling relation}
\label{sec:WLSclRel}

Measuring the tangential shear of galaxies lying behind galaxy clusters is a powerful tool for estimating the cluster mass. 
As with all mass proxies, the weak-lensing inferred mass is also a biased mass proxy; however, the factors contributing to the bias are under control and can be accurately calibrated using simulations with knowledge of the survey-specific background source distribution. 
The general form of the scaling relation that relates the weak-lensing inferred mass with the true mass \citep[see][]{Grandis2021a} required in Equation~\eqref{eq:Mwl_given_M_z}, is described by the following equation:

\begin{equation}
\left\langle \log \frac{M_{\rm WL}}{M_p} \bigg| M, z \right\rangle = b(z) + b_{M} \log \left( \frac{M}{M_p} \right)
\label{eq:WLbias_given_M_z}
.\end{equation}

\noindent Here, $b(z)$ is the redshift-dependent evolution of the weak-lensing bias as a function of redshift. It is computed by interpolating simulations snapshots at redshifts: $0.24$, $0.42$, $0.68$, and $0.95$.

The calibration approach is first presented in \citet{Grandis2021a}, and consists of Monte Carlo simulations, where each single iteration consists of three steps:
1) starting from surface mass density maps of massive halos from the cosmological hydro-dynamical TNG300 simulations \citep{Pillepich2018s, Marinacci2018, Springel2018, Nelson2018, Naiman2018, Nelson2019}, and the properties of the cluster sample and the WL surveys, we predict a catalog with realistic synthetic shear profiles;
2) we use the same expression $P(\hat{g}_t | M_{\rm WL}, \hat{z})$ described in Sect.~\ref{sec:wl--mass-cali}, which depends on the dataset used, in our case DES, KiDS, and HSC, evaluated on the synthetic shear profiles, to assign WL masses in the simulation, and
3) we finally fit the relation given above between WL mass and halo mass, recovering a WL bias values $b_l$ for each snapshot, and a global mass trend $b_\text{M}$.

As introduced in \citet{Grandis2021a}, the true novelty of this approach is that the simulation above can be repeated order 1000 times while varying the inputs within their uncertainties, thus effectively propagating all WL related systematic uncertainties to a distribution on the WL bias values $b_l$ and $b_\text{M}$. To compress this information and ingest it into the cosmological pipeline as priors, we derive the mean WL bias $\mu_l$. The scatter of the Monte Carlo realizations around this mean is well captured by two principal components $PC_{1,2,l}$ \citep{Grandis2024}. For the mass trend we report mean $\mu_{b_{M,}}$ and standard deviation $\sigma_{b_{M, i}}$. We similarly fit the scatter in the WL mass, finding that its redshift trend is well described by one principal component.

Since we use the weak lensing shear in our mass calibration from the three independent surveys, DES, KiDS, HSC, we calibrate the count rate to mass scaling relations jointly. We note that the parameters $b(z)$ and $b_M$ are survey dependent since the specific properties of each survey affect the reconstructed weak lensing mass slightly differently, see \citet{Grandis2024} and \citet{Kleinebreil2024}. However, since only the mean relation and uncertainty around it are survey-dependent, we express $b(z)$ using the set of parameters $\mu_{b, i}(z)$, $PC_{1, i}(z)$, $PC_{2, i}(z)$, $\mu_{b_{M, i}}$, and $\sigma_{b_{M, i}}$ with $i \in \{ \textrm{DES, KiDS, HSC} \}$. To emulate the scatter in the WL bias simulations, we introduce three additional random normal parameters, $A_{\rm WL}$, $B_{\rm WL}$, and $C_{\rm WL}$, which are independent of the surveys employed. In practice, $b(z)$ and $b_M$ can be expressed as:

\begin{align}
b(z) & = \mu_{b, i}(z) + A_{\rm WL} \, PC_{1, i}(z) + B_{\rm WL} \, PC_{2, i}(z), \nonumber \\
b_M & = \mu_{b_{M, i}} + C_{\rm WL} \, \sigma_{b_{M, i}},
\end{align}

\noindent where $\mu_{b}, i (z)$, $PC_{1, i}(z)$, and $PC_{2, i}(z)$, the survey-dependent average relation. The two principal components, $PC_{1, i}(z)$, and $PC_{2, i}(z)$, are the values computed in the simulations used for interpolation at the specific cluster redshift. For $b_M$, no interpolation is required, but written in this form allows us to have only one free parameter for the three surveys, as the three surveys have slightly different mass trends in their WL biases.

As mentioned above, in the same simulation setup, we can also derive the redshift and mass trend of the scatter in WL mass, which is given by:

\begin{equation}
\log \sigma_{M_{\rm WL}}^2 = s(z) + s_{M} \log \left( \frac{M}{M_p} \right)
\label{eq:EWL}
,\end{equation}

\noindent where the simulations have shown that the mass component of the scatter is consistent with zero; we fix $s_{M} = 0$ in further analysis. For the redshift evolution of the scatter, we follow a similar procedure for the redshift evolution of the bias $b(z)$. As before, we take the values at simulation snapshots and interpolate. We use the survey-specific mean value and principal component $\mu_{s, i}(z)$ and $PC_i(z)$ and a random normal parameter $D_{\rm WL}$ and express $s(z)$ as:

\begin{equation}
s(z) = \mu_{s, i}(z) + D_{\rm WL} \, PC_i(z)
,\end{equation}

\noindent where the average scatter for each survey and principal component ($\mu_{s, i}(z)$ and $PC_i(z)$) are computed in the simulation snapshots and used in interpolation. This formalism allows distinguishing the average survey-dependent components from the simulation-dependent scatter around them, dominated by simulation uncertainty. It leaves us with four free parameters related to WL to fully describe the systematic and statistical uncertainty of the bias.

This method effectively assumes that the WL mass measurement systematics are correlated between the three surveys. This is true for the low redshift cases, where the hydrodynamical modeling uncertainties represent almost the entirety of the systematic uncertainty on the estimated weak lensing masses. At higher redshifts, each survey is dominated by a different systematic: DES by the photometric source redshift uncertainty \citep[see][Sect.~4.5]{Grandis2024}, HSC by the shape measurement uncertainty \citep[see][]{Kleinebreil2024}, and KiDS by the cluster member contamination model \citep[see][Appendix A]{Kleinebreil2024}. All these systematics might be partially correlated. Assuming that the WL mass measurement systematics are correlated effectively increases the systematics error budget. Thus, our analysis strategy provides a conservative upper bound on the systematics.

\subsection{Richness-mass scaling relation}
\label{sec:opticalscaling}

Although the X-ray count rate is our main observable, richness measurements of the detected sources are required to distinguish the contamination (from an AGN or RS) from the cluster number counts. 
Similarly to previous studies in the literature, optical confirmation and richness determination are useful for assessing the contamination level of the sample and improving its purity to produce state-of-the-art cosmological analysis \citep{Bocquet2024}. In \citet{Kluge2024}, we develop a concept of a redshift-dependent richness cut by combining optical and X-ray properties of the ICM to statistically model both clusters and contamination simultaneously. This method comes at the cost of introducing richness as another observable but with the benefit of self-consistent modeling of clusters and contaminants in the sample.
In the likelihood term in Equation~\eqref{eq:Poisson_log_like_final}, the relation between richness and count rate is explicitly expressed in the third term. We modify the optical richness to mass scaling relation in \citet{Saro2015, Bleem2020, Grandis2020, Grandis2021b}, which we require as Equation~\eqref{eq:rich_given_M_z}, as follows:

\begin{equation}
 \left\langle \log \lambda | M, z \right\rangle = \log A_{\lambda} + b_\lambda(z) \, \log \left( \frac{M}{M_p} \right) + C_{\lambda} \, \log \left( \frac{1+z}{1+z_{p}} \right)
 \label{eq:richness_to_mass}
,\end{equation}

\noindent while the slope of the cluster mass term $b_\lambda(z)$ can be expanded as:

\begin{equation}
 b_\lambda(z) = B_{\lambda} + D_{\lambda} \, \log \left( \frac{1+z}{1+z_{p}} \right)
,\end{equation}

\noindent where $A_{\lambda}$ is the normalization of the scaling relation, $B_{\lambda}$ is the mass slope, $C_{\lambda}$ is the redshift evolution, and $D_{\lambda}$ is the redshift evolution of the mass slope. 

In this formulation, marginalizing over mass allows us to account for the correlation between count rate and richness and their uncertainties. It is crucial to calibrate the scaling relation between richness and count rate well to achieve precise and accurate cosmological results. However, we expect from our mathematical framework that the best-fit parameters of the mass-to-richness scaling relation and the parameters of the mass-to-count rate scaling relation degenerate with each other.

The best-fit parameters of the richness-M scaling relations are provided in Table~\ref{tab:scaling_fit}, and their posteriors are shown in Fig.~\ref{fig:LCDM_scaling}. In the case of the richness-to-mass scaling relation, Equation~\eqref{eq:richness_to_mass}, the best-fit value of the mass slope, $B_{\lambda} = 1.01^{+0.02}_{-0.03}$, is consistent with the majority of reported findings in the literature within $1\sigma$ level, e.g., 
$B_\lambda$ = $0.99^{+0.06}_{-0.07} \pm 0.04$ and $B_\lambda$ = $1.07^{+0.22}_{-0.20}$ in the CODEX sample \citep[][respectively]{Capasso2019, IderChitham2020}, 
$B_\lambda$ = $1.02\pm0.08$ in SPT sample \citep{Bleem2020},
$B_\lambda$ = $1.028^{+0.043}_{-0.037} \pm 0.04$ and $B_\lambda$ = $1.03 \pm 0.10$ in the DES Y1 and SPT samples \citet{Costanzi2021, Grandis2021b}, $B_\lambda$ = $1.05^{+0.13}_{-0.12}$ in \citet{Chiu2023a} in the eFEDS sample within the stated uncertainities. Our measurements are 8$\sigma$ away from the value reported by the DES collaboration on Y1 cosmology analysis, where they find $B_\lambda$ = $0.737 \pm 0.028 \pm 0.004$ \citep{McClintock2019, Abbott2020}\footnote{In the paper they reported the inverse of this number: $\frac{1}{B_\lambda} = 1.356 \pm 0.051 \pm 0.008$.}. The tension in the richness slope is identified as the driving factor for the discrepancy between cosmological results obtained from DES Y1 WL calibrated cluster number counts and cosmic shear and galaxy auto- and cross-correlations of the same data \citep{Abbott2020}.

We find a sample lower scatter in the $\lambda-M$ scaling relation of $0.3\pm0.03$ compared to the relation between $C_{R}-M$. This should somehow not mislead since, as mentioned, all the parameters of the mass-to-richness and the mass-to-count scaling relations are degenerate, see Fig.~\ref{fig:LCDM_scaling}. Furthermore, it is reported that the scatter of the mass-to-richness scaling relation is better behaved in an X-ray-selected sample than in an optically-selected sample \citep{Saro2015, Bleem2020, Grandis2021b, Giles2022}.

The redshift evolution term has a best-fit value of $C_{\lambda}=-0.49^{+0.14}_{-0.15}$ shows a slight redshift trend in our sample. The best-fit value is consistent with the results from the CODEX sample for a sample with spectroscopic member galaxies $\lambda>3$ \citep[$-0.26^{+0.23}_{-0.24}$;][]{Capasso2019}; however, their results are consistent with no-redshift evolution, while ours indicate a redshift evolution at 3$\sigma$ level. In the same CODEX sample, the negative trend in redshift evolution becomes stronger as the authors change the sample with clusters with a larger number of member galaxies, for instance, $\lambda>20$ \citep[$-1.00^{+0.49}_{-0.56}$;][]{Capasso2019}. In the DES sample\citep{McClintock2019}, the evolution trend is consistent with our measurements within the uncertainties. In their sample, no significant redshift evolution was observed $-0.30 \pm0.30\pm 0.06$ for a sample with richness cut of $\lambda>40$. In contrast, the DES/SPT analyses reported in \citet{Costanzi2021} $0.95\pm0.3$ is in tension with our results at the $>5\sigma$ level when the SPT number counts are included in the analysis, while consistent with our finding at $<2\sigma$ level, $-0.02\pm0.34$, when only SPT observable-to-mass relation is used \citep{Costanzi2021}. 
The reported redshift trends or slopes on the richness-mass relations depend on the cuts applied to the optically selected samples because of the rapid change in contamination fraction as a function of richness \citep{Abbott2020}. Our sample relies mostly on X-ray selection, which, after optical cleaning, remains very little contaminated; further study is needed to understand better the effect of different selection effects on the best-fit values of this relation and their comparisons. 

We note that the cluster richness is a sum of the color-based membership probability of the galaxies in the field \citep{Rykoff2014}. In \citet{Kluge2024}, we show that the \eromapper returns different richness measurements because of the different luminosity thresholds adopted for the combination in the different optical filter bands, for instance, $grz$, $griz$, $grzw1$, due to variations in membership probabilities. Despite these differences, \citet{Kluge2024} demonstrate that there is a simple constant scaling factor between these richnesses, for example, $\lambda_{grz} = S_{\rm norm} \lambda_{griz} = S_{\rm norm}^\prime \lambda_{grzw1}$, without any obvious dependence on redshift. This result is also supported by the reported comparison of the richness measured independently in the literature. For the \erass\ clusters detected in the common area of the DES~Y1 survey \citep{Abbott2020}, the comparisons between the richness of \erass1\ clusters measured in the $grz$ band and of DES~Y1 clusters in the $grizy$ band shows a scaling factor that has no dependency on redshift; as such the observed relation is $\lambda_{grz} = 0.79 \lambda_{\rm DES~Y1}$ \citet[see][for further details]{Kluge2024}. Therefore, this demonstrates that all the comparisons of richness to mass scaling relations we have performed above are reasonable because richness scaling affects only the normalization of the scaling relations, not the slope or redshift dependence.

\section{Cosmological constraints}
\label{sec:results}

This section discusses the cosmology parameters constrained by \erosita\ cluster abundance measurements compared to and combined with data from other cluster surveys and cosmological probes. In addition, our pipeline allows us to test different cosmological models beyond the \lcdm concordance model. This current state-of-the-art cosmological model describes the formation and evolution of our Universe. In its name, \lcdm, indicates the two primary ingredients of this cosmological model; $\Lambda$ implies the dark energy introduced after the discovery of the accelerated expansion of the Universe \citep{Riess1998, Perlmutter1998}, while CDM stands for cold dark matter. When combined, the dark sector comprises 95\% of the energy density of the Universe \citep{Planck2020}. In the \lcdm framework, the parameters that describe the Universe are the Hubble constant $H_0$, the total matter density $\Omega_{\mathrm{m}}$ (including dark matter and baryons), the spectral index of the primordial power spectrum $n_s$, the baryon density $\Omega_{\mathrm{b}}$, the amplitude of the matter fluctuations $\sigma_8$, and the reionization optical depth $\tau$. The other cosmological parameters can be derived from this baseline set of parameters, for instance, the age of the Universem $t_0$, the critical density of the Universem $\rho_c$, and a combination of $\Omega_{\mathrm{m}}$ with $\sigma_8$ along the principal direction of the degeneracy between the two, $S_8$. In our baseline model, the flat \lcdm, we fit for $\Omega_{\mathrm{m}}$, $\sigma_8$\footnote{To be more precise, in the code, we do not fit $\sigma_8$ but $A_s$, the normalization of the power spectrum, and $\sigma_8$ is a derived from the constraints on $A_s$, see Appendix~\ref{app:sigma8}.}, $H_0$, $n_s$, and $\Omega_{\mathrm{b}}$.

In the fits providing the constraints beyond the \lcdm model, where the dark energy equation of state or the summed neutrino masses are allowed to vary, namely, are not fixed to $-1$ and 0 as in the \lcdm model. In the $w$CDM cosmology, we allow the dark energy equation of state to be free and leave summed neutrino masses fixed to 0~eV. In the $\nu$CDM cosmology, we allow summed neutrino masses to be free and leave the dark energy equation of state fixed to $-1$. We also provide joint constraints on the equation of the state of dark energy and the sum of neutrino masses in the flat $\nu w$CDM cosmological model.

\begin{figure}[h!]
 \centering
 \includegraphics[width=0.47\textwidth]{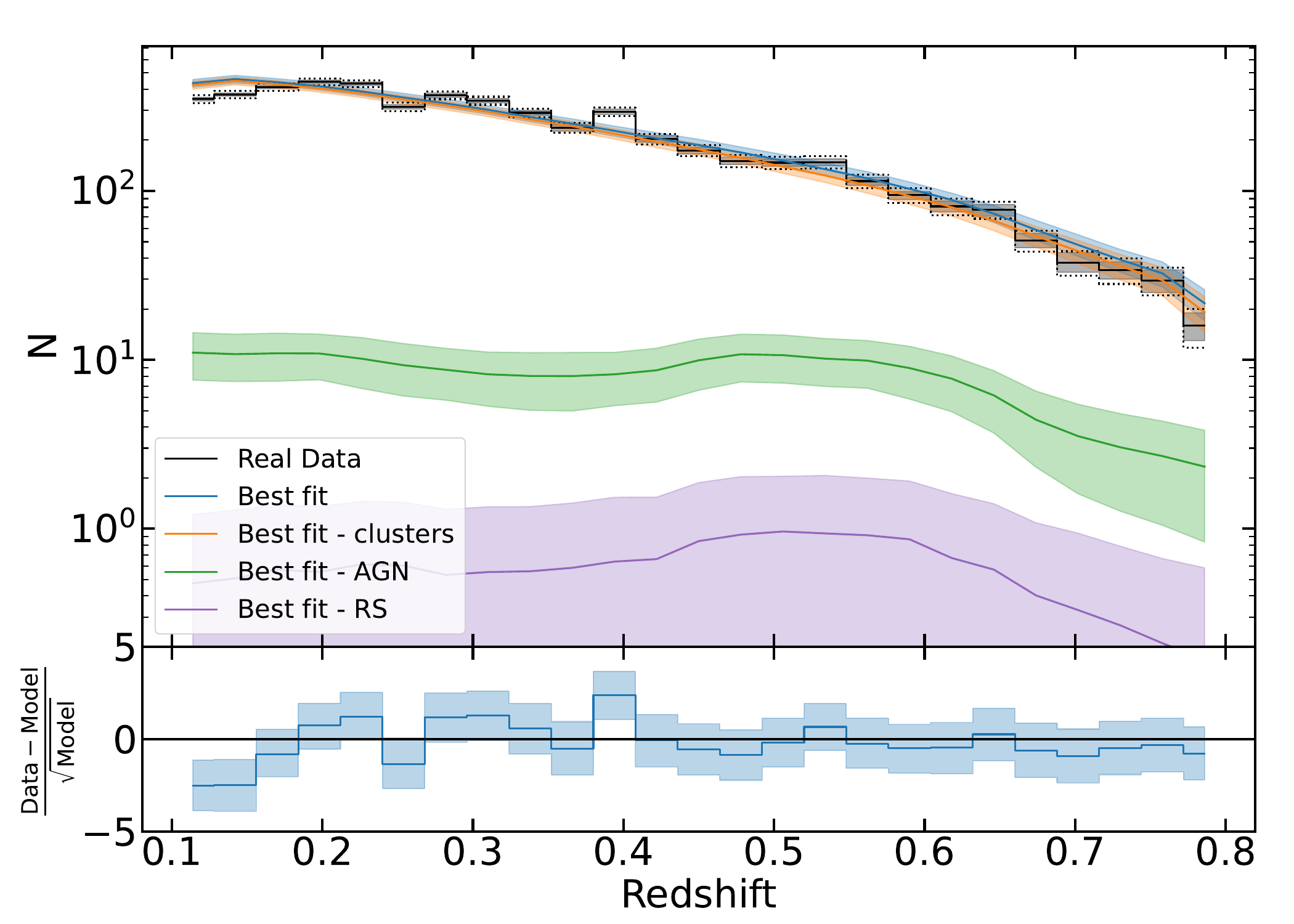}
 
 \includegraphics[width=0.47\textwidth]{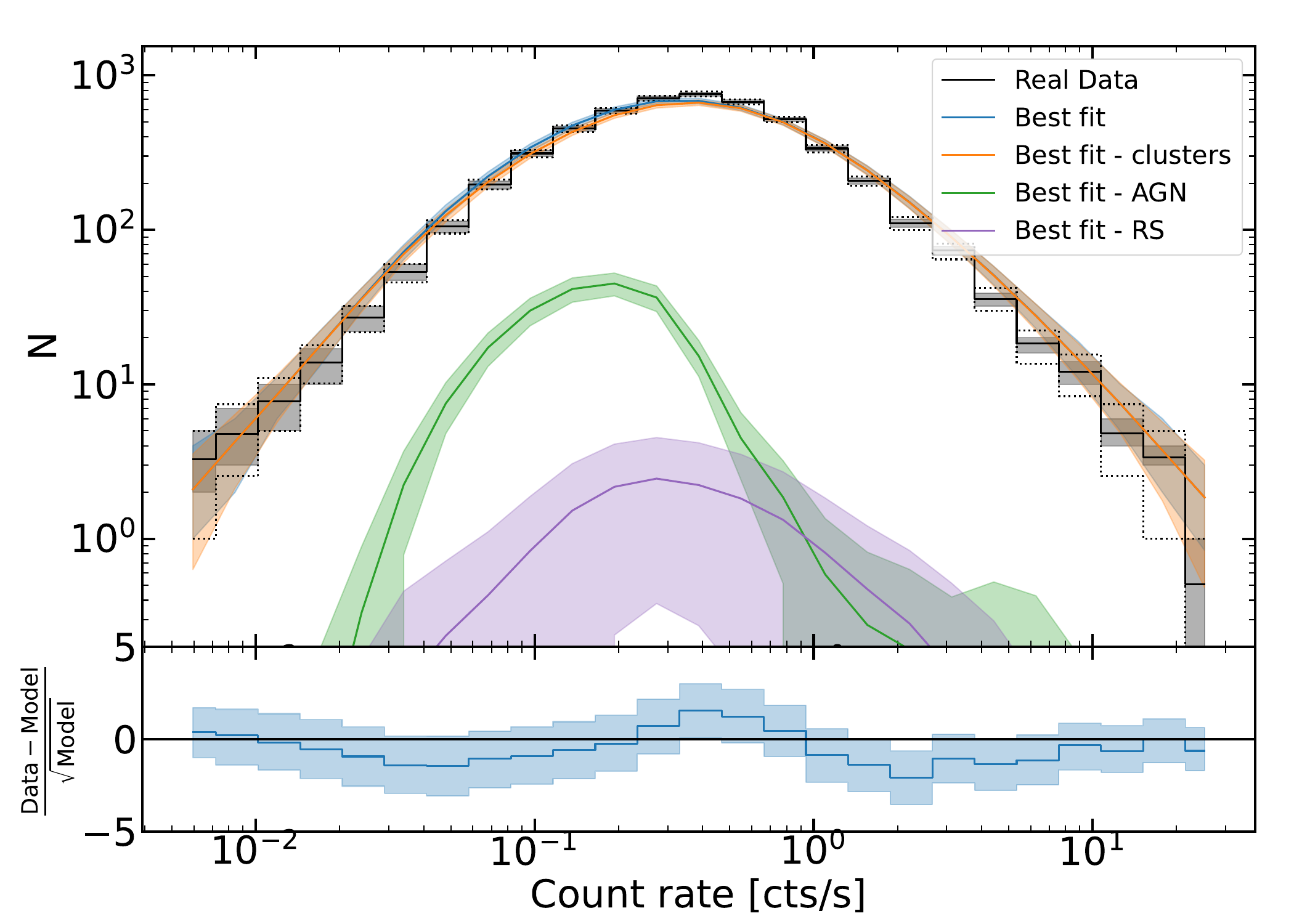}
 
 \includegraphics[width=0.47\textwidth]{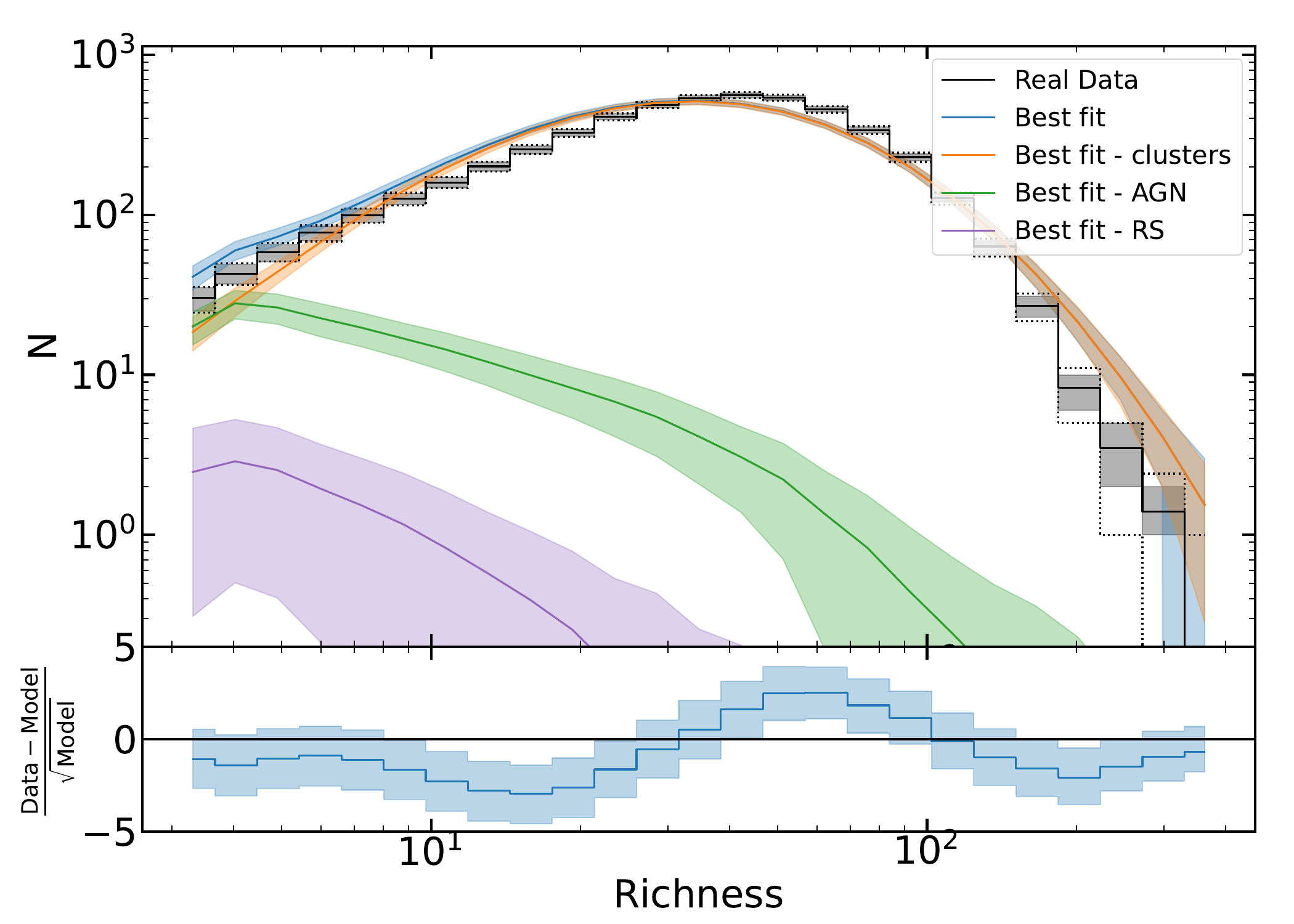}
 
 \caption{Distribution of clusters (black) as a function of redshift (top panel), count rate (middle panel), and richness (bottom panel) compared with the best fitting model prediction of the same quantities (blue) and the contribution to the total model from clusters (orange), AGN contamination (green), and RS contamination (purple). The black solid lines represent the data distribution, the shaded grey area the uncertainty due to measurement errors, and the black dotted lines indicate the Poisson uncertainty on the number of objects in each bin.
 In all figures, the bottom panel shows the Poisson Pearson residuals and the 2$\sigma$ contours
 of the best-fit model. 
 } 
 \label{fig:goodness_of_fit}
\end{figure}
\begin{figure*}
 \centering
 \includegraphics[width=\textwidth]{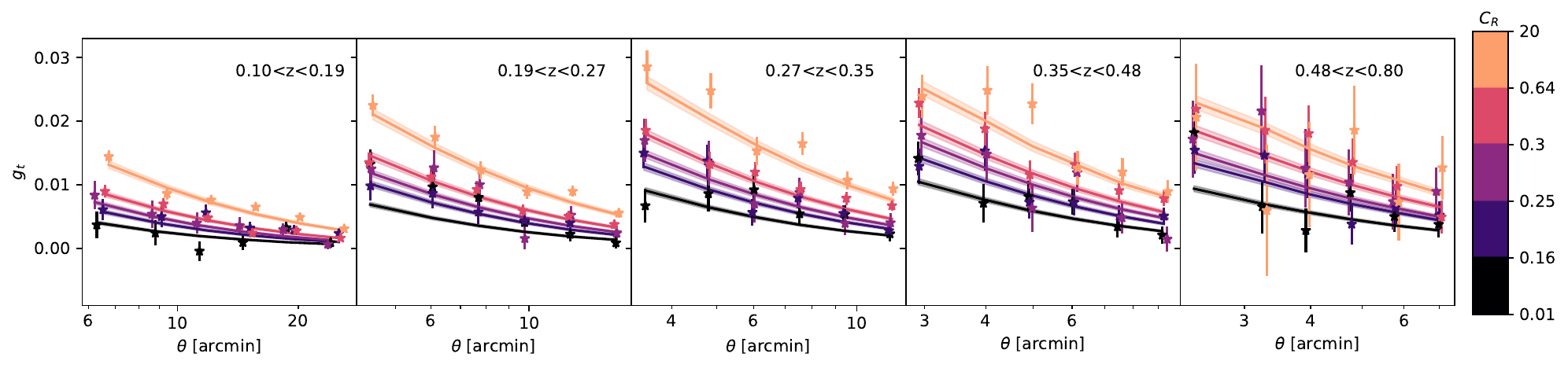}

 \includegraphics[width=\textwidth]{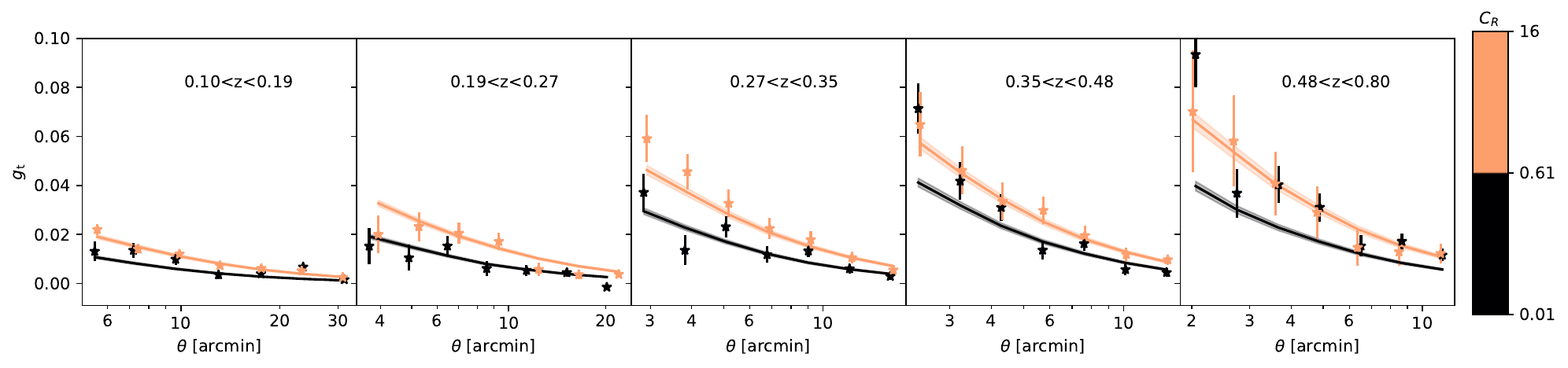}

 \includegraphics[width=0.585\textwidth]{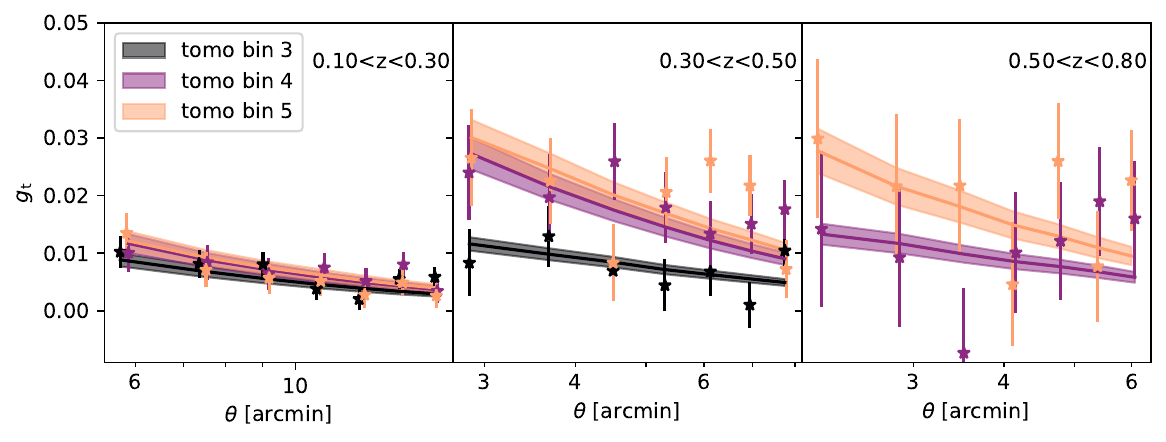}
 
 \caption{Goodness of fit for the three WL surveys, DES (top panel), HSC (middle panel), and KiDS (bottom panel). Each panel shows a stacked observed profile in the count rate and redshift bin (stars), compared with the best-fitting weak lensing shear profiles according to mass calibration, i.e., the weak lensing masses based on the count rates of the clusters and calibrated scaling relations. The color bar indicates the count rate binning applied to DES and HSC, while tomographic bins for KiDS. For a more extensive description of the weak lensing goodness of fit, see \citet{Grandis2024}, and \citet{Kleinebreil2024}.}
 \label{fig:shear_fit_goodness}
\end{figure*}
\subsection{Goodness of fit}

In principle, determining the goodness of fit of the various cosmology models is not a trivial process in our analysis. The cosmology likelihood is not a simple distribution but comprises various components following different distributions. For instance, the component for the cluster number counts in the global likelihood (Sect.~\ref{sec:likelihood}) should follow a Poisson distribution, while the likelihoods of the follow-up observables, such as weak lensing shear and optical richness, follow a log-normal distribution. Adding the likelihood for the mixture model complicates the process even further. 

To assess the goodness of fit of the standard \lcdm cosmological model to the cluster abundance, richness, and count-rate distributions, we performed 1000 weighted\footnote{We use the \texttt{ultranest}\citep{Buchner2021} provided weights to ensure that the points sampled from the posterior distribution provide a non-negligible weight to the sampled posterior} realizations from best-fitting parameters extracted from the posterior distribution and calculate the $\chi^{2}$ value. This simulation step is identical to the generation of the mock observations described in the validation phase; see Sect.~\ref{sec:validation} for further details. Overall, the total simulated number of objects in each bin of the observables is relevant in assessing the goodness of fit. In Fig.~\ref{fig:goodness_of_fit}, we provide the distribution of the observables employed in the analysis: count rate, redshift, and richness, and the corresponding best-fit models. Our methodology also allows us to simultaneously distinguish the relative contributions of the clusters and the contamination (AGN and RS) to the overall total model. For illustration purposes, the Pearson residuals are widely used in assessing the goodness of fit in the fits of the data with Poissonian nature statistics, which is also shown in Fig.~\ref{fig:goodness_of_fit}. As evident from the figure, the total model shown in the shaded region in orange produces a good fit to the data with $\chi^2=30.3$ for the redshift distribution, $\chi^2=24.5$ for the count rate distribution, and $\chi^2=71.5$ for richness distribution, all three with 26 degrees of freedom (d.o.f.).

The best-fit shear model derived from the posterior of the scaling relation between count rate and cluster mass is shown on the observed stacked shear data from the DES (top panel), HSC (middle panel), and the KiDS (bottom panel) of Fig.~\ref{fig:shear_fit_goodness}. The stacking procedure is described in detail in Sect.~5.2 of \citet{Grandis2024} for DES, and Sect.~4.3 of \citet{Kleinebreil2024} for KiDS and HSC. We find $\chi^{2}$ values of $\chi^{2}_\text{DES} = 173.2^{+43.2}_{-27.6}$ (150 d.o.f.), $\chi^{2}_\text{HSC} = 140.0^{+3.2}_{-6.2}$ (70 d.o.f.), and $\chi^{2}_\text{KiDS} = 60.2_{-5.6}^{+9.7}$ (63 d.o.f) for fits to the shear data from DES, HSC, and KiDS data, respectively. Note that the upper and lower limits in the $\chi^2$ distributions are obtained from the higher 1$\sigma$ (upper edges of the shaded region in Fig.~\ref{fig:shear_fit_goodness}) and the lower 1$\sigma$ (lower edge of the shaded region in Fig.~\ref{fig:shear_fit_goodness}) of the predicted mean shear profile, respectively.
We note that HSC does not yield a good fit to our data. We note from the middle panel in Fig.~\ref{fig:shear_fit_goodness}) that this is mostly due to the highest redshift low count rate bins. To assess the robustness of our results and conclusions below, we have checked that running the analysis without the inclusion of HSC does not affect our results.
We note that we find good agreement between shear measurements and WL-only mass calibration of \erass\ clusters in the lensing analyses from three surveys we utilize \citep[][Sects.~4.1, and 4.2]{Kleinebreil2024}.
In short, we show that HSC, DES, and KiDS WL are consistent both at the shear level and at the WL-calibrated scaling relation level \citep{Kleinebreil2024}.

Our strategy in analyzing cluster number counts involves utilizing weak lensing tangential shear information that allows us to simultaneously fit and constrain cosmological parameters and the parameters of the scaling law between cluster mass and observables. We perform extensive checks on the influence of the individual WL survey measurements in our analysis to ensure that these surveys do not introduce a significant bias in the scaling relations or fit for the cluster number counts presented in the companion papers; \citet{Grandis2024} and \citet{Kleinebreil2024}. To this end, the scaling relation between count rate to mass is tested on the individual weak lensing datasets by setting the first two terms in the likelihood, in Equation~\eqref{eq:Poisson_log_like_final}, to zero and fitting only the last two terms. 
We find that the overall model is consistent between the three surveys, and the scaling relation has enough degrees of freedom (see Table~\ref{tab:scaling_fit} for free parameters) to be essentially fixed by the mass calibration. 
Furthermore, we highlight that using three independent WL surveys simultaneously safeguards us against possible hidden systematics from any individual WL survey.

\subsection{Constraints on the \lcdm cosmology}
\begin{figure*}
 \sidecaption
 \includegraphics[width=12cm]{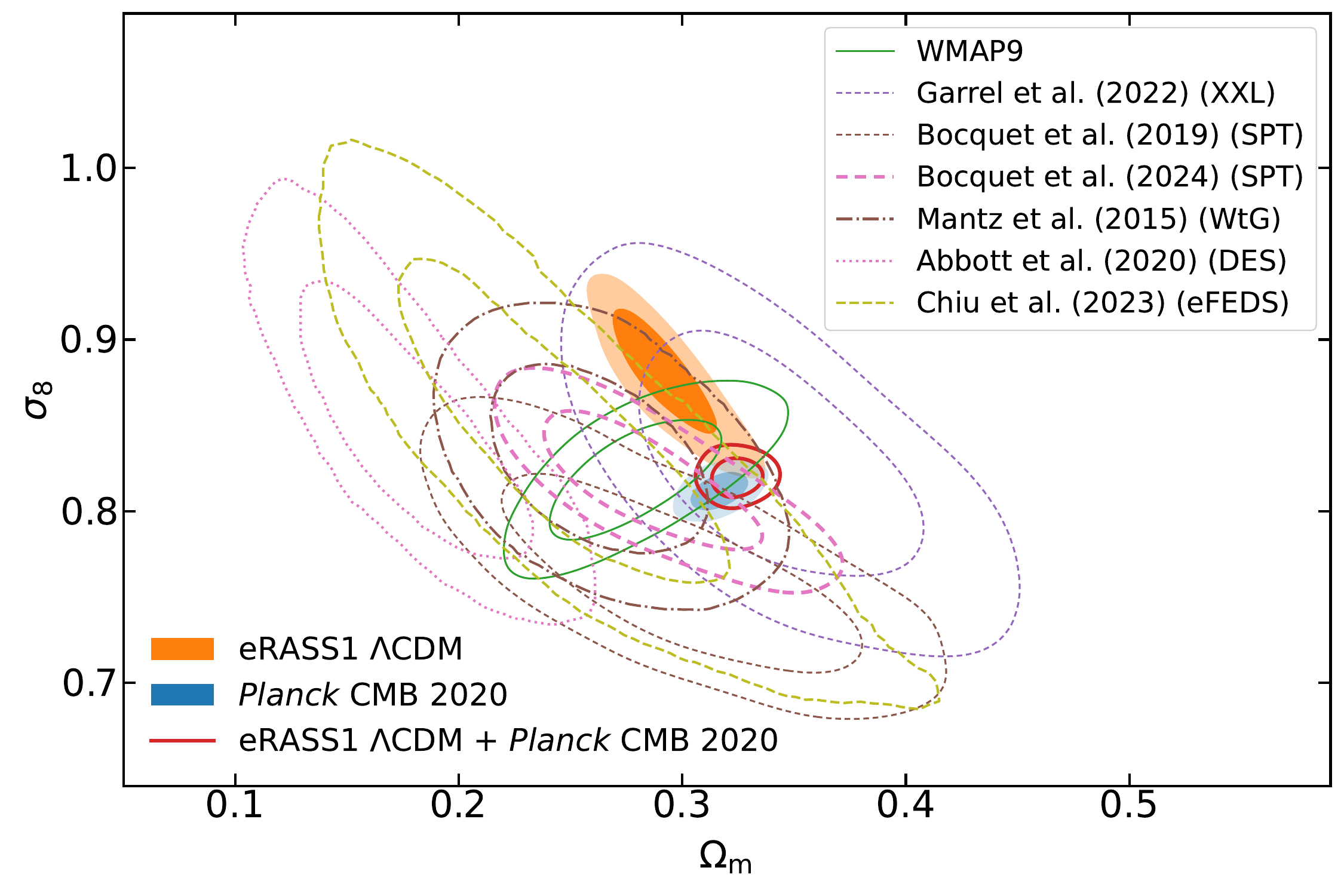}
 \caption{Posterior distributions on the parameters $\Omega_{\mathrm{m}}$ and $\sigma_8$ from the \lcdm fit on \erass data shown in orange. In blue, we show the \planck CMB~2020 constraints without combination with BAO and SNe~Ia \citet{Planck2020}. In red, we show the combination of \erass with \planck CMB. As a comparison, we also present previous results from similar cluster surveys that employ weak lensing shear data in their mass calibration, e.g., WtG \citep{Mantz2015}, DES \citep{Abbott2020}, SPT-SZ \citep{Bocquet2019}, eFEDS \citep{Chiu2023a}, XXL \citep{Garrel2022}, and SPT \citep{Bocquet2024} surveys. \\ \\ \\ \\ }
 \label{fig:LCDM_omegam_sigma8}
\end{figure*}

 \begin{figure*}
\sidecaption
  \includegraphics[width=12cm]{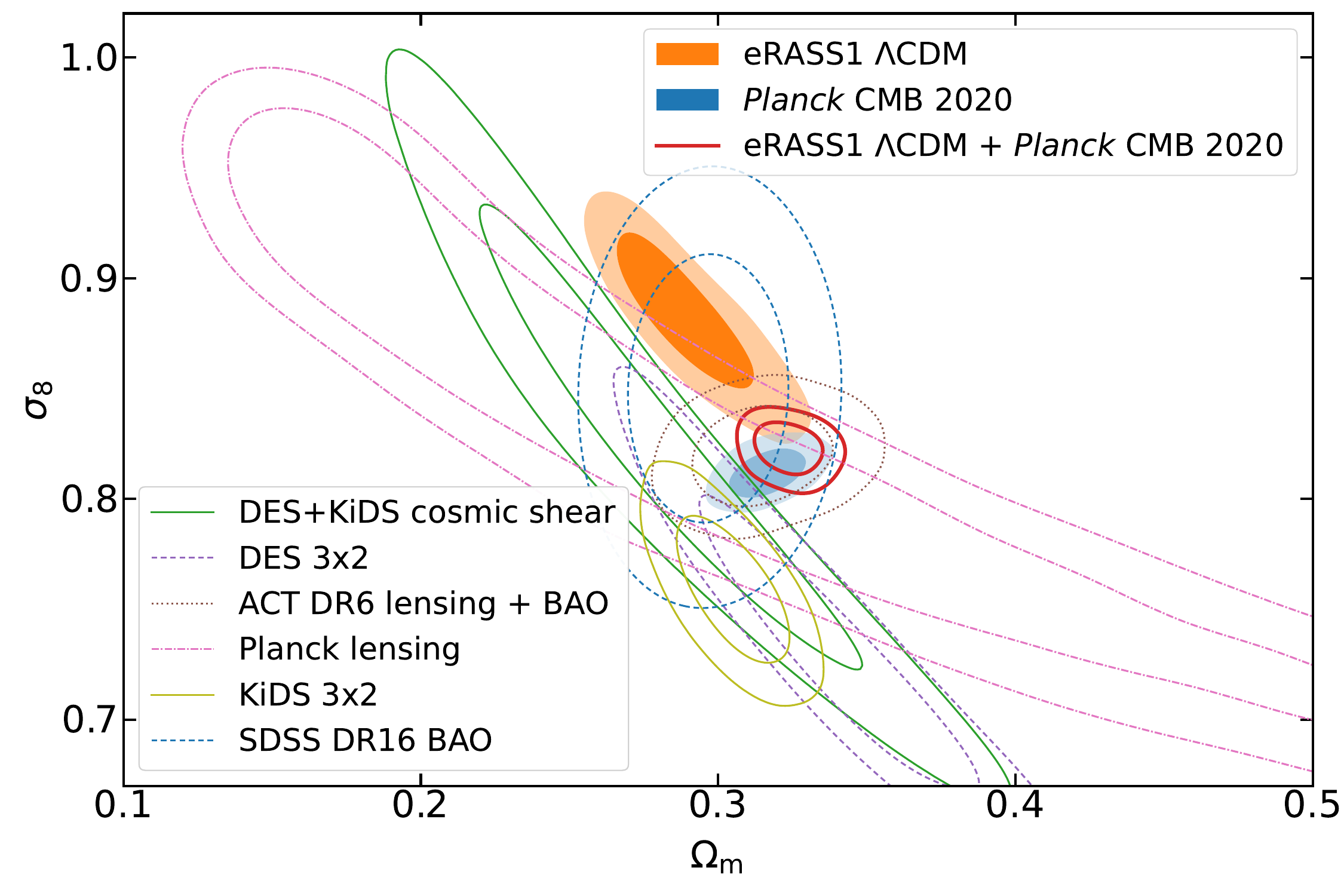}
     \caption{Posterior distributions on $\Omega_{\mathrm{m}}$ and $\sigma_8$ from the \lcdm fit to \erass data are shown in orange. Similarly, the \planck CMB~2020 best-fit contours and \erass with \planck CMB are shown in blue and red \citep{Planck2020}. This figure provides a comparison of \erass with other cosmology probes, e.g., the cosmic shear constraints from the joint analysis of DES and KiDS data \citep{DESKiDS2023}, DES 3x2 clustering constraints \citep{Abbott2021}, KiDS 3x2 constraints \citep{Heymans2021}, results from the combined ACT lensing and BAO \citep{Qu2023}, Planck lensing analyses \citep{Planck2020b}, and SDSS BAO RSD \citep{Alam2021}. \\ \\ \\ \\ }
     \label{fig:LCDM_omegam_sigma8_shear}
\end{figure*}

Cluster number counts can provide powerful constraints on the cosmological parameters, especially on $\Omega_{\mathrm{m}}$ and $\sigma_8$, when the selection effects are adequately taken into account, and the bias in the mass calibration process is well-controlled.
In our baseline \lcdm cosmology analysis, we freeze the dark energy equation of state to $w=-1$, and the remaining best-fit parameters, $\Omega_{\mathrm{m}}$ and $\sigma_8$, and $S_8$ are:

\begin{align}
 \Omega_{\mathrm{m}} &= 0.29^{+0.01}_{-0.02},\nonumber \\
 \sigma_8 &= 0.88\pm 0.02, \nonumber \\
 S_8 &= 0.86\pm 0.01. 
 \label{eq:LCDM_parameters}
\end{align}

\noindent The posterior distributions of $\Omega_{\mathrm{m}}$ and $\sigma_8$ from \erosita\ only analysis are shown in contours in orange in Fig.~\ref{fig:LCDM_omegam_sigma8}. In the same figure, we also provide a comparison of our results with the other cluster number count experiments utilizing the WL mass calibration in the literature, e.g., WtG \citep{Mantz2015}, SPT-SZ \citep{Bocquet2019}, XXL \citep{Garrel2022}, and eFEDS \citep{Chiu2023a}. We find consistent results within $2-3\sigma$ with the results in the literature, except for the cosmology results from DES cluster abundance measurements \citet{Abbott2020}. In \citet{Abbott2020}, authors report a value of $\sigma_8= 0.85^{+0.04}_{-0.06}$ consistent with our result, whereas their best-fit $\Omega_{\mathrm{m}}=0.179^{+0.031}_{-0.038}$ measurement is significantly lower than our value. A similar pattern is observed in comparisons of DES with the results from other experiments based on cluster counts. \citet{Costanzi2021} argues that the observed difference could be due to a biased mass-richness scaling relation. 
Compared to the previous cluster number count experiments mentioned above \erass cosmology has the advantage of having a significantly larger number of clusters, an overall better understanding of the selection function, and, compared with optical surveys, an observable that we expect to be more closely linked to the actual virialized halos instead of line-of-sight projections. Furthermore, we have a mass reconstruction accuracy of up to a few percent, robustly established using the method proposed in \citet{Grandis2021a}. We can then calibrate our masses with state-of-the-art wide area WL surveys (KiDS, DES, HSC) with a high S/N.

We compare our findings from cluster abundances with the recently published results from other cosmology probes, for instance, CMB measurements from \planck CMB \citep{Planck2020}, the joint cosmic shear constraints from DES and KiDS \citep{DESKiDS2023}, the $3\times2$ correlation constraints from DES \citep{Abbott2021}, lensing measurements from ACT DR6 + BAO \citep{Qu2023}, and lensing measurements from Planck \citep{Planck2020b} in Fig.~\ref{fig:LCDM_omegam_sigma8_shear}. Both $\sigma_8$ and $\Omega_{\mathrm{m}}$ parameter measurements from the \erosita\ cluster abundances agree well within $2\sigma$ confidence interval with CMB \citep{Planck2020b} and lensing measurements \citep{Abbott2021, Qu2023, Planck2020b}; however, they are in slight tension with the constraints from the cosmic shear measured from the combination of DES and KiDS data \citep{Abbott2021, DESKiDS2023}.

The correlation and the correlation coefficients between the free parameters in the fit are shown in Figs.~\ref{fig:LCDM_degeneracy} and \ref{fig:LCDM_correlation} in Appendix~\ref{app:tests}. We find that the terms representing the slopes ($B_{X}$ and $B_{\lambda}$) and the redshift evolution ($G_{X}$ and $C_{\lambda}$) of the scaling relations between X-ray observable and richness are correlated with $\Omega_{\rm m}$. In contrast, these terms are anti-correlated with $\sigma_8$. Interestingly, $S_8$ shows a strong correlation with the amplitude of the scaling relations ($A_X$ and $A_{\lambda}$), but we do not observe any correlation between $S_8$ and the mass slope and the redshift evolution. Notably, none of the best-fit cosmological parameters are degenerate with the scatter terms in scaling relations. This indicates that the observables employed in this work do not need to be tightly correlated with cluster mass, as only the mass distribution at the population level plays a determining role in cosmology. In our forward modeling approach, it is crucial to statistically reproduce the observables at a population level; therefore, the X-ray count rate, despite being a high-scatter mass proxy, is a powerful observable because of its role in the selection process.
To calculate the consistency of our result with CMB measurements, we 1) resampled our MCMC chains and the \planck CMB~2020 chains to obtain the same number of points; 2) took the differences between the chains and assume that it is distributed according to a multivariate normal distribution with data inferred mean and covariance. This allows for a calculation of the $\chi^2 = \bar{\mu}^T C^{-1} \bar{\mu}$; and 3) we computed the survival function\footnote{The survival function complements one of the cumulative distribution functions} of the $\chi^2$ distribution.

This method allows for an immediate estimate of the departure from the best fit in terms of its standard deviation. We find that our measurements on $\Omega_{\mathrm{m}}$ and $\sigma_8$ agree with the \citet{Planck2020}'s CMB measurements (TT, TE, EE+low$l$+lowE combination; we use \planck CMB~2020 hereafter) at the $2.3\sigma$ confidence level. 

To break the degeneracy between $\Omega_{\mathrm{m}}$ and $\sigma_8$ and have a fair comparison with the \planck CMB~2020 constraints, we combine the two probes given the consistency between the two. The combination of \erosita\ cluster abundances and \planck CMB~2020 chains yields the following cosmological parameters:
\begin{align}
 \Omega_{\mathrm{m}} &= 0.32\pm0.01, \nonumber \\
 \sigma_8 &= 0.82\pm0.01, \nonumber \\
 S_8 &= 0.85\pm0.01. 
\end{align}

We show the combined results in thick red contours in Figs.~\ref{fig:LCDM_omegam_sigma8} and~\ref{fig:LCDM_omegam_sigma8_shear}. Compared to the previous \erosita-only results in Equation~\eqref{eq:LCDM_parameters}, we see a shift to a higher value of $\Omega_{\mathrm{m}}$ and to a lower value for $\sigma_8$, a significant improvement on our constraints on the $\Omega_{\mathrm{m}}$ and $\sigma_8$ parameter plane. This is because the \planck CMB~2020 constraints on $\Omega_{\mathrm{m}}$ and $\sigma_8$ are tight and degenerate along a direction orthogonal to the $S_8$ direction. As a result, the combined best-fit parameters are at the parameter space where these two constraints intersect. It is worth noting that \planck CMB~2020 \lcdm constraints are obtained with a different set of assumptions compared to ours, namely, the sum of neutrino masses are fixed in their fits to 0.06~eV, the lower bound on the neutrino masses from oscillation experiments \citep{Tanabashi2018}. In our \lcdm analysis, we set the neutrino mass to 0~eV. Figure~\ref{fig:eRASS1_omegam_sigma8} in Appendix~\ref{app:tests} shows the negligible impact of fixing the summed neutrino on our constraints to a lower limit of 0.06~eV obtained from the ground-based experiments \citep{Tanabashi2018}.

\begin{figure*}
 \sidecaption
 \includegraphics[width=12cm]{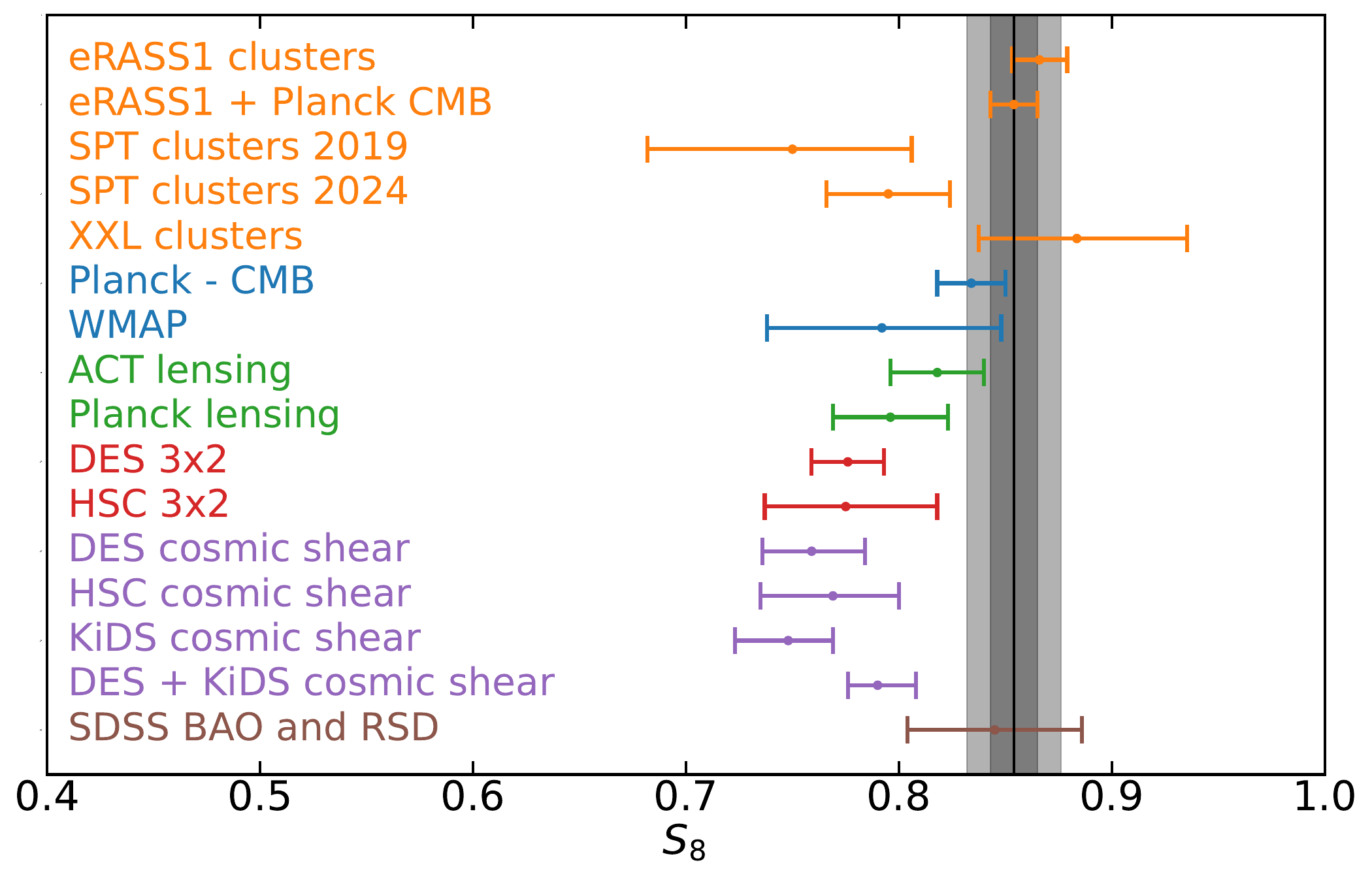}
 \caption{Comparison of the constraints on $S_8 = \sigma_8 \times \sqrt{\Omega_{\mathrm{m}}/ 0.3}$ with literature results 
 from SPT-SZ \citep{Bocquet2019}, 
 from SPT-SZ, SPTpol ECS, and SPTpol 500d cluster counts \citep{Bocquet2024}, 
 from XXL cluster number counts and clustering \citep{Garrel2022},
 from \planck\ CMB~2020 \citep{Planck2020}, 
 from WMAP CMB \citep{Hinshaw2013}, 
 from ACT lensing \citep{Qu2023}, 
 from ACT lensing \citep{Madhavacheril2023}, 
 from \planck\ lensing \citep{Planck2020b}, 
 from DES 3x2 correlation function \citep{Abbott2021}, 
 from HSC 3x2 correlation function \citep{Sugiyama2023}, 
 from DES cosmic shear \citep{Amon2022, Secco2022}, 
 from HSC cosmic shear \citep{Li2023}, 
 from KiDS cosmic shear \citep{vandenBusch2022}, 
 from joint DES and KiDS cosmic shear \citep{DESKiDS2023}, 
 from SDSS BAO and RSD \citep{Alam2021}. 
 All the errors shown are at the $1\sigma$ level.
 The vertical line and grey shaded areas represent our measurement's location and $1-2\sigma$ confidence level.
 }
 \label{fig:S8_comparison}
\end{figure*}

A more direct way to compare our constraints with the results from the CMB analysis is to measure the $S_8$, which combines the constraints on parameters $\Omega_{\mathrm{m}}$ and $\sigma_{8}$. Our data constrains the parameter $S_8$ with high precision due to the direction and orientation of the degeneracy between these two parameters. The best-fit $\alpha_{\rm best}$ value that minimizes the standard deviation of $\sigma_8\times({\Omega_{\mathrm{m}}} / {0.3})^\alpha$ along the direction of our degeneracy, is $\alpha_{\rm best} = 0.42\pm0.02$. Our best-fit value is close to 0.5, the canonical value of $\alpha$ used in the definition of $S_8$ in \citet{Planck2020}. The cluster cosmology experiments, e.g., \citet{Bocquet2019} reports $\alpha_{\rm best} = 0.2$ for the SPT-SZ sample, while \citet{Chiu2023a} finds $\alpha_{\rm best} = 0.3$ for the eFEDS clusters, a smaller value than what we find in this work. On the other hand, the power spectrum analysis of the shear measurements of the HSC field produces an $\alpha_{\rm best}$ value of 0.45, consistent with our results \citet{Hikage2019}.
Regarding the $S_8$ constraints, we obtain a degeneracy between $\sigma_8$ and $\Omega_{\mathrm{m}}$, which is almost perfectly aligned with the slope in the $S_8$ definition. This makes the constraints on $\sigma_8$ tighter in our experiment than other surveys. Similarly, the recent SPT constraints show the slope $\alpha_{\rm best} \sim 0.25$ \citep{Bocquet2024}, and indeed their optimal parameter combination $\sigma_8\times({\Omega_{\mathrm{m}}} / {0.3})^{0.25}$ measure a relative uncertainty very close to the uncertainty we obtain on $S_8$ in the \erass sample. The degeneracy reported in other experiments does not have a direction in line with the $S_8$ definition.

Figure~\ref{fig:S8_comparison} shows our constraint on the $S_8$ parameter and its comparisons with the selected results from experiments utilizing cluster abundances and weak lensing mass calibration in the literature \citep{Bocquet2019, Garrel2022}, cosmic shear \citep{Amon2022, Secco2022, Li2023, vandenBusch2022, DESKiDS2023}, clustering \citep{Abbott2021, Sugiyama2023, Qu2023}. 
Our measurement is consistent with results from CMB measurements with \planck and WMAP \citep{Hinshaw2013, Planck2020}, ACT and \planck lensing constraints \citep{Qu2023, Madhavacheril2023}, the HSC clustering results \citep{Sugiyama2023}, and the cluster count measurements based on the XXL survey within $2\sigma$ confidence interval \citep{Planck2020, Garrel2022}. However, it is in significant tension ($\geq3\sigma$ level) with constraints from other probes, in particular from cosmic shear from DES, HSC, KiDS surveys individually, and the combined KiDS and DES cosmic shear measurements \citep{Amon2022, Secco2022, vandenBusch2022, Li2023}, DES $3\times2$ clustering \citep{Sugiyama2023}, and the SDSS BAO and redshift space distortions (RSD) \citep{Alam2021} measurements. Given the consistency of our results with the CMB measurements, we combine our results with \planck 2020 CMB chains. The results of this combined analysis result in a negligible improvement on $S_8$ constraints, which can be seen in the second data point from the top on Fig.~\ref{fig:S8_comparison}. This behavior can be easily explained that the constraints on the $S_8$ from \erass are already tight; therefore, the combination of the parameters does not significantly improve precision.

As the error budget on our measurements is smaller than the systematic uncertainties due to the assumption of the halo mass function, we employ the \citet{Tinker2008} halo mass function because of its simple semi-analytical form. To test the effect of the assumption on the halo mass function on our results, we fit the \lcdm to our data with the \citet{Tinker2010} and \citet{Despali2016} halo mass functions. The results of the comparisons are shown in Fig.\ref{fig:tests_hmf}. We observe a slight decrease in the parameter $\sigma_{8}$ from $0.88\pm0.02$ to $0.86\pm0.02$ when the \citet{Tinker2010} and \citet{Despali2016} halo mass functions are used. The shift is statistically consistent within $1\sigma$ intervals with the \citet{Tinker2008} results. The change in the halo mass function does not significantly impact the parameter $\Omega_{\mathrm{m}}=0.30^{+0.01}_{-0.02}$ respectively.
This shows that the choice of the halo mass function does not yet limit our cosmological constraints in this work. As the \erosita\ survey gets deeper and the constraints on the cosmological parameters get tighter, the uncertainties related to the halo mass function will become a dominant source of systematics and should therefore  be treated with caution (see tests in Appendix~\ref{hmf:tests} and Figure~\ref{fig:tests_hmf}).

\subsection{Constraints on the $\nu$CDM cosmology}

Despite their small size and mass, the left-handed neutrinos influence the formation of large-scale structure by hindering the formation of small-scale haloes and leaving their imprints on the gravitational collapse process. The relative number of low-mass haloes in cluster abundance measurements can shed light on the summed masses of three left-handed neutrino species. With the discovery of neutrino oscillations between different flavors, electron, muon, and tau neutrinos, it has become clear that neutrino eigenstates must have non-zero mass \citep{Fukuda1998, Ahmad2002}. The theoretical predictions indicate a hierarchical relation between the mass eigenstates of the three species of neutrinos; however, their ordering in mass remains elusive. Based on the normal neutrino mass hierarchy model, the third neutrino mass eigenstate, associated with the tau neutrino, is the heaviest among the three. This is followed by the eigenstate of the muon neutrino, which has an intermediate mass, and that of the electron neutrino, namely, $m_\tau > m_\mu > m_e$. The ordering is swapped in the inverted mass hierarchy model such that $m_\tau < m_\mu < m_e$. Recent constraints of their summed mass show that the lower limit provided by the oscillation experiments depends on the assumed underlying hierarchy model \citep{Qian2015, Esteban2020}, with a lower limit of $\sum m_{\nu} > 0.059\mathrm{~eV}$ for normal mass hierarchy and $\sum m_{\nu} > 0.101\mathrm{~eV}$ for the inverted hierarchy models \citep{Tanabashi2018, Athar2022}. Constraining the summed mass to $<0.1\mathrm{~eV}$ implies that the inverted model is excluded. The ground-based experiments through beta decay of tritium imply the sum to be $\sum m_{\nu} < 1.1\mathrm{~eV}$ at 90\% confidence level, a narrow range for the allowed summed mass \citep{Aker2019}. The space-based \emph{Planck} CMB measurements provide a similar upper limit of $\sum m_\nu < 0.26~{\rm ~eV}$ at 95\% confidence level \citep{Planck2020}.

\begin{figure*}
 \sidecaption
 \includegraphics[width=12cm]{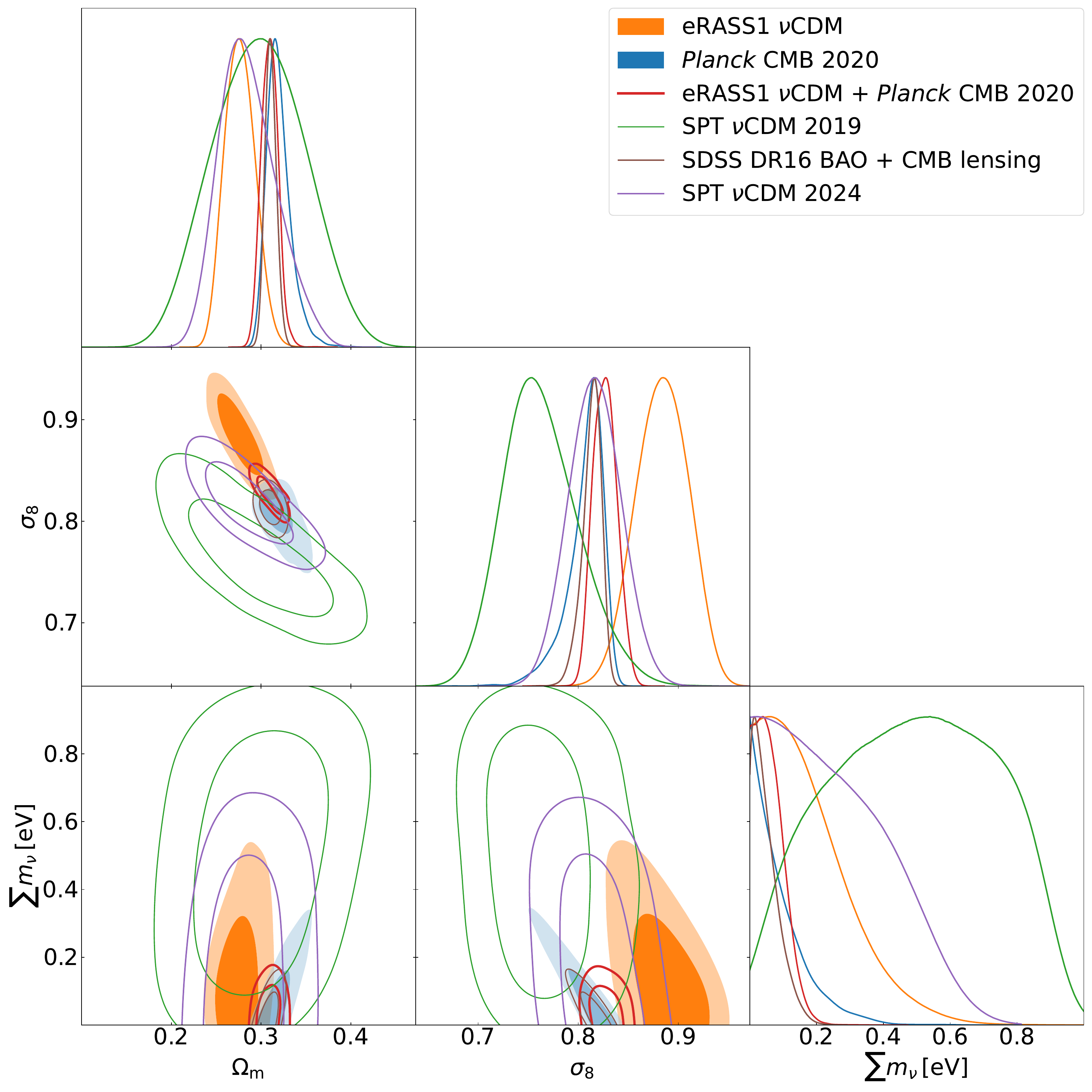}
 \caption{Best-fit values for $\nu$CDM cosmological model from \erass cosmological experiment (orange), \planck CMB~2020 data (blue), and combination of \erosita and \planck CMB (red) are shown. The SPT-SZ cluster abundances and SDSS BOSS DR16 BAO + CMB lensing results are shown in green and brown \citep{Bocquet2019, Alam2017}. SPT recent constraints from \citet{Bocquet2024} are added in purple. \erass cosmological experiment places the very tight constraints on the sum of the neutrino masses to be $\sum m_\nu < 0.43 {\rm eV}$ from cluster number counts alone. \vspace{4cm} }
 \label{fig:nuCDM}
\end{figure*}

As we probe the largest collapsed objects in the Universe, galaxy clusters, the cluster number counts can be used to constrain the summed masses of the neutrinos. To do so, we allow the sum of neutrino masses to be free in the cosmology pipeline, with a uniform prior of $\mathcal{U}(0 {\rm ~eV}, 1 {\rm ~eV})$. 
Including massive neutrinos in cosmological analysis requires a fine-tuned interaction with the choice of the halo mass function. As shown in \citet{Costanzi2013}, a few changes to the use of the matter power spectrum are fundamental to properly make use of \citet{Tinker2008} halo mass function in the presence of massive neutrinos.
As the \texttt{pyccl} 3.0.0 library of \citet{Chisari2019pyCCL} available on GitHub\footnote{https://github.com/LSSTDESC/CCL/tree/master} does not implement the prescriptions of \citet{Costanzi2013}, therefore accounting for corrections to the power spectrum, matter density, and, finally, to the halo mass function, we modify the local version of \texttt{pyccl} by editing several hardcoded parts of code to allow the user to employ \citet{Costanzi2013} prescription.
We stress that our lower limit is set to 0~eV instead of a value of 0.059~eV \citep{Tanabashi2018} adopted by the \planck CMB~2020 analysis. As the lower limit on the neutrino mass depends on the assumptions of the neutrino mass model, we do not make any prior assumptions about its value and adopt uniform priors on its value to avoid this parameter from affecting our posterior distributions. 
The cosmological constraints with free neutrino mass components are:

\begin{align}
 \Omega_{\mathrm{m}} &= 0.28^{+0.02}_{-0.02}, \nonumber \\
 \sigma_8 &= 0.88\pm 0.03,\nonumber \\
 S_8 &= 0.85\pm 0.02, \nonumber \\
 \sum m_\nu &< 0.43 {\rm eV} \quad (95\% {\rm \ CL}).
 \label{eq:nuCDM_parameters}
\end{align}

These results are visualized in Fig.~\ref{fig:nuCDM}. The upper limit to the sum of neutrino masses is $\sum m_\nu < 0.43~{\rm eV}$. The addition of the \citep{Costanzi2013} correction to the power spectrum relaxes the upper limits on neutrino masses from 0.22~${\rm eV}$ to 0.43~${\rm eV}$, as expected.  Our results represent the tightest limits on the sum of neutrino masses from cluster abundance experiments; for instance, the SPT-SZ sample results in an upper limit of $< 0.74\mathrm{~eV}$ \citep[95\% confidence interval][]{Bocquet2019}, while the updated SPT results yield $< 0.55\mathrm{~eV}$ \citep[95\% confidence interval][]{Bocquet2024}. 
We verify that the values of the $\Omega_{\mathrm{m}}$ and $\sigma_8$ remain statistically consistent when $\sum m_\nu$ is allowed to be non-zero, see Fig.~\ref{fig:eRASS1_omegam_sigma8}. 
We find that \erass and \planck CMB $\nu$CDM parameters are consistent at the 2.0$\sigma$ level.

Given the excellent agreement with \emph{Planck} CMB measurements, the resulting cosmological parameters of two probes can be combined in a statistically meaningful way, enabling a much tighter measurement of the impact of massive neutrinos on both the formation and evolution of the large-scale structure and on the primordial density field. As performed in the $\Lambda CDM$ analysis, we combine our results with the \planck CMB 
constraints to break the degeneracy between 
$\Omega_{\mathrm{m}}$ and $\sigma_8$. We obtain:

\begin{align}
 \Omega_{\mathrm{m}} &= 0.31\pm 0.01, \nonumber \\
 \sigma_8 &= 0.83\pm 0.01, \nonumber \\
 S_8 &= 0.84\pm 0.01, \nonumber \\
 \sum m_\nu &< 0.14 {\rm ~eV} \quad (95\% {\rm \ CL})
\end{align}

Simultaneously fitting our measurements with \planck CMB~2020 likelihood chains yield an upper limit of $\sum m_\nu < 0.14{\rm ~eV}$, consistent with the results in the literature results with marginally consistent with the inverted mass hierarchy model, which requires $\sum m_\nu > 0.101{\rm ~eV}$, where just 18\% of the \erosita - \planck combined samples are above this limit and consistent with the inverted hierarchical model.

As a final step, we combined \erass cluster abundance measurements via an importance sampling with \planck CMB and with the lower limits from ground-based oscillation experiments \citep{Tanabashi2018}. In the case of a normal mass hierarchy scenario, the summed masses of $\sum m_\nu = 0.09_{-0.02}^{+0.04}\mathrm{~eV}$; 
while assuming the inverted mass hierarchy model, we obtain summed masses of 
$\sum m_\nu = 0.12_{-0.02}^{+0.03}\mathrm{~eV}$. 
It is interesting to point out (from Fig.~\ref{fig:neutrino_hierarchy} in Appendix~\ref{app:neutrino}) that the constraints on the mass of the lightest neutrino eigenstates are similar:
for both mass hierarchy we obtain $m_{\rm light} = 0.017^{+0.016}_{-0.010}\mathrm{~eV}$ (68\% confidence intervals) or with an upper limit of $0.044\mathrm{~eV}$.

\subsection{Constraints on the $w$CDM cosmology}
\label{sec:wCDM}

The late-time acceleration of the expansion of the Universe is one of the major discoveries of modern physics for the standard \lcdm cosmology model, where the cosmological constant $\Lambda$ drives the late-time acceleration \citep{Riess1998, Perlmutter1998}. Several independent experiments from Supernovae Ia \citep{Riess1998, Astier2006, Kowalski2008}, CMB \citep{Komatsu2011, Planck2020}, and BAO \citep{Beutler2011, Alam2017, Alam2021} produced constraints on fractional energy density of the dark energy to the total energy of the Universe to be $\sim$67-70\%. Despite the intensive experimental effort invested into constraining the significant contribution of dark energy, its nature remains elusive. One of the main science drivers of \erosita\ is to place constraints on its equation of state using the cluster abundances. We define $w$, the equation of state, as the ratio of the pressure of dark energy to its energy density ($w = p / \rho$) and assume that its density evolves with $\rho \propto a^{-3(1+w)}$ in an isotropic Universe. Previous dark energy equation state measurements point to $w\sim-1$ with no evolution with redshift \citep{Astier2006, Beutler2011, Alam2017, Planck2020}. In this section, we consider an extension to the standard \lcdm by allowing the equation of state of dark energy to $w$ to vary.

\begin{figure*}
 \sidecaption
 \includegraphics[width=12cm]{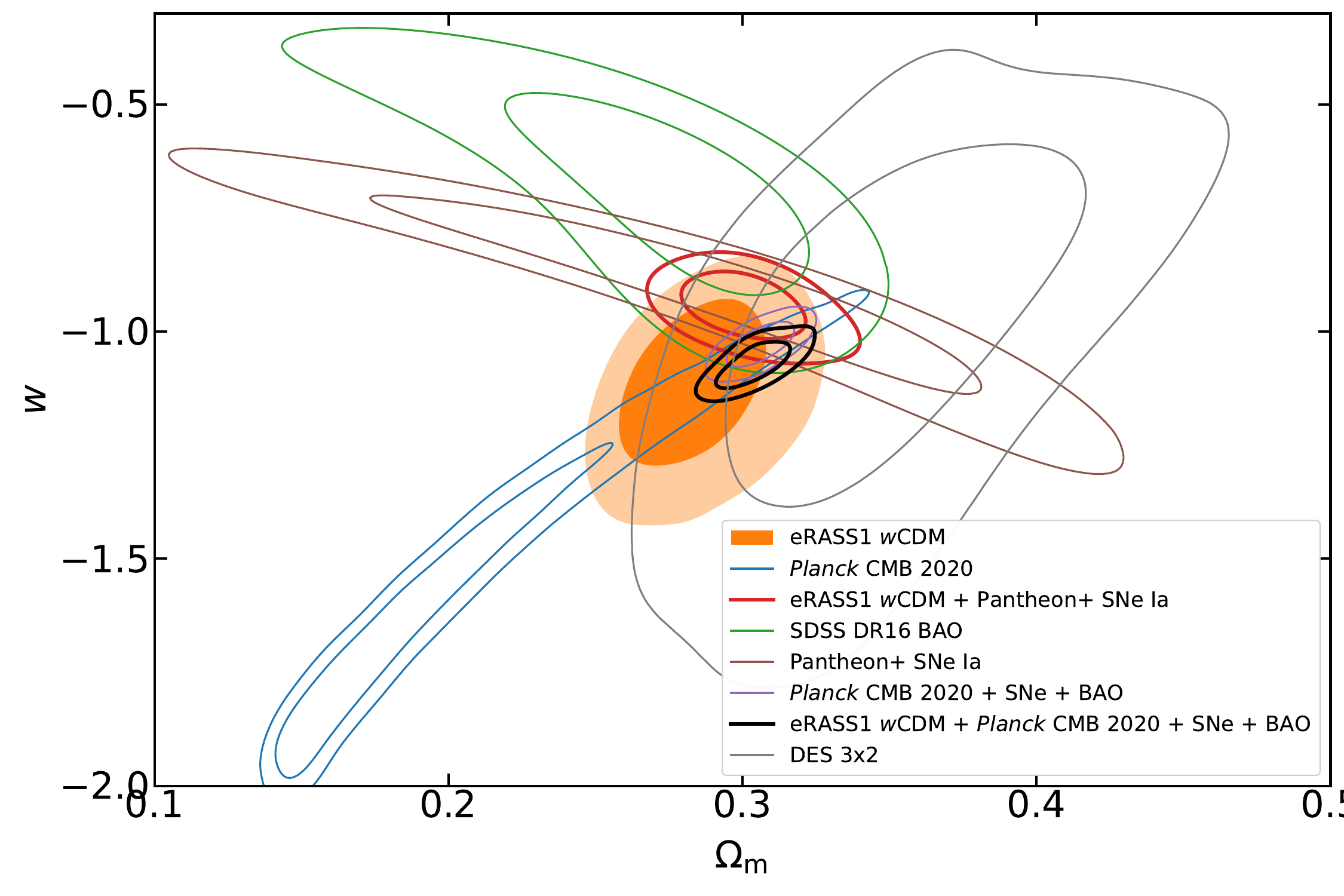}
 \caption{Best-fit contours of $w$ and $\Omega_\mathrm{m}$ in the $w$CDM cosmological model from \erosita experiment (orange), \planck CMB data (blue), SDSS BOSS DR16 BAO constraints (green), Pantheon SNe Ia constraints (brown), DES 3x2 constraints (grey) are shown. We also show the resulting combination of \erosita and Pantheon SNe Ia (red), the combination of \planck CMB~2020 with Pantheon SNe Ia and BAO (purple), and finally, the combination of \erosita with \planck CMB~2020, Pantheon SNe Ia and SDSS BOSS DR16 BAO (black). \erass experiment can place tight constraints on the dark energy equation of state parameter without any need to combine with other cosmological experiments. \vspace{2cm}}
 \label{fig:wCDM}
\end{figure*}

\begin{figure}
 \centering
 \includegraphics[width=0.5\textwidth]{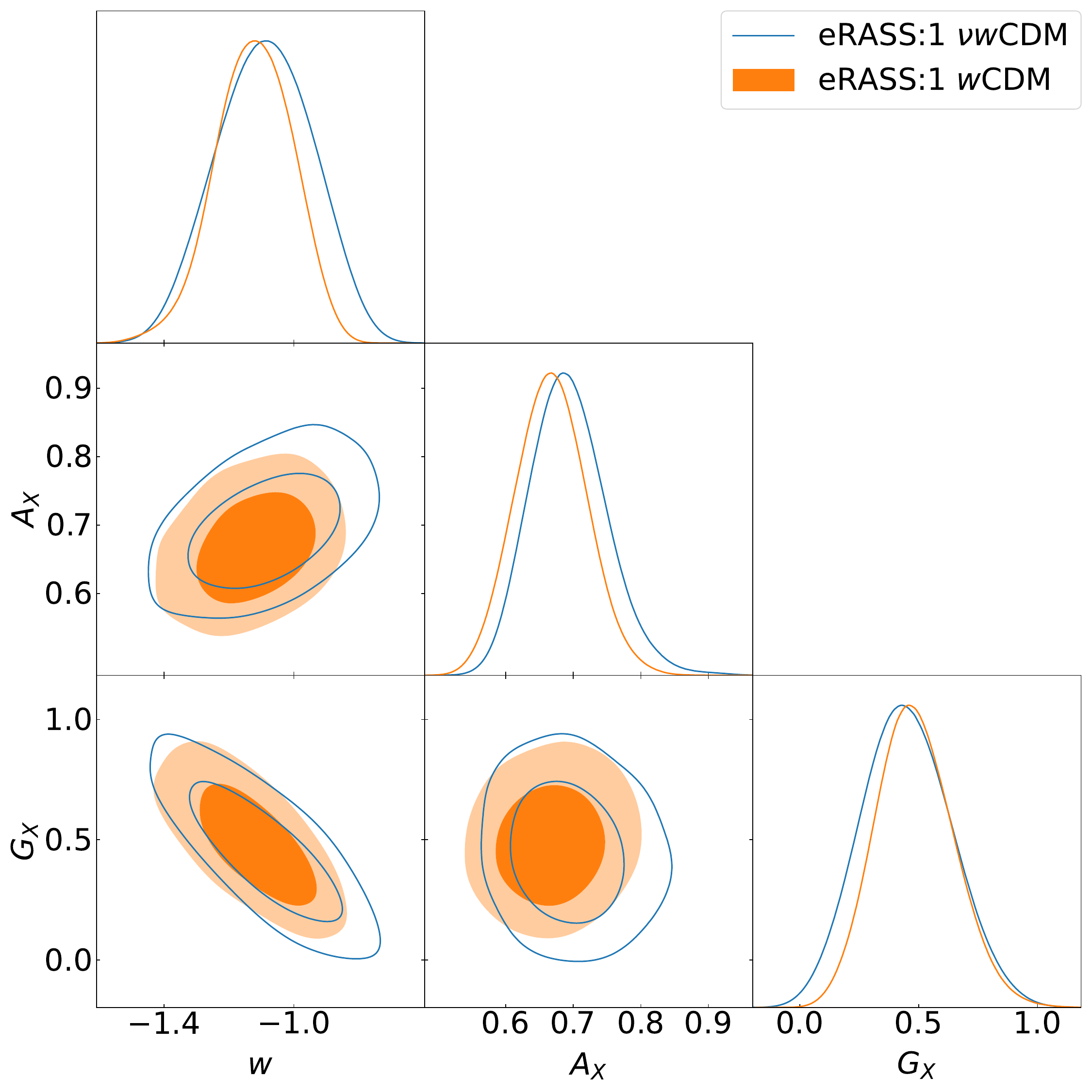}
 \caption{Correlation between $w$ and the X-ray count rate to mass scaling relation normalization and redshift evolution. The correlation for the $w$CDM case is shown in orange and blue for the $\nu w$CDM case.}
 \label{fig:w_degeneracy}
\end{figure}

To constrain the value of $w$, we set the sum of neutrino masses back to zero ($\sum m_\nu = 0\mathrm{~eV}$), and we allow the $w$ to be free with a uniform prior $\mathcal{U}(-2, -0.33)$ in the $w$CDM model. Naturally, a change in the equation of the state affects both the background and the hierarchical growth of structure as it changes the expansion history of the Universe and the cosmological distances converted from the redshift measurements. We obtain the best-fit parameters as:

\begin{align}
 \Omega_{\mathrm{m}} &= 0.28\pm 0.02, \nonumber \\
 \sigma_8 &= 0.88\pm0.02, \nonumber \\
 S_8 &= 0.86\pm0.02, \nonumber \\
 w &= -1.12 \pm 0.12.
 \label{eq:wCDM_parameters}
\end{align}

Our fits appear to prefer $w$ to be $-1.12 \pm 0.12$, shown in Fig.~\ref{fig:wCDM}. consistent with the results from \planck CMB \citep{Planck2020}, DES~Y3 3x2~pt \citep{Abbott2021}, Pantheon SNe Ia \citep{Scolnic2018, Scolnic2022, Brout2022}, DES Y5 SNe Ia \citep{Abbott2024}, and SDSS BOSS DR16 BAO \citep{Alam2017}, at $1-2\sigma$ level. The best-fit parameters of the scaling relations, the mixture model, and the correlation between parameters of this model are consistent with the baseline \lcdm case.

Given the consistency between \erosita and Pantheon SNe Ia results on $w$ at 0.5$\sigma$ level and the orthogonal direction of the degeneracy between $\Omega_\mathrm{m}$ and $w$, we combine two datasets to break degeneracies and obtain;
\begin{align}
 \Omega_\mathrm{m} = 0.30\pm0.01, \nonumber \\
 \sigma_8 = 0.87\pm0.02, \nonumber \\
 S_8 = 0.87\pm0.01, \nonumber \\
 w = -0.95^{+0.05}_{-0.04}. 
\end{align}

Compared with our results, the parameters $\Omega_{\mathrm{m}}$, $\sigma_8$, and $S_8$ are not significantly affected by the inclusion of Pantheon SNe Ia data.
However, our best fitting $w$ shifts to a slightly higher value of $-0.95^{+0.05}_{-0.04}$, consistent with the \erosita measurements with a significantly increased precision. $w$ remains consistent with the standard \lcdm value of -1 at 1$\sigma$ level.

Dark energy is responsible for a late-time expansion of the Universe and, therefore, has a negligible effect on the CMB signal. The constraining power of \planck CMB data on the dark energy equation of state is quite limited. We find that \erass constraints on $w$ are consistent $w$ at the 1.4$\sigma$ and $1.8\sigma$ level with \planck CMB~2020 and SDSS BOSS DR16 BAO results, respectively. 
Given the consistency, we combine \erass chains with both Pantheon SNe, \planck CMB, and SDSS BAO data. Including these two extra constraints helps improve the precision of the parameters significantly. We measure;

\begin{align}
 \Omega_{\mathrm{m}} &= 0.31\pm 0.01, \nonumber \\
 \sigma_8 &= 0.84\pm 0.01, \nonumber \\
 S_8 &= 0.85\pm 0.01, \nonumber \\
 w &= -1.06\pm 0.03. 
\end{align}

When combined with the geometrical probes, such as BAO, SNe Ia, and CMB, our data constrain the best-fit value of the dark energy equation of state to be $w= -1.06\pm 0.03$ with 3\% precision, thus improving the \erosita only results by a factor of $\sim$3, and consistent with $w = -1$ at 2$\sigma$ level. The shift on the best-fit \erosita parameters is less than 1$\sigma$. In particular, $\Omega_{\mathrm{m}}$ is unchanged, $\sigma_8$ and $S_8$ shifts to a slightly lower value.

\subsection{Constraints on the $\nu w$CDM cosmology}

In connection with the previous sections, to test the effect of the neutrino masses on the dark energy equation of state, we allow both $w$ and $\sum m_\nu$ to vary in our fits. The priors are set to uniform distributions for $\mathcal{U}(-2, -0.33)$ on $w$, and $\mathcal{U}(0 {\rm ~eV}, 1 {\rm ~eV})$ for $\sum m_\nu$. We find:

\begin{figure*}
 \sidecaption
 \includegraphics[width=12cm]{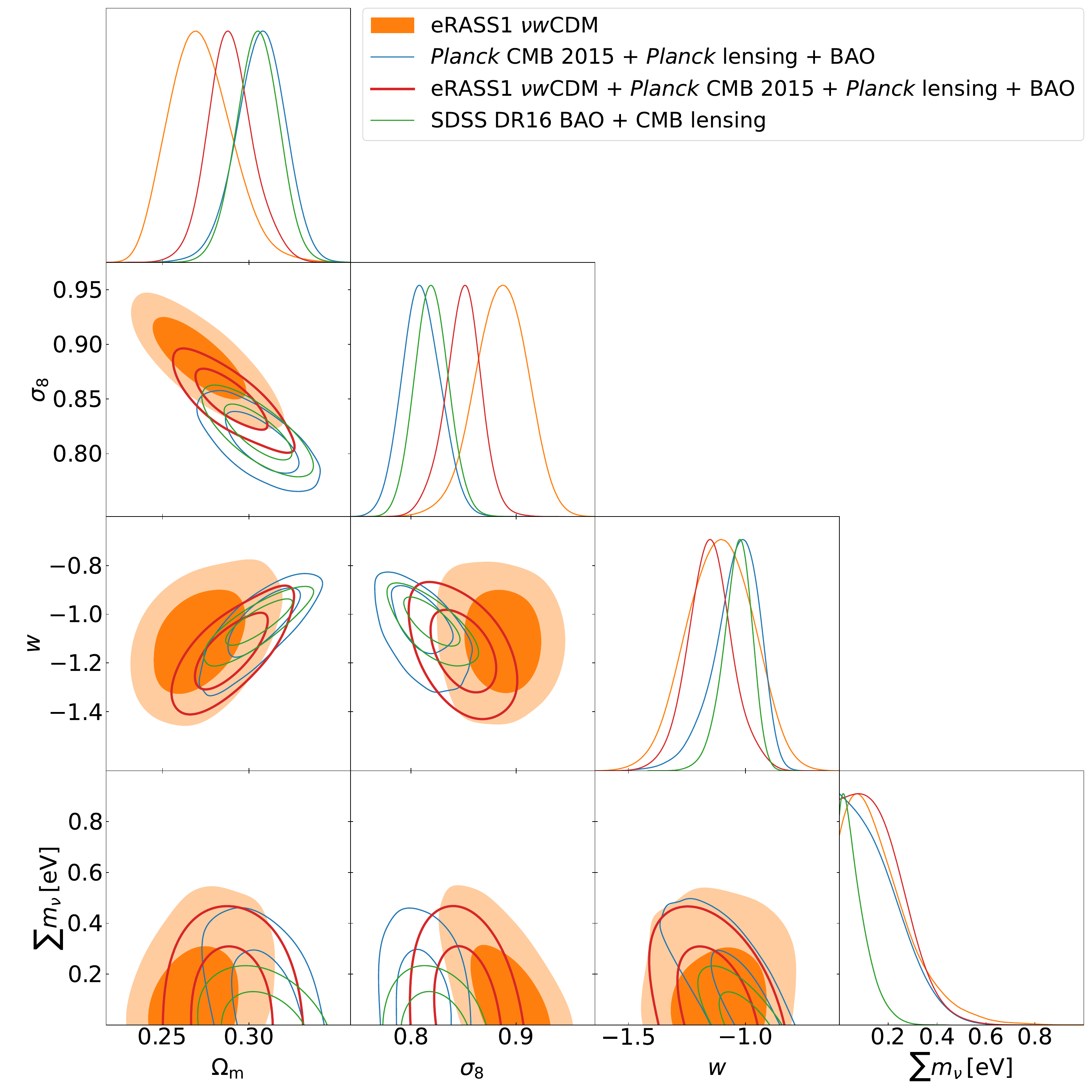}
 \caption{Best-fit values for $\nu w$CDM cosmological model from \erosita experiment (orange), \planck CMB 2015 data (blue), and combination of \erosita and \planck CMB 2015 (red). The best-fit parameters are consistent with the canonical \lcdm model. \vspace{4cm}}
 \label{fig:nuwCDM}
\end{figure*}
\begin{align}
 \Omega_{\mathrm{m}} &= 0.27\pm0.02, \nonumber \\
 \sigma_8 &= 0.89\pm0.03, \nonumber \\
 S_8 &= 0.84\pm 0.02, \nonumber \\
 w &= -1.11\pm 0.14, \nonumber \\
 \sum m_\nu &<0.44 {\rm ~eV} \quad (95\% {\rm \ CL}).
 \label{eq:nuwCDM_parameters}
\end{align}

This case combines the two previous cosmologies: as for the $\nu$CDM and $w$CDM case, our results in the $\Omega_{\mathrm{m}}$ - $\sigma_8$ parameter space are unchanged. The degeneracy that $w$ has with normalization and redshift evolution of the scaling relation remains the same as that seen in Fig.~\ref{fig:w_degeneracy}.

Due to the lack of public \planck CMB-only likelihood chains on the simultaneous fits on the $w$ and $\sum m_\nu$ in Planck CMB~2020, we combine our results with the readily available CMB+lensing+BAO in \citet{Planck2016b} chains. Relying on the 1.6$\sigma$ agreement with our results, we find:

\begin{align}
 \Omega_{\mathrm{m}} &= 0.29
\pm 0.01, \nonumber \\
 \sigma_8 &= 0.85\pm 0.02, \nonumber \\
 S_8 &= 0.83\pm 0.01, \nonumber \\
 w &= -1.15\pm 0.10, \nonumber \\
 \sum m_\nu &<0.37 {\rm ~eV} \quad (95\% {\rm \ CL}).
\end{align}

 In this case, the constraints on the $w$ move closer to the value $-1$, with the summed neutrino masses with an upper limit of $\sum m_\nu <0.37 \, \mathrm{eV}$. Compared with the other combinations with CMB results, we notice that the improvements are modest, which is expected since the cosmological constraints in \citet{Planck2016b} are weaker than what is published in the latest Planck CMB~2020 work. Our best-fit result on $w$ is closer to the best-fit value in the SPT-SZ+\planck combination, where the authors report $w=-1.12 \pm0.21$ \citep{Bocquet2019}. In the same work, the reported sum of neutrino masses is $\sum m_{\nu}=0.50\pm0.24\mathrm{~eV}$ in the same combination of datasets, whereas we find tighter upper limits on $\sum m_{\nu}$ in this work.

\subsection{Model comparison}

In this paper, we fit our \erass dataset with four different cosmological models, \lcdm, $\nu$CDM, $w$CDM, and $\nu w$CDM. We compare the goodness of fit for these different models to assess whether the data prefer one of the cosmological models in this section.
To achieve this, we use the Bayes factor \citep{Jeffreys1961, Trotta2008, Buchner2014}, which, in our case, becomes available with the fitting sampler we use in our analysis. The \texttt{ultranest} method gives the Bayesian evidence\footnote{The Bayesian evidence is defined as $Z = \int L(D|\theta) \pi(\theta) d\theta$, where $L(D|\theta)$ is the likelihood and $\pi(\theta)$ is the prior.} directly as an output. Therefore, it is straightforward to use it to compute the Bayes factor\footnote{The Bayes factor is the ratio of the Bayesian evidence of two different models, $B = Z_{M1} / Z_{M_2}$} and Jeffreys scale to compare different cosmology fits \citep{Jeffreys1961}. More complex models tend to be flexible since they can accommodate a broader range of possible data with a larger number of free parameters; however, that flexibility is automatically penalized by Bayesian inference in this framework and ends up giving us an objective estimator. Specifically, since a more complex model predicts that the observed data will be drawn from a more extensive range of possibilities, a model with extra free parameters will have a larger volume over which to integrate to compute the Bayesian evidence and, therefore, it may be disfavoured.

In Table~\ref{tab:evidence}, we show the Bayesian evidence obtained in the various models we test with the \erass dataset. We find no evidence for a clear preference among the models we test since all the measured bets-fit parameters have values similar to that of the standard \lcdm case, see Equations~\eqref{eq:LCDM_parameters},~\eqref{eq:nuCDM_parameters},~\eqref{eq:wCDM_parameters}, and \eqref{eq:nuwCDM_parameters}.

\begin{table}[t]
 \centering
 \caption{Bayesian evidence for the cosmological models tested.}
 \begin{tabular}{ c c }
 \hline
 \hline
 Cosmology & Bayesian evidence \\
 \hline
 \lcdm & $-11419.6 \pm 0.7$ \\
 \lcdm with $\sum m_\nu = 0.06\mathrm{~eV}$ & $-11418.3 \pm 0.5$ \\
 $\nu$CDM & $-11418.8 \pm 0.8$ \\ 
 $w$CDM & $-11419.5 \pm 0.5$ \\
 $\nu w$CDM & $-11421.2 \pm 0.6$ \\
 \hline
 \hline
 \end{tabular}
 \tablefoot{To interpret this table, we need the Jeffreys scale \citep{Jeffreys1961}: if the difference between the two is greater than 5, then there is significant evidence to prefer the model with the largest Bayesian evidence; if the difference between the two is between 2.5 and 5, then there is weak preference for the model with the largest Bayesian evidence; otherwise, the models are statistically indistinguishable. }
 \label{tab:evidence}
\end{table}
\section{Discussion}
\label{sec:discussion}

\subsection{Methodology}

This paper presents the cosmological parameters obtained from the first All-Sky Survey of \erosita by studying the abundance of clusters of galaxies detected in the survey. The tight constraints on cosmology we present in this work result from statistical power with the employment of the largest pure \erosita-ICM selected cluster sample and the well-controlled systematics related to mass calibration and selection modeling in our analysis. Below, we discuss several significant improvements and critical aspects that allow us to constrain cosmology with cluster number counts with well-understood and characterized systematics. 

 \textbf{Contamination modeling:} The purity level of the cosmology subsample can reach up to 94\% level after the X-ray extent likelihood \extlike cut on the X-ray selection on the cluster sample and optical confirmation process performed as described in \cite{Bulbul2024, Kluge2024}. The remainder of the contamination is modeled then through the novel mixture method (see Sect.~\ref{sec:mixture}), a forward model for eliminating the contamination in statistical samples, considering the X-ray and optical properties of the contaminants and clusters modeled through the simulations in our analysis. This effort made it possible to maintain the maximally acceptable \extlike cut, namely, $>6 $, without reducing the number of clusters in our sample. Figure~\ref{fig:tests} in Appendix~\ref{app:tests} shows the impact on our analysis of the contamination modeling. It is worth noting that the contamination also impacts the mass calibration process. When considering a population of detected clusters at a fixed count rate, the weak lensing signal from the source is partly due to contaminants along the line of sight. The contamination reduces the expected mass since they have, on average, lower tangential shear than actual halos. This depends on the count rate affecting the overall induced mass bias. This means that non-negligible contamination will also bias the slope and scatter of the scaling relations.
 In the present and future \erosita cosmology analyses, the mixture model is necessary since it allows us to account for and correct these effects in our likelihood and, therefore, marginalize them out to obtain the presented precision cosmological constraints.

\textbf{Selection function:} The completeness of the \erosita\ survey is estimated from the state-of-the-art realistic twin simulations of the sky \citep{Comparat2020, Seppi2022}. The simulations include the most relevant components to an X-ray survey, namely, the detector properties, instrumental noise, and both resolved and unresolved astrophysical sources. Our selection function model constructed from these realistic simulations depends on a limited number of parameters due to a trade-off between modeling complexity and cosmology- and/or model-dependency. The reformulation of the selection function modeling into a classification problem enabled us to leverage the efficiency of Gaussian process classifiers; their robust mathematical framework and quick computational response make them well-suited for our analysis. The full construction of the selection function is presented in an accompanying paper by \citet{Clerc2024}. 
 In this analysis, we model the detection probability of clusters as a function of count rate, redshift, and sky position through the selection function. Although robust, our selection function is calibrated with a single cosmological setup using realistic \erass simulations of the sky. 
 This makes our overall analysis marginally dependent on the underlying cosmological setup.
 However, the cluster count rate profiles at fixed global count rate value ingested in the selection function are representative enough \citep{Comparat2020, Seppi2022}, and the choice of the X-ray observable makes this marginal dependence quite negligible. 
 Additionally, the tests performed on the selection function indicate that the assumptions of cosmology do not significantly alter our probability distribution and our results on cosmology (see Sect.~\ref{sec:validation} and \citet{Clerc2024}).
 
 \textbf{Weak-lensing mass calibration and bias:} In the past decade, a cutting-edge approach has been developed for cosmology with cluster abundances utilizing weak lensing measurements in the mass calibration process to minimize the bias on cosmological parameters due to hydrostatic equilibrium assumption. The most crucial drawback of hydrostatic masses is the large impact of unaccounted nonthermal pressure support due to baryonic physics in the form of shocks, bulk motion, turbulence, and magnetic field \citep{Dolag1999, Dolag2005, Rasia2006, Vazza2009, Burns2010, Lau2009}. On the other hand, the weak lensing signal, although not bias-free, depends on the total enclosed mass within an aperture; therefore, it is a more reliable measurement. For this reason, mass calibration using weak lensing shear data is the preferred choice in cosmology once the small biases are calibrated via numerical simulations \citep{Hoekstra2012, vonderLinden2014, McClintock2019}. Triaxiality \citep{Corless2007, Meneghetti2010}, substructures along the line-of-sight \citep{Hoekstra2001, Becker2011}, and miscentering can bias weak lensing shear estimation in the three independent surveys (DES, KiDS, and HSC) we employed in this analysis. However, we obtain a strong and realistic proxy for the weak lensing bias by calibrating the shear measurements with the N-body or hydrodynamic simulations. We account for the total bias in our cosmology framework, as described in our supporting papers for weak lensing follow-up by \citet{Kleinebreil2024} and \citet{Grandis2024}.

 \textbf{Forward modeling approach:} In our analysis, we build a Bayesian forward modeling approach to account for the position-dependent detection chain of \erosita. \erass survey strategy is designed such that the exposure time varies with the location across the sky with significantly higher exposures close to the north and south ecliptic poles, which affects our detection efficiency \citep[see Fig.~16 in][]{Predehl2021}. Additionally, galactic absorption from atomic and molecular hydrogen \citep{HI4PI2016, Willingale2013} and \erosita's detector and sky background \citep{Freyberg2020} also varies significantly with sky position. This means our ability to detect clusters and accurately measure their count rate is not uniform but position-dependent. In our work, we integrate the variation in sky position in the implementation in all components of our analysis, for instance, in the selection function while modeling the uncertainty of the observables, predicting our contamination level via mixture modeling, and the forward-modeled likelihood. 
 Finally, Bayesian hierarchical forward modeling is the approach of choice when trying to understand a complex system based on robust theoretical considerations, which has strong predictive power in understanding the dependence of the data on input parameters. Cosmology studies with cluster abundances are well suited to employ modern advanced statistical approaches, such as the Bayesian hierarchical forward modeling approach, especially with state-of-the-art computing systems that allow for fast evaluation of complex models.
 One possible drawback of such modeling, compared to machine learning, for instance is the somewhat fixed relation between mass and observable through the scaling relation and its scatter around it. We have tested the impact of changing a single scatter value model with a mass and redshift-dependent scatter in Appendix~\ref{app:Systematics on scaling relation modeling}, showing that our conclusions are effectively unchanged.

\subsection{$S_8$ Measurements}

The discrepancy in the inferred value of the amplitude of matter fluctuation between this analysis and the results from cosmic shear, see Fig.~\ref{fig:S8_comparison}, merits special attention. 
We use WL data to reconstruct our cluster sample's mass distribution accurately. Comparing this with cosmological predictions of halo mass function \citep{Tinker2008} provides constraints on the cosmological parameters. 
Cosmic shear experiments use the same WL data but study its correlation function of large and intermediate scales, which is also measured with percent accuracy from the same optical data. Both approaches are thus similarly affected by the challenges in measuring galaxy shapes and photometric redshift. The key difference comes from the theoretical uncertainties that must be accounted for when predicting the WL signals for cosmological inference. 

While cosmic shear can be predicted very accurately at large scales, predictions at intermediate scales of the order of 10~Mpc could be affected by several astrophysical systematics, as discussed in detail in \citet{Mandelbaum2018c} and \citet{Huterer2023}. Among these is the modeling of the intrinsic alignment of source galaxy shapes with the lensing potential, the non-linear matter power spectrum calculation, and the impact that baryon feedback and its associated hydro-dynamical modeling uncertainties have on the cosmic shear correlation functions.

These concerns do not impact the halo mass function derived from the linear fluctuation matter power spectrum, as collapsed halos retain information on the linear fluctuation at the epoch when detached from background expansion. 
Weak lensing measurements of galaxy clusters are also unaffected by these: the halos dominate the line of sight potential and are thus the major source of the weak lensing signal. We deliberately limit our analysis to scales smaller than $3.2 / (1+z) $~Mpc/h to ensure that we measure only the clusters' potential and are thus not affected by the intermediate scale issues that affect the cosmic shear measurements.

We stress that the systematics related to the theoretical foundation of cluster number counts, cluster detection, selection effects, and mass calibration are well understood and accounted for in our analysis. Cluster number counts rely on limited assumptions on mass function, which can be computed analytically \citep{Press1974}. Detection through X-rays is one of the most reliable methods for locating clusters and for constructing pure cluster catalogs without any major bias because the X-ray signal from the hot ICM is proportional to the square of the electron number density and, therefore, the density contrast of a galaxy cluster is enhanced, and projection effects are minimized. The strict cuts we apply on the X-ray detection chain and the utilization of their follow-up optical observations can eliminate any leftover contamination in the cluster samples, reducing the contamination to $<6\%$. The remaining contamination is modeled through our mixture method in-situ \citep{Bulbul2024, Kluge2024}. The selection function is produced from the end-to-end simulations of the \erosita\ footprint and is well understood and tested against the mocks \citep{Comparat2019, Seppi2022, Clerc2024}.

Weak lensing measurements required for mass calibration of galaxy clusters share with cosmic shear analysis some observational systematics, like photo-z calibration and shape measurement \citep{Kleinebreil2024, Grandis2024}.
However, cluster mass calibration through weak lensing observations is significantly less affected by the physical systematics that impact cosmic shear (intrinsic alignment of the source galaxies, the non-linear calculation of the power spectrum). 
For instance, the difference between WL masses calibrated on gravity-only simulations and hydro-dynamical simulations is 2\% \citep{Grandis2021a}, while the difference between the matter power spectra between gravity-only and hydrodynamical simulations, but also among different hydro-dynamical simulations, is the order of 10 \%. 
We have further tested in the validation process that the effect on the mock observations from additional biases is small ($<1\sigma$ on the cosmological constraints) and does not significantly impact our results from cluster counts (see Sect.~\ref{sec:validation}).
In short, weak lensing results are remarkably robust for small scales around massive halos but start to be more challenging to model accurately for
 intermediate scales averaged over large sky areas because of the different sources of weak lensing signals.
In light of these arguments, WL-calibrated number counts of X-ray-selected clusters may well be regarded as more robust than cosmic shear analyses.

\subsection{Constraints on the total mass of neutrinos}

The effect of massive neutrinos on cluster abundance has been explored in this work. Particularly, they affect the primordial power spectrum and the assembly of large-scale structures \citep{Costanzi2013, Zeng2019}. 
Massive neutrinos will not cluster on scales smaller than their free-streaming length due to their high thermal velocities. With non-zero masses, neutrinos are part of the total matter field and contribute to its smoothing. The global effect is, therefore, the dumping of structure formation compared to a standard \lcdm cosmological model with massless neutrinos.

In the literature, measurements of summed neutrino masses from individual cosmological experiments are generally not very tight. For example, \citet{Bocquet2019} using the \texttt{SPT} cluster sample found $\sum m_\nu < 0.74 \, \mathrm{eV}$ at 95\% confidence level, and even CMB results from \citet{Planck2020} yield $\sum m_\nu < 0.26 \, \mathrm{eV}$ at 95\% confidence level.
The tightest constraints on the neutrino masses are obtained via combinations of multiple probes. For instance, combining Lyman-$\alpha$ forest results and CMB measurements give $\sum m_\nu < 0.15 \, \mathrm{eV}$ \citep{Rossi2015}, while BAO and CMB together yields $\sum m_\nu < 0.15 \, \mathrm{eV}$ at 95\% confidence level \citep{Vagnozzi2017}. More recent results from the \planck CMB and BOSS BAO measurements provide slightly tighter upper limits of $\sum m_\nu < 0.14 \, \mathrm{eV}$ at 95\% confidence level \citep{Tanseri2022}.

The results in our work from \erosita\ are competitive by themselves since we obtain $\sum m_\nu < 0.43 \, \mathrm{eV}$ at 95\% confidence level, close to a level that is competitive with the constraints from \planck CMB~2020.
However, as in the literature, it is the combination of these two state-of-the-art cosmological probes that we obtain very tight constraints of $\sum m_\nu < 0.14 \, \mathrm{eV}$ at 95\% confidence level.
We stress that although the neutrino hierarchy mass problem remains unsolved, our findings bring us remarkably close to a breakthrough. In fact, our result combined with \planck CMB~2020 rule out at a 82\% confidence level the inverted hierarchy model, which requires $\sum m_\nu > 0.101 \, \mathrm{eV}$. As we move forward to the future cosmology with the exploitation of deeper \erosita surveys, we will significantly improve the constraints on neutrino summed masses, possibly with higher confidence in discriminating against different neutrino mass hierarchy models of neutrino eigenstates.
Furthermore, measuring the mass of the lightest neutrino eigenstate (combining our results with oscillation experiments) with an order of magnitude better precision will guide the upcoming ground- and space-based experiments to finally pin down the neutrino masses.

\subsection{Constraints on the dark energy equation of state}

Accurate measurements of the dark energy equation of state $w$ as the ratio between pressure and density are extremely important in our effort to understand the nature of this elusive Universe component responsible for the late-time accelerated expansion of the Universe \citep{Riess1998, Perlmutter1998}. 
As discussed in \citet{Planck2020}, we also adopt the parameterized post-Friedmann (PPF) model \citep{Fang2008} to allow the dark energy equation of state to cross the \lcdm value of $-1$.
When this crossing happens (i.e., $w < -1$), dark energy is said to be considered ``phantom'', and many mechanisms can explain the crossing \citep[see, for example,][for extensive theoretical background on the topic]{Vikman2005, Kunz2006, Amendola2018}.

As shown in Sect.~\ref{sec:wCDM} \erosita\ cluster abundances slightly prefer a phantom equation of state, i.e., $w<-1$, as in the case of the \planck CMB analysis \citep{Planck2020}. Still, the value we find, $w = -1.12 \pm 0.12$ , is consistent at 1$\sigma$ level with the canonical \lcdm value of $-1$, and is also consistent with previously reported literature results from 
DES~Y3 3x2~pt \citep[$w = -0.98^{+0.32}_{-0.20}$, ][]{Abbott2021}, 
SPT clusters \citep[$w = -1.55 \pm 0.41$, ][]{Bocquet2019}, 
SDSS eBOSS DR16 BAO \citep[$w = -0.69 \pm 0.15$, ][]{Alam2021},
Pantheon SNe Ia \citep[$w = -1.09 \pm 0.22$, ][]{Scolnic2018},
and
Pantheon+ SNe Ia \citep[$w = -0.90 \pm 0.14$, ][]{Brout2022}. 
We note that differently from these other cosmological experiments, our analysis strategy using \erosita\ selected clusters enables the constrain of all the main late-time parameters with high precision at the 10\% percent level without the need to combine with additional cosmological experiments to break the degeneracies between the measured parameters.
The addition of external cosmological experiments enables much tighter constraints on the dark energy equation of state, as we find $w = -0.95_{+0.05}^{-0.04}$ when combining with Pantheon+ SNe Ia \citep{Brout2022}, and $w = -1.06 \pm 0.03$ when \planck CMB \citep{Planck2020} and SDSS BOSS DR16 BAO \citep{Alam2017} are included.

Shown clearly in Fig.~\ref{fig:w_degeneracy}, $w$ is slightly correlated with the amplitude of the X-ray count rate to mass scaling relation, and it is significantly anti-correlated with the redshift evolution of the X-ray count rate to mass scaling relation, $G_X$. This highlights the need for a new cluster sample spanning a larger redshift span, with corresponding WL mass measurements, to accurately calibrate scaling relations out to high redshifts $z\sim1$. Deeper \erosita\ surveys will expand our redshift coverage and improve count rate measurement with higher S/N data. The cosmology analyses performed with future \erosita-based cluster catalogs will allow us to constrain the dark energy equation of state with higher precision.

\subsection{Future improvements}

The critical aspects summarized above allow us to place stringent constraints on the standard \lcdm cosmology, the neutrino masses, and the dark energy equation of state. Although we fully account for all known systematics effects that could lead to significant uncertainties compared to the statistical precision of \erass sample in our analysis, we will improve our methodology by considering the following additional systematics to realize precision cosmology with the future deeper \erosita surveys. The halo mass function employed in this work, the semi-analytical \citet{Tinker2008} mass function, is constructed for the standard Flat-\lcdm analyses. However, we utilize it even when the total mass of neutrinos $\sum m_{\nu}$ or dark energy equation of state $w$ parameters are varied. Our future work will use the halo mass function emulators with consistent physics to overcome this caveat. 

Additionally, the \citet{Tinker2008} halo mass function is calibrated on the dark matter-only simulations and, therefore, does not include any baryonic physics. It has been shown that baryons can impact the halo mass function, especially when the low-mass and low-redshift haloes, such as galaxy groups, are included in cosmology analyses \citep{Bocquet2016}. In our analysis, the baryonic effects are negligible due to our sample's limited number of low-mass haloes. We estimate that roughly 12\% of the sources in the cosmology catalog have a mass smaller than $10^{14} M_\odot$. Following the method established in the cleaning of the group catalog in \citet{Bahar2024} to identify the group size haloes, we find that our sample has a 30\% spurious fraction below masses M$<10^{14}$~M$_{\sun}$, which will not be included in deeper \erosita data. Despite including these spurious sources, we do not apply further cuts to avoid modifications in the selection function and allow our mixture model to model the contaminants. This effect is completely absorbed in the weak lensing mass bias calibrated in the hydrodynamical simulations. 
However, the baryonic contribution will become a critical bias as we include low-mass haloes in cosmology analysis as the depth of the \erosita survey increases. Indeed, at low redshifts, the limiting factor for our WL mass calibration is the residual uncertainty from baryonic physics.
In the future, the inclusion of low mass halos in the samples will be handled either with a modification to the halo mass function or by further absorbing this effect in the WL bias calibration, as discussed in \citet{Grandis2021a} and \citet{Grandis2024}. 

Further, the effect of the super-sample covariance \citep{Hu2003} on the recovered contour levels of cosmological parameters are negligible in our analysis with \erass sample. 
Following the methods of \citet{Valageas2011} and \citet{Clerc2024}, we estimate the ratio between the sample variance and the Poisson shot noise and found that the sample variance is just a fifth of the Poisson shot noise, whereas an SPT-like sample has a ratio of a tenth. 
The \erass super-sample covariance is small and can be neglected in this analysis. This analysis is also consistent with the reported values in the literature. In particular, \citet{Fumagalli2021} clearly showed that for large area surveys with a mass limit around few $10^{14} M_{\odot}$, such as SPT and \erass, the results from a Poisson likelihood are not affected by the addition of super-sample covariance. In SPT cosmology analyses \citep{Bocquet2019, Bocquet2024}, the authors also ignore the effect of sample variance.
\citet{Fumagalli2021} also showed that the deeper \erosita surveys with a pseudo-flux-limited selection function detecting 200k objects are significantly affected by super-sample covariance. Therefore, as the \erosita survey depth increases and the mass limit of the cluster sample decreases, super-sample covariance will play an essential role in the total error budget in precision cosmology achieved by future \erosita surveys. We will, therefore, fully account for super-sample covariance in our cosmology analyses in the near future.

Calibrating the point-spread function and mirror effective area of \erosita\ is an ongoing effort \citep[see][]{Merloni2024}. In a companion paper by \citet{Bulbul2024}, we observe a flux discrepancy in comparisons of \erosita and \chandra data for extended sources. We stress that this observed difference does not affect our cosmology results. Increasing or decreasing the count rate measurements with a constant factor for all clusters will only shift the normalization of the scaling relations between the count rate and cluster mass and does not alter the slope; therefore, it does not influence our results in a significant way.

The covariant intrinsic correlation between three observables, count rate, weak lensing mass, and richness, and the uncertainties on the selection function \citep{Clerc2024} will be considered in future \erosita cosmological analysis. Significant improvements in both memory usage and likelihood computation time are needed to achieve both of these goals.

Thus far, \erosita completed four All-Sky surveys and 40\% of the fifth survey before it was placed in a safe mode \citep[see][for further details]{Merloni2024}. To project the results from this work to future \erosita\ cosmology work with deeper surveys and \erass's impact compared to other X-ray cluster surveys, utilizing weak lensing observations in their mass calibration, we compute the figure of merit (FOM), as presented in \citet{Garrel2022}. We calculate the area inside the 1$\sigma$ contour in the $\Omega_{\mathrm{m}}$~-~$\sigma_8$ plane using Green's theorem. We find that \erass\ provides an improvement on the cosmological parameters by a factor of 9.4, 8.1, 6.6, 2.7, 5.4, and 5.6 over the latest results from the XXL survey, eFEDS, SPT-SZ, SPT-SZ-polECS-pol500, WtG, DES surveys, respectively \citep{Garrel2022, Chiu2023a, Bocquet2019, Bocquet2024, Mantz2015, Abbott2020}. Overall, \erosita\ provides a factor of 5 to 10 over the published cosmology results from cluster abundances in the literature. This exercise demonstrates the revolutionary power of \erosita in cluster cosmology.

\section{Summary}

\label{sec:summary_conclusions}

In this work, we present the constraints on the cosmological parameters from the clusters of galaxies detected in the first \erosita's All-Sky survey in the western Galactic hemisphere through galaxy cluster mass function. We employ the X-ray count rate of the 5259 confirmed clusters of galaxies, the largest pure ICM-selected sample to date, in the redshift range of 0.1--0.8. The photometric redshifts are measured self-consistently using the Legacy Survey South DR10 data covering a 12,791~deg${^2}$ area. The mass indicator X-ray observable, count-rate, is measured from the \erass survey by reprocessing the X-ray data with advanced analysis, including rigorous background subtraction, the correction for the Galactic absorption, and the sophisticated ICM modeling. The mass calibration between the count rate and weak lensing shear signal from the three large area surveys (DES, HSC, and KiDS) with overlapping footprints of 4968~deg${^2}$ is utilized in this work. In this common area, we use the shear information of a total of 2348 clusters of galaxies in mass calibration likelihood to fine-tune the mass to X-ray observable relation. The sample's selection effects and left-over contamination are fully accounted for in the probability distribution obtained from the \erosita's digital twin in a Bayesian inference forward modeling procedure. Our strategy, combined with higher statistical precision ensured by the \erosita All-Sky survey, leads to better-controlled systematic uncertainties than those available in previous literature works. Thus, it allows for reliable and stringent constraints to be placed on the cosmological parameters.

We built scaling relations between the X-ray count rate and the weak lensing mass for the clusters in the common footprint with DES, KiDS, and HSC. We find that the slope and normalization of the relation are consistent with the studies in the literature utilizing the count rate and luminosity at the $1\sigma$ level. Finally, we highlight that the scaling relation scatter is orthogonal to all constrained cosmological parameters, indicating that reducing the mass-to-count rate scatter (while affecting the mass reconstruction) does not improve our cosmological results significantly because it is the precision (and not the value) that matters.

From the \erass sample and cosmological analysis presented in this paper, we derived the $\sigma_8=0.88\pm 0.02$ with 3\% precision and $\Omega_{\mathrm{m}}=0.29^{+0.01}_{-0.02}$ with $\sim5\%$ precision for the \lcdm scenario. To remove the degeneracy between $\sigma_8$ and $\Omega_{\mathrm{m}}$ and to have a fair comparison between CMB results, we combined these two parameters and measure the $S_{8}=0.86\pm 0.01$ parameter with a 2\% precision. Our results, from a late-time probe, are competitive and consistent with \planck CMB~2020 measurements \citep{Planck2020}, an early-time probe, not confirming the previously reported $S_8$ tension between early and late-time probes \citep{Heymans2021, Huterer2023}. However, we observe a $>3\sigma$ discrepancy on the $S_{8}$ parameter measured in some of the cosmic shear, clustering, and BAO/RSD experiments in the literature \citep{Amon2022, Secco2022, vandenBusch2022, Li2023, Sugiyama2023, Alam2021}.

To test the effect of left-handed neutrinos, we fit an additional set of models to \erass data where the total neutrino mass was allowed to vary. The upper limit to the sum of neutrino masses is $\sum m_\nu < 0.43\mathrm{~eV}$ (95\% upper limit), tighter than the upper limits provided by CMB measurements, for instance, $\sum m_\nu <0.66\mathrm{~eV}$ by \citep{Komatsu2011} and $\sum m_\nu < 0.26\mathrm{~eV}$ by \citet{Planck2020} (all 95\% confidence level), and much tighter than previous cluster count experiments, e.g., SPT-SZ sample \citep{Bocquet2019}, place an upper limit of $\sum m_\nu < 0.74\mathrm{~eV}$ (95\% upper limit) on the sum of the neutrino mass. We finally combine \erass cluster abundance measurements with Planck CMB~2020 chains and find that the constraints on the total mass of neutrinos improve to $\sum m_\nu < 0.11\mathrm{~eV}$. This result represents the tightest bound on the total mass of the left-handed neutrinos to date, and it excludes the inverted hierarchy mass model with a 93\% level. When we combine the \erass cluster abundance measurements with \planck CMB and the neutrino oscillation experiments \citep{Tanabashi2018}, we measured the summed masses to be $\sum m_\nu = 0.09_{-0.02}^{+0.04}\mathrm{~eV}$ and $\sum m_\nu = 0.12_{-0.02}^{+0.03}\mathrm{~eV,}$ assuming a normal and inverted mass hierarchy scenario for neutrino eigenstates. 

We also explored several other cosmological models with \erass cluster abundances and their combination with \planck CMB or Pantheon+ SNe~Ia measurements. Allowing the dark energy equation of state parameter $w$ to be free, we find a best-fit measurement of $w = -1.12\pm0.12$ with 10\% precision, consistent at 1$\sigma$ with the canonical value of $-1$ of the \lcdm model. This measurement is a significant improvement over the constraints placed by the previous cluster abundance experiments with the SPT-SZ survey \citep[$w=-1.55\pm0.41$, ][]{Bocquet2019}. This measurement highlights the remarkable constraining power of \erosita of all the late-time parameters, particularly $\Omega_{\rm m}$, $\sigma_8$, and $w$, without needing an external cosmology experiment to break the degeneracies. When the likelihood is combined with \planck CMB and other geometrical probes, such as BAO, SNe Ia, to break parameter degeneracies, the best-fit value does not change significantly; however, the uncertainties improve significantly ($ w = -1.06 \pm 0.03$) to a 3\% level.
Leaving both $w$ and $\sum m_{\nu}$ free in our fits does not change the cosmological parameters significantly; the upper limit on the summed neutrino masses becomes 0.23~eV, while the equation of state parameter remains consistent with the canonical value ($w = -1.09\pm 0.15$). Including the information from the \planck CMB 2016 data does not significantly improve our constraints on this case.

Finally, we compare all the tested cosmology models against each other in a Bayesian formalism to assess if a preferred model by the \erass data exists. However, we have found no clear preference for a model more complex than the canonical-\lcdm. Furthermore, we studied the correlation between cosmological and scaling relation parameters, finding that $S_8$ depends only on the normalization of the scaling relation, while $\Omega_{\rm m}$ and $\sigma_8$ rely mainly on the mass slope and the redshift evolution of the scaling relation. 

With this work, we have established the reliability of the cluster abundance experiments in precision cosmology. The cosmological constraints obtained from the first \erosita All-Sky Survey performed within the first six months of operations provide a factor of 5 to 9 improvements over the previous results from similar cluster surveys in the literature. 
Future \erass cosmology analyses, enabled by deeper \erosita\ surveys, will allow for higher precision measurements to be drawn for the fundamental cosmological parameters over \erass as the number of clusters increases, and the redshift span of the sample widens.
The robustness of the cluster cosmology experiment demonstrated here corroborates and lends support to many future large experiments, such as \emph{Euclid} and \emph{Vera C Rubin} Observatory studies.
To enable a fair comparison of our cosmological results from cluster abundances with other probes, we provide the chains for all our fits online in  \texttt{GetDist} \citep{getdist} readable format\footnote{The link   has become available after acceptance of the paper.}.

\begin{acknowledgement}

The authors thank the referee for helpful and constructive comments on the draft. 
The authors thank Prof. Catherine Heymans for her help and availability for the implementation of the blinding strategy and her useful comments on the draft and Prof. Anja von der Linden for her comments on the population modeling.

This work is based on data from eROSITA, the soft X-ray instrument aboard SRG, a joint Russian-German science mission supported by the Russian Space Agency (Roskosmos), in the interests of the Russian Academy of Sciences represented by its Space Research Institute (IKI), and the Deutsches Zentrum f{\"{u}}r Luft und Raumfahrt (DLR). The SRG spacecraft was built by Lavochkin Association (NPOL) and its subcontractors and is operated by NPOL with support from the Max Planck Institute for Extraterrestrial Physics (MPE).

The development and construction of the eROSITA X-ray instrument was led by MPE, with contributions from the Dr. Karl Remeis Observatory Bamberg \& ECAP (FAU Erlangen-Nuernberg), the University of Hamburg Observatory, the Leibniz Institute for Astrophysics Potsdam (AIP), and the Institute for Astronomy and Astrophysics of the University of T{\"{u}}bingen, with the support of DLR and the Max Planck Society. The Argelander Institute for Astronomy of the University of Bonn and the Ludwig Maximilians Universit{\"{a}}t Munich also participated in the science preparation for eROSITA.

The eROSITA data shown here were processed using the \esass software system developed by the German eROSITA consortium.
\\

V. Ghirardini, E. Bulbul, A. Liu, C. Garrel, S. Zelmer, and X. Zhang acknowledge financial support from the European Research Council (ERC) Consolidator Grant under the European Union’s Horizon 2020 research and innovation program (grant agreement CoG DarkQuest No 101002585). N. Clerc was financially supported by CNES. T. Schrabback and F. Kleinebreil acknowledge support from the German Federal
Ministry for Economic Affairs and Energy (BMWi) provided
through DLR under projects 50OR2002, 50OR2106, and 50OR2302, as well as the support provided by the Deutsche Forschungsgemeinschaft (DFG, German Research Foundation) under grant 415537506.
V.G. also thanks Margherita Ghirardini for her invaluable joyful support.

\\

The Legacy Surveys consist of three individual and complementary projects: the Dark Energy Camera Legacy Survey (DECaLS; Proposal ID \#2014B-0404; PIs: David Schlegel and Arjun Dey), the Beijing-Arizona Sky Survey (BASS; NOAO Prop. ID \#2015A-0801; PIs: Zhou Xu and Xiaohui Fan), and the Mayall z-band Legacy Survey (MzLS; Prop. ID \#2016A-0453; PI: Arjun Dey). DECaLS, BASS and MzLS together include data obtained, respectively, at the Blanco telescope, Cerro Tololo Inter-American Observatory, NSF’s NOIRLab; the Bok telescope, Steward Observatory, University of Arizona; and the Mayall telescope, Kitt Peak National Observatory, NOIRLab. Pipeline processing and analyses of the data were supported by NOIRLab and the Lawrence Berkeley National Laboratory (LBNL). The Legacy Surveys project is honored to be permitted to conduct astronomical research on Iolkam Du’ag (Kitt Peak), a mountain with particular significance to the Tohono O’odham Nation.

\\

Funding for the DES Projects has been provided by the U.S. Department of Energy, the U.S. National Science Foundation, the Ministry of Science and Education of Spain, the Science and Technology FacilitiesCouncil of the United Kingdom, the Higher Education Funding Council for England, the National Center for Supercomputing Applications at the University of Illinois at Urbana-Champaign, the Kavli Institute of Cosmological Physics at the University of Chicago, the Center for Cosmology and Astro-Particle Physics at the Ohio State University, the Mitchell Institute for Fundamental Physics and Astronomy at Texas A\&M University, Financiadora de Estudos e Projetos, Funda{\c c}{\~a}o Carlos Chagas Filho de Amparo {\`a} Pesquisa do Estado do Rio de Janeiro, Conselho Nacional de Desenvolvimento Cient{\'i}fico e Tecnol{\'o}gico and the Minist{\'e}rio da Ci{\^e}ncia, Tecnologia e Inova{\c c}{\~a}o, the Deutsche Forschungsgemeinschaft, and the Collaborating Institutions in the Dark Energy Survey.

The Collaborating Institutions are Argonne National Laboratory, the University of California at Santa Cruz, the University of Cambridge, Centro de Investigaciones Energ{\'e}ticas, Medioambientales y Tecnol{\'o}gicas-Madrid, the University of Chicago, University College London, the DES-Brazil Consortium, the University of Edinburgh, the Eidgen{\"o}ssische Technische Hochschule (ETH) Z{\"u}rich, Fermi National Accelerator Laboratory, the University of Illinois at Urbana-Champaign, the Institut de Ci{\`e}ncies de l'Espai (IEEC/CSIC), the Institut de F{\'i}sica d'Altes Energies, Lawrence Berkeley National Laboratory, the Ludwig-Maximilians Universit{\"a}t M{\"u}nchen and the associated Excellence Cluster Universe, the University of Michigan, the National Optical Astronomy Observatory, the University of Nottingham, The Ohio State University, the OzDES Membership Consortium, the University of Pennsylvania, the University of Portsmouth, SLAC National Accelerator Laboratory, Stanford University, the University of Sussex, and Texas A\&M University.

Based on observations made with ESO Telescopes at the La Silla Paranal Observatory under programme IDs 177.A-3016, 177.A-3017, 177.A-3018 and 179.A-2004, and on data products produced by the KiDS consortium. The KiDS production team acknowledges support from: Deutsche Forschungsgemeinschaft, ERC, NOVA and NWO-M grants; Target; the University of Padova, and the University Federico II (Naples).

N.O. and S.M. acknowledge JSPS KAKENHI Grant Number JP19KK0076.

The Hyper Suprime-Cam (HSC) collaboration includes the astronomical communities of Japan and Taiwan, and Princeton University. The HSC instrumentation and software were developed by the National Astronomical Observatory of Japan (NAOJ), the Kavli Institute for the Physics and Mathematics of the Universe (Kavli IPMU), the University of Tokyo, the High Energy Accelerator Research Organization (KEK), the Academia Sinica Institute for Astronomy and Astrophysics in Taiwan (ASIAA), and Princeton University. Funding was contributed by the FIRST program from the Japanese Cabinet Office, the Ministry of Education, Culture, Sports, Science and Technology (MEXT), the Japan Society for the Promotion of Science (JSPS), Japan Science and Technology Agency (JST), the Toray Science Foundation, NAOJ, Kavli IPMU, KEK, ASIAA, and Princeton University.

This paper makes use of software developed for the Large Synoptic Survey Telescope. We thank the LSST Project for making their code available as free software at http://dm.lsst.org

This paper is based [in part] on data collected at the Subaru Telescope and retrieved from the HSC data archive system, which is operated by Subaru Telescope and Astronomy Data Center (ADC) at NAOJ. Data analysis was in part carried out with the cooperation of Center for Computational Astrophysics (CfCA), NAOJ. We are honored and grateful for the opportunity of observing the Universe from Maunakea, which has the cultural, historical and natural significance in Hawaii.

I. C. acknowledges the support from the National Science and Technology Council in Taiwan (Grant NSTC 111-2112-M-006-037-MY3).

\\

This work made use of the following Python software packages: 
SciPy\footnote{https://scipy.org/} \citep{Virtanen2020SciPy}, 
Matplotlib\footnote{https://matplotlib.org/} \citep{Hunter2007matplotlib}, 
Astropy\footnote{https://www.astropy.org/} \citep{Astropy2022}, 
NumPy\footnote{https://numpy.org/} \citep{Harris2020numpy},
CAMB \citep{Lewis2011CAMB},
pyCCL\footnote{https://github.com/LSSTDESC/CCL} \citep{Chisari2019pyCCL},
GPy\footnote{https://github.com/SheffieldML/GPy} \citep{gpy2014},
climin\footnote{https://github.com/BRML/climin} \citep{bayer2015climin},
ultranest\footnote{https://github.com/JohannesBuchner/UltraNest/} \citep{Buchner2021}

\end{acknowledgement}

\bibliographystyle{aa}
\bibliography{references}

\begin{appendix}

\section{Mixture model}
\label{app:mixture}
We start by employing our mixture model on the first term of Equation~\eqref{eq:Poisson_log_like_final}, by explicitly dividing the three populations that comprise our sample; clusters, RSs, and AGNs, as:

\begin{align}
& \frac{dN_{tot}}{d\hat{C}_R d\hat{z} d\hat{\mathcal{H}}_i} 
P(I | \hat{C}_R, \hat{z} ,\hat{\mathcal{H}}_i) = 
\frac{dN_C}{d\hat{C}_R d\hat{z} d\hat{\mathcal{H}}_i} P(I | \hat{C}_R, \hat{z}, \hat{\mathcal{H}}_i) 
\nonumber \\
& \quad \quad + \frac{N_C}{(1-f_{\rm RS}-f_{\rm AGN})} \, \bigg( f_{\rm RS} \, P(\hat{C}_R, \hat{z}, \hat{\mathcal{H}}_i | {\rm RS}) \nonumber \\ 
& \quad \quad + f_{\rm AGN} \, P(\hat{C}_R, \hat{z}, \hat{\mathcal{H}}_i | {\rm AGN}) \bigg)
\label{mix:dN}
\end{align}

\noindent where $P(\hat{C}_R, \hat{z}, \hat{\mathcal{H}}_i | {\rm RS})$ and $P(\hat{C}_R, \hat{z}, \hat{\mathcal{H}}_i | {\rm AGN})$ are the probability density function (PDF) for the contaminants that represent RSs and AGNs, respectively, while the first term, 
$\frac{dN_C}{d\hat{C}_R d\hat{z} d\hat{\mathcal{H}}_i} P(I | \hat{C}_R, \hat{z}, \hat{\mathcal{H}}_i)$,
represents the distribution of our selection observable for the detected clusters.

On the second term of Equation~\eqref{eq:Poisson_log_like}, we use the relation between these three populations when computing the total number of detected objects as,

\begin{equation}
 N_{tot} = \frac{N_{\rm C}}{1-f_{\rm RS}-f_{\rm AGN}}
 \label{mix:N}
\end{equation}

\noindent where

\begin{equation}
N_{\rm C} = \iiiint \frac{dN_{\rm C}}{d\hat{C}_R d\hat{z} d\hat{\lambda} 
 d\hat{\mathcal{H}}_i} 
P(I | \hat{C}_R, \hat{z}, \hat{\mathcal{H}}_i) 
\Theta(\hat{\lambda} > 3) d\hat{C}_R d\hat{z} d\hat{\lambda} d\hat{\mathcal{H}}_i
\end{equation}

On the third term in Equation~\eqref{eq:Poisson_log_like}, when using the optical information, we write

\begin{align}
P(\hat{\lambda} | \hat{C}_R, \hat{z}, \hat{\mathcal{H}}_i, I) &= 
\bigg( 1 - f_{\rm RS} \, \frac{P(\hat{C}_R, \hat{z}, \hat{\mathcal{H}}_i | {\rm RS}, I)}{P(\hat{C}_R, \hat{z}, \hat{\mathcal{H}}_i, I)} \nonumber \\
& \quad \quad - f_{\rm AGN} \, \frac{P(\hat{C}_R, \hat{z}, \hat{\mathcal{H}}_i | {\rm AGN}, I)}{P(\hat{C}_R, \hat{z}, \hat{\mathcal{H}}_i, I)} \bigg) 
\nonumber \\
& \quad \quad \times \ P(\hat{\lambda} | \hat{C}_R, \hat{z}, \hat{\mathcal{H}}_i, I, C) \nonumber \\
&\quad + f_{\rm RS} \, \frac{ P(\hat{\lambda}, \hat{z} | {\rm RS}, I) \, P(\hat{C}_R, \hat{\mathcal{H}}_i | {\rm RS}, I)}{P(\hat{C}_R, \hat{z}, \hat{\mathcal{H}}_i, I)} \nonumber \\
&\quad + f_{\rm AGN} \, \frac{P(\hat{\lambda}, \hat{z} | {\rm AGN}, I) \, P(\hat{C}_R, \hat{\mathcal{H}}_i | {\rm AGN}, I)}{P(\hat{C}_R, \hat{z}, \hat{\mathcal{H}}_i, I)}
\end{align}

Finally, on the last term in Equation~\eqref{eq:Poisson_log_like}, when using the weak lensing information;

\begin{align}
P(\hat{g}_{t} | \hat{C}_R, \hat{z}, \hat{\mathcal{H}}_i, I) &= 
\bigg( 1 - f_{\rm 
 RS} \, \frac{P(\hat{C}_R, \hat{z}, \hat{\mathcal{H}}_i | {\rm RS}, I)}{P(\hat{C}_R, \hat{z}, \hat{\mathcal{H}}_i, I)} \nonumber \\
& \quad \quad - f_{\rm AGN} \, \frac{P(\hat{C}_R, \hat{z}, \hat{\mathcal{H}}_i | {\rm AGN}, I)}{P(\hat{C}_R, \hat{z}, \hat{\mathcal{H}}_i, I)} \bigg) \nonumber \\
& \quad \quad \times \ P(\hat{g}_{t} | \hat{C}_R, \hat{z}, \hat{\mathcal{H}}_i, I, C) 
\nonumber \\
 + \ f_{\rm RS} \, & \frac{ P(\hat{g}_{t} | {\rm RS}, I) \, P(\hat{z} | {\rm RS}, I) \, P(\hat{C}_R, \hat{\mathcal{H}}_i | {\rm RS}, I)}{P(\hat{C}_R, \hat{z}, \hat{\mathcal{H}}_i, I)} \nonumber \\
 + \ f_{\rm AGN} \, & \frac{P(\hat{g}_{t} | \hat{z}, {\rm AGN}, I) \, P(\hat{z} | {\rm AGN}, I) \, P(\hat{C}_R, \hat{\mathcal{H}}_i | {\rm AGN}, I)}{P(\hat{C}_R, \hat{z}, \hat{\mathcal{H}}_i, I)}.
\end{align}

To estimate the overall total contamination in the sample, we use the PDF of count rate and sky position information, that is: $P(\hat{C}_R, \hat{\mathcal{H}}_i | {\rm AGN}, I)$, and $P(\hat{C}_R, \hat{\mathcal{H}}_i | {\rm RS}, I)$ of the AGN and RS populations, obtained from our realistic \erass\ simulations \citep{Seppi2022, Comparat2020}. We obtain the AGN and RS PDF in redshift and richness \citep{Kluge2024}, i.e., $P(\hat{\lambda}, \hat{z} | {\rm AGN}, I)$ from the \eromapper\ run on locations of the \erass\ point sources \citep{Kluge2024, Merloni2024} and $P(\hat{\lambda}, \hat{z} | {\rm RS}, I)$ from the \eromapper\ run at random lines of sight. 

Finally, we use the predicted AGN shear profiles reported in the \citet{Comparat2023}. In contrast, we assume that the shear profile is zero at all radii for the RS component. Therefore the shear signal is consistent with a noise fluctuation, as elaborated in \citet{Grandis2024}, Sect.~5.1.1.

\clearpage
\section{Additional tests on systematics}
\label{app:tests}
\subsection{Relation between scatter and normalization}

The reported normalization parameters for the X-ray count rate to mass scaling relation reported in \citet{Chiu2022, Chiu2023a} are in tension with our results. In particular, the difference in normalization is due to the observed degeneracy between the normalization and the scatter. \citet{Chiu2022, Chiu2023a} assume an informative prior on the count rate mass scatter, $\sigma_\text{X}= 0.30 \pm 0.08$ derived by \citet{grandis19} from the luminosity and temperature scaling relations measured by \citet{Bulbul2019}. This choice artificially breaks the amplitude scatter degeneracy. We show this in the following Fig.~\ref{fig:Ax_scatter} and the selection of the sample \citep{Bahar2022}. For a more detailed comparison of the scaling relation parameters, we refer to \citet{Kleinebreil2024} and \citet{Grandis2024}.

\begin{figure}[h!]
 \centering
 \includegraphics[width=0.5\textwidth]{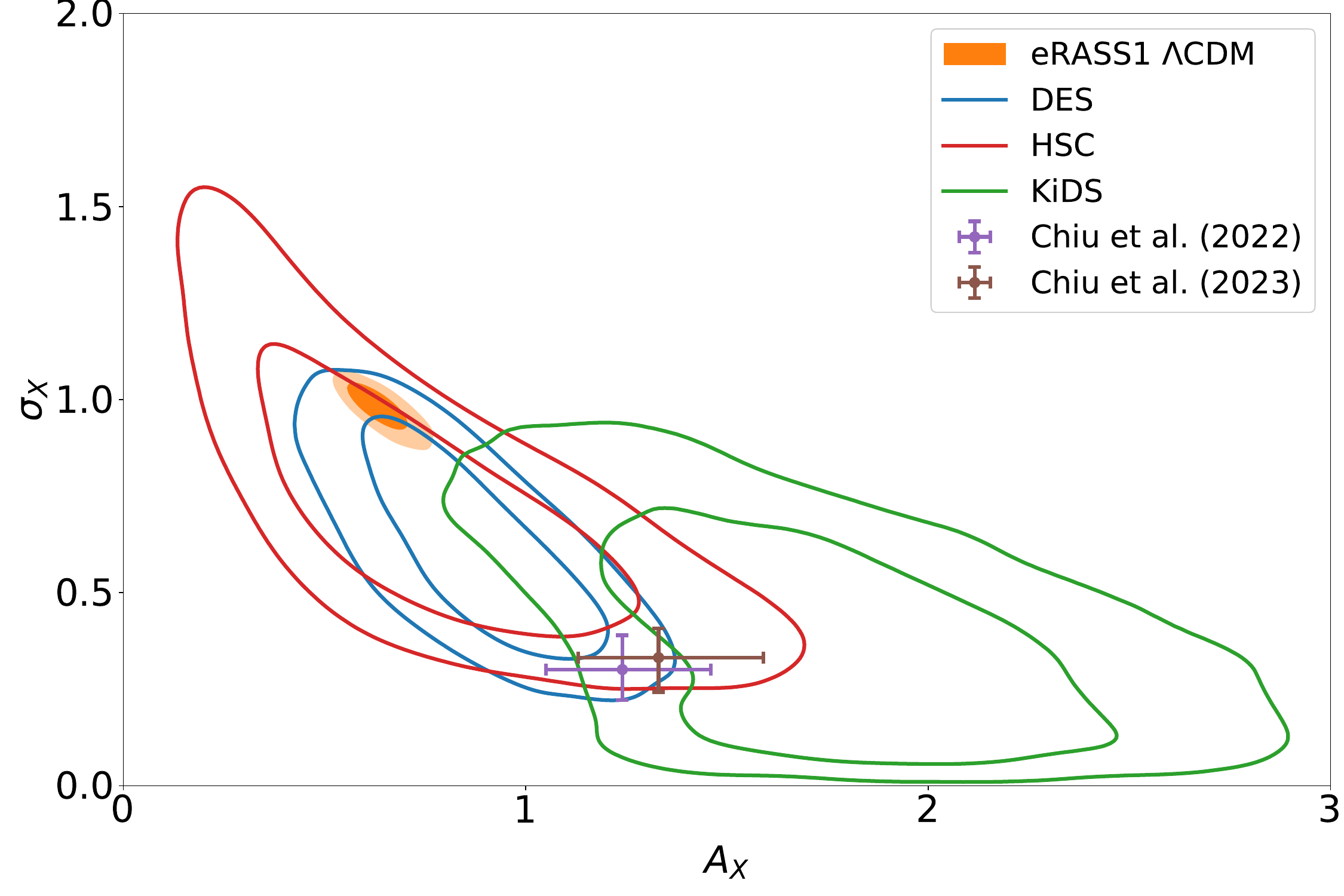}
 \caption{Degeneracy between scatter and amplitude of the count rate to mass scaling relation. We show our obtained value and the comparison with individual mass calibration from DES, KiDS, and HSC as reported in \citet{Kleinebreil2024, Grandis2024}. The previous measurements from the eFEDS survey presented in \citet{Chiu2022} and \citet{Chiu2023a} are shown with crosses.}
 \label{fig:Ax_scatter}
\end{figure}

\FloatBarrier

\subsection{Relation between cosmology and scatter parameters}
In the following figures, Fig.~\ref{fig:LCDM_degeneracy} and \ref{fig:LCDM_correlation}, we show the posterior and correlation between the cosmology parameters in the fit and the X-ray scaling relation parameters. 

\begin{figure}[h]
 \centering
 \includegraphics[width=0.5\textwidth]{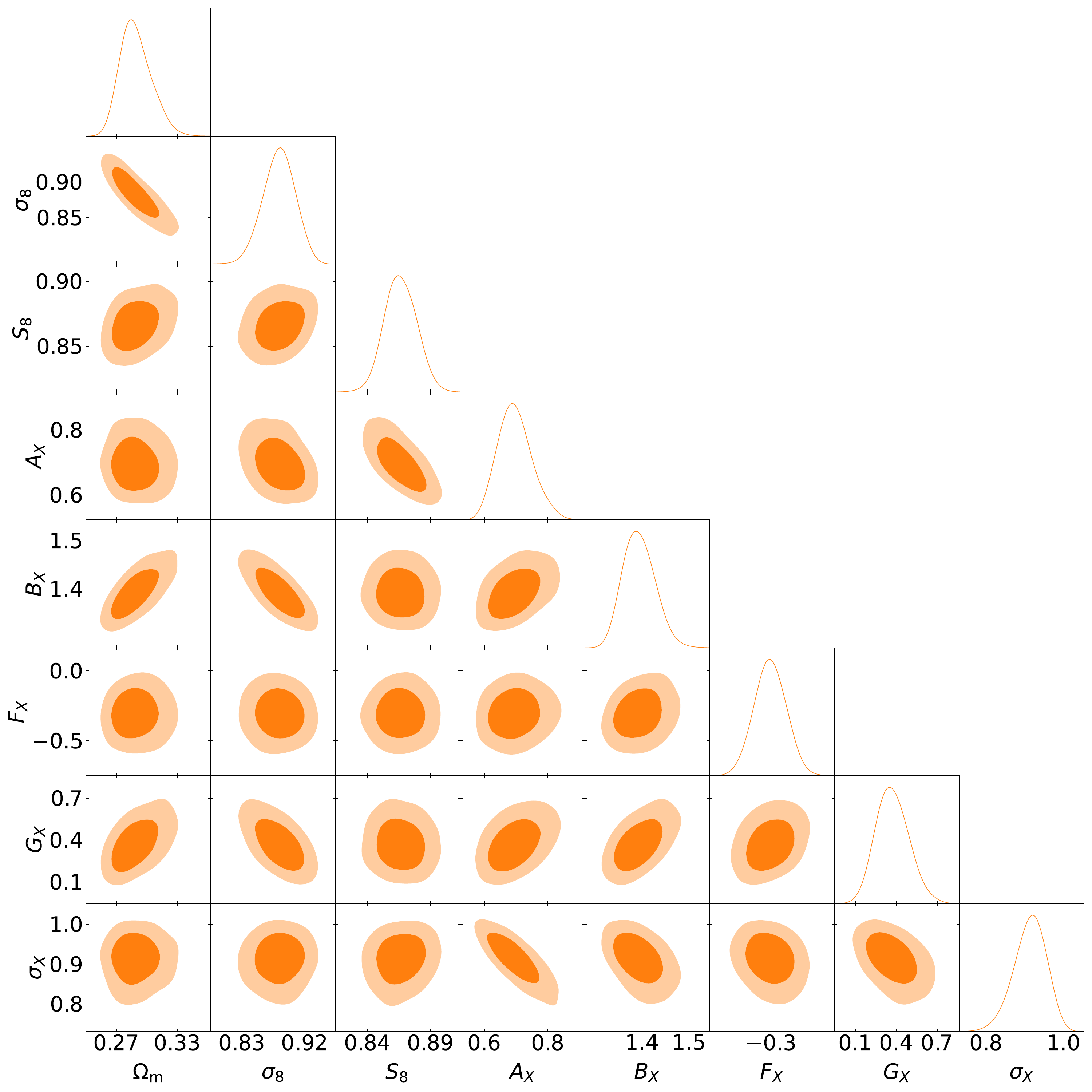}
 \caption{Degeneracies between cosmological parameters and X-ray count rate scaling relation parameters. It is interesting to point out that the scaling relation parameters affect the degeneracy between $\Omega_{\rm m}$ and $\sigma_8$. }
 \label{fig:LCDM_degeneracy}
\end{figure}

\begin{figure}[h]
 \centering
 \includegraphics[width=0.5\textwidth]{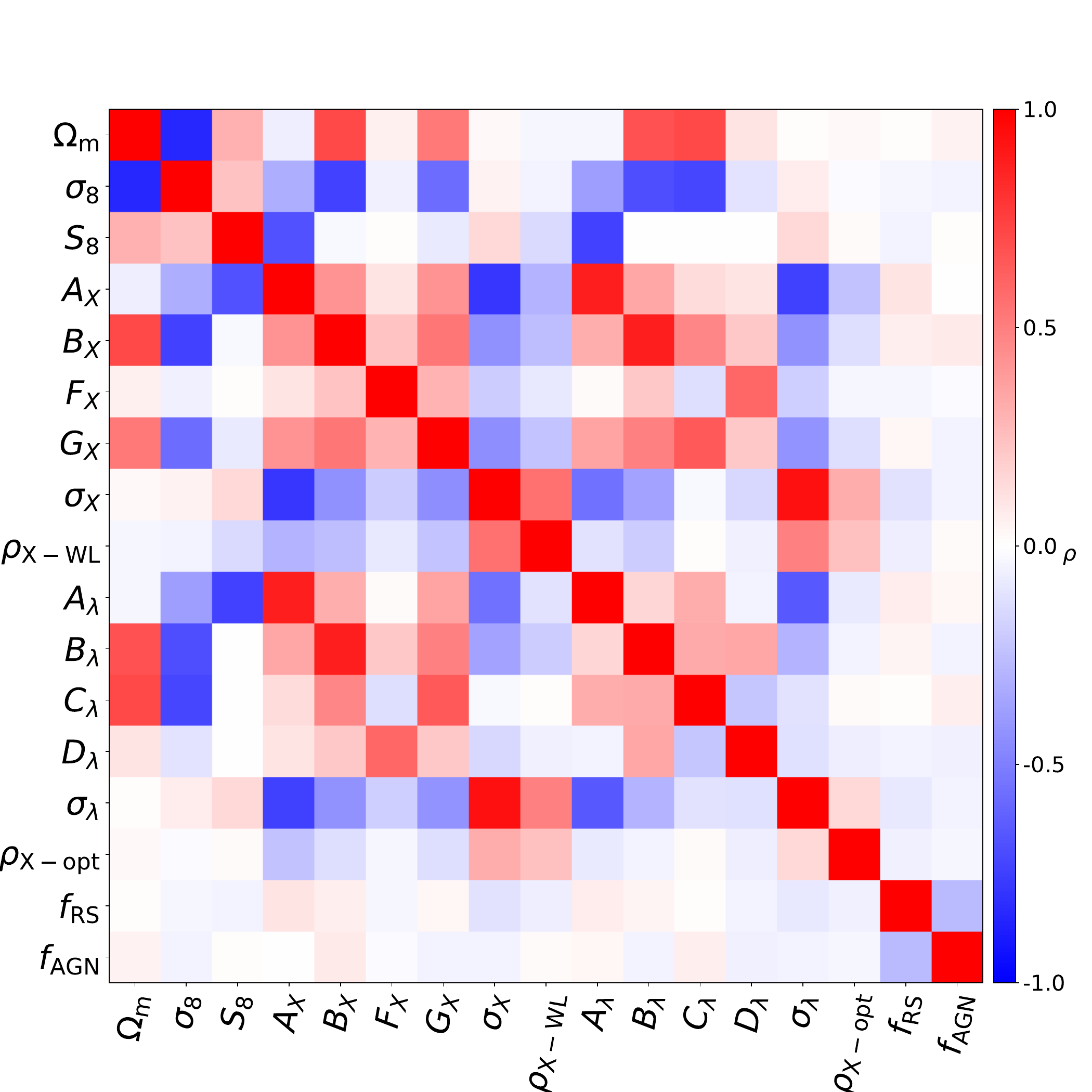}
 \caption{Correlation between fitted parameters. }
 \label{fig:LCDM_correlation}
\end{figure}

\FloatBarrier
\clearpage

\subsection{$\Omega_\mathrm{m}$ vs $\sigma_8$ in the several models fitted to our dataset}

In this paper, we have fitted for several different cosmological models. 
All of these extensions have in common the starting point in the standard vanilla \lcdm cosmological model.
We have fitted the \lcdm, and then we have allowed several parameters to be free. 
Particularly dark energy equation of state $w$ and summed neutrino masses $\sum m_\nu$. 
Finally, for comparison purposes with \lcdm fit in \planck analysis \citep{Planck2020}, we have also fitted the \lcdm model with $\sum m_\nu$ fixed to 0.06 eV.
It is important to show that these extra free parameters do not significantly affect the base robust parameters we are fitting and constraining in the \lcdm: $\Omega_\mathrm{m}$ vs $\sigma_8$.
We show this test in Fig.~\ref{fig:eRASS1_omegam_sigma8}.

\begin{figure}[h]
 \centering
 \includegraphics[width=0.5\textwidth]{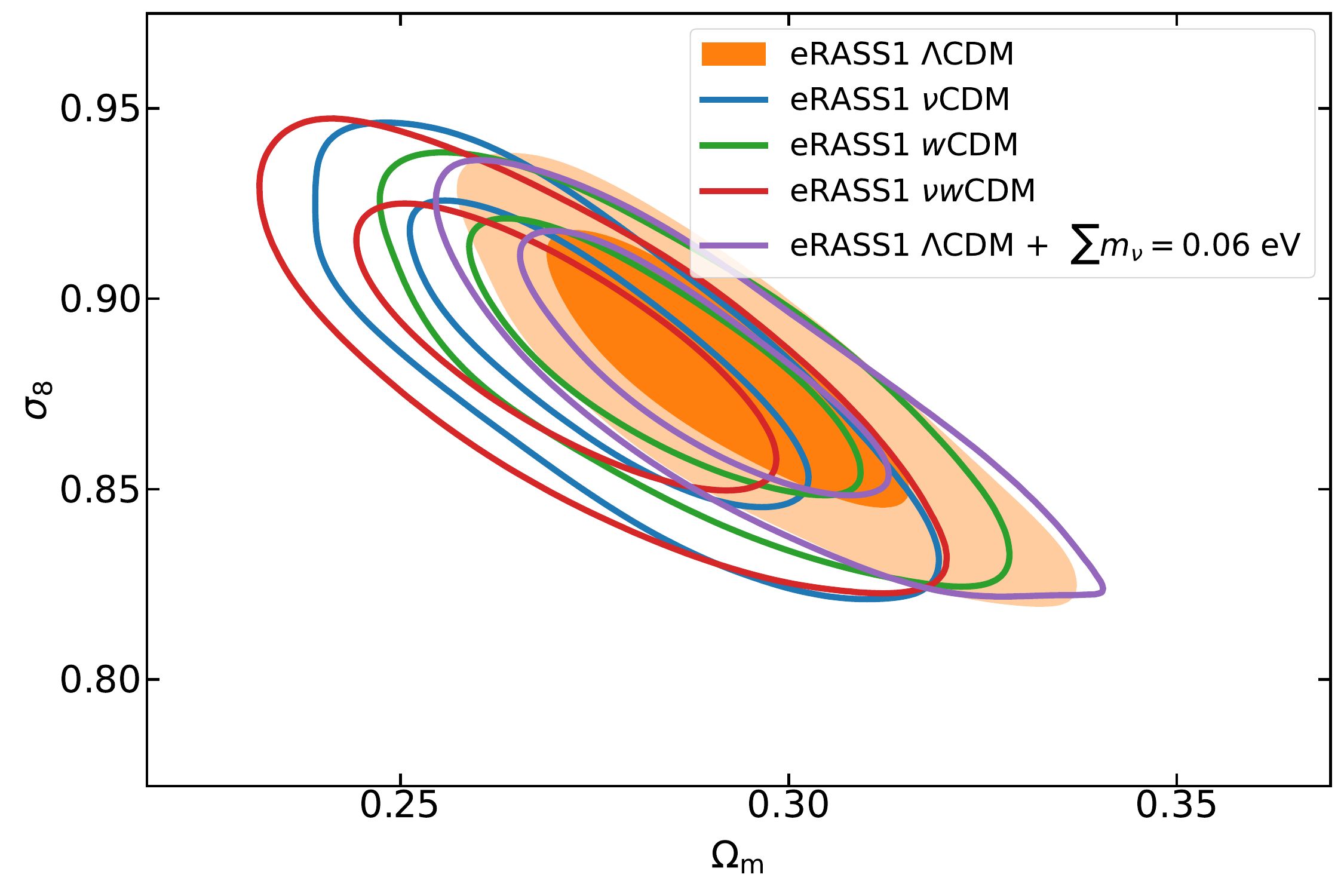}
 \caption{Posterior on $\Omega_{\mathrm{m}}$-$\sigma_8$ from the \lcdm, $\nu$CDM, $w$CDM, $\nu w$CDM, and \lcdm with $\sum m_\nu = 0.06\mathrm{~eV}$ fits from our \erass cluster dataset. We see the nice overlap in the parameter space that indicates that our fit in the $\Omega_{\mathrm{m}}$-$\sigma_8$ plane is stable against modification to the standard \lcdm cosmological model.}
 \label{fig:eRASS1_omegam_sigma8}
\end{figure}

\FloatBarrier

\subsection{Tests of the effect on $\Omega_\mathrm{m}$ vs $\sigma_8$ from our modeling choices}
\label{hmf:tests}

In this paper, we have applied several cuts to the datasets, for instance, $0.1<z<0.8$, EXT\_LIKE$>6$, we have chosen a specific halo mass function in the \citet{Tinker2008}, and we have employed the mixture model. The following two figures show the effect of our constraints on $\Omega_\mathrm{m}$ vs $\sigma_8$ in \lcdm cosmological model.
In Fig.~\ref{fig:tests} in Appendix~\ref{app:tests}, we show the impact of not applying the mixture model, cutting the data in redshift at 0.6 instead of 0.8, and changing the EXT\_LIKE cut to a slightly more conservative value of 10. We see changes within the 1$\sigma$ level for all these cases. 
In Fig~\ref{fig:tests_hmf}, we show the effect of changing the halo mass function from \citet{Tinker2008} to either \citet{Tinker2010} or \citet{Despali2016}. 
We further tested the effects of systematic uncertainties on the halo mass function by applying the formalism introduced in \citet{Costanzi2019} and used in the literature \citep{Abbott2020, Bocquet2024} to correct \citet{Tinker2008} model adding two nuisance parameters, $q_{\rm HMF}$ and $s_{\rm HMF}$, which allow for extra degrees of freedom for both normalization and slope of the halo mass function. We used the following priors on these parameters: $q_{\rm HMF} \sim \mathcal{N}(1, 0.035)$ and $s_{\rm HMF} \sim \mathcal{N}(0, 0.02)$ as in the analysis of \citet{Bocquet2024}. 
We can see that $\Omega_\mathrm{m}$ is unchanged, while $\sigma_8$ is lowered by around 1$\sigma$. This indicates that change in mass function does not significantly impact our constraints; however, in future \erosita cosmological analysis, we will need to pay much more care and attention to our modeling choice regarding the halo mass function.

\begin{figure}[h]
 \centering
 \includegraphics[width=0.5\textwidth]{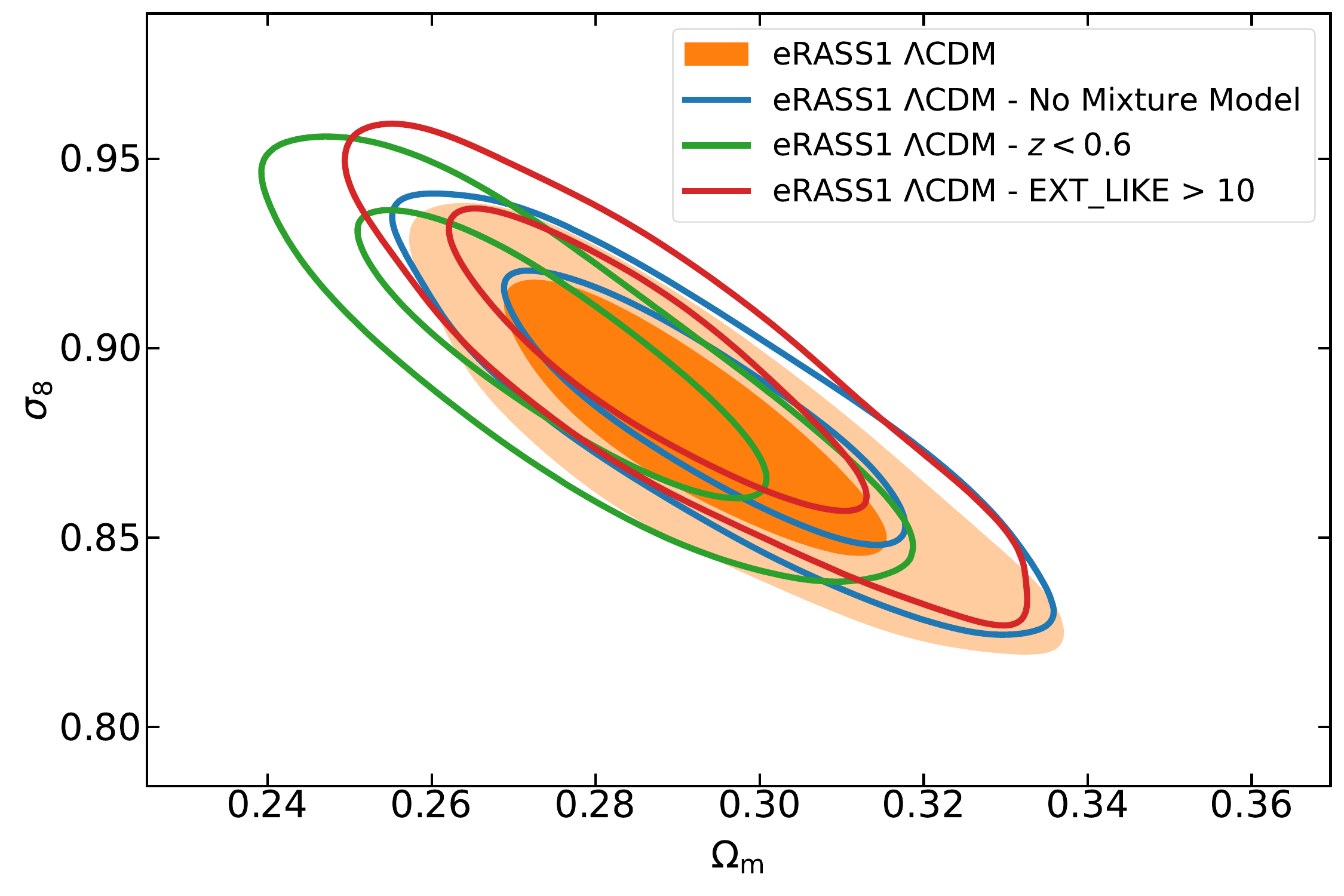}
 \caption{Comparison of our main \lcdm results (orange) with several tests: switching off the mixture model likelihood of Sect.~\ref{sec:mixture} (blue), restricting the redshift range $z < 0.6$. 
 }
 \label{fig:tests}
\end{figure}

\begin{figure}[h]
 \centering
 \includegraphics[width=0.5\textwidth]{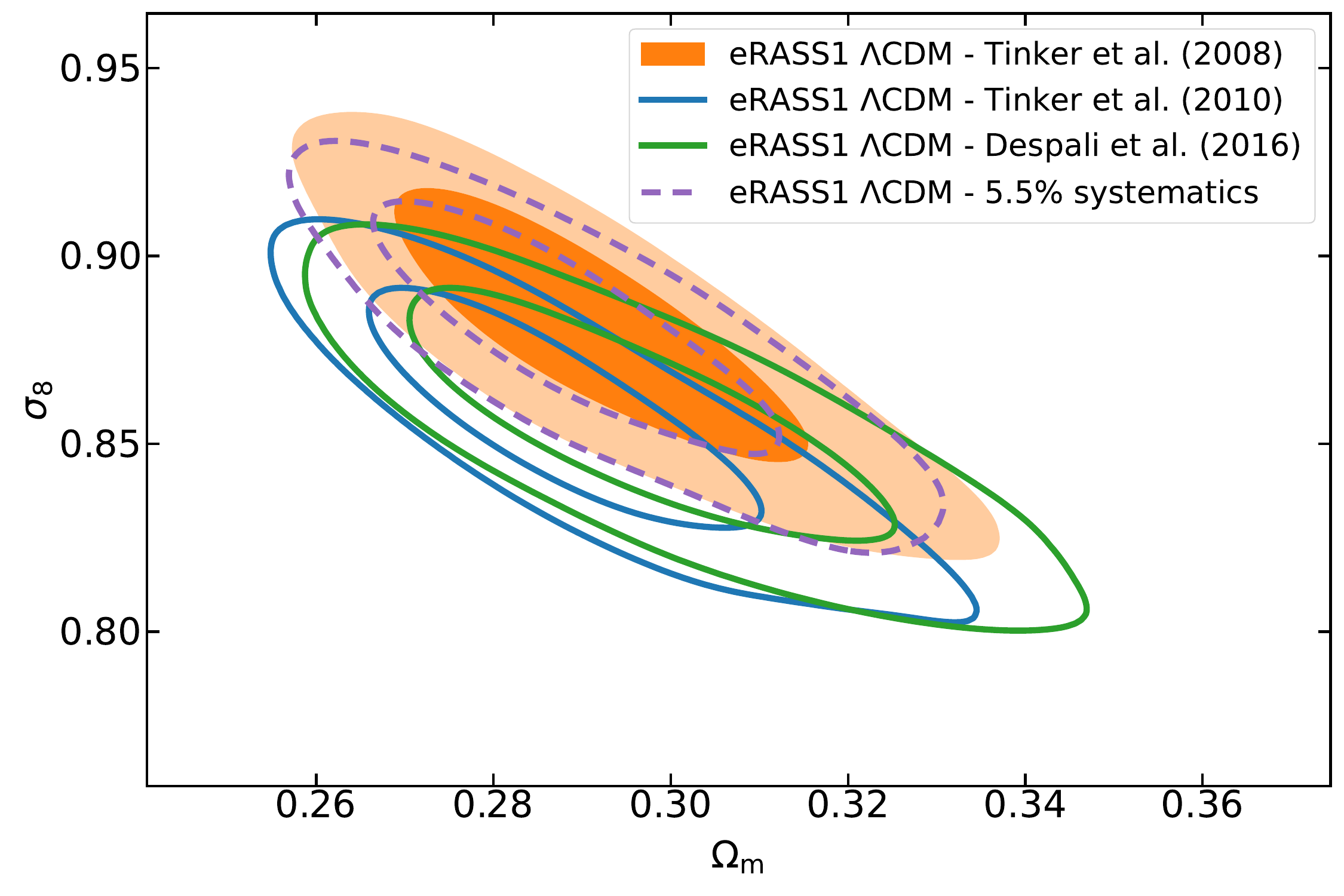}
 \caption{Robustness of our main \lcdm results (orange) against changes in the halo mass function: \citet{Tinker2008} (blue), and \citet{Despali2016} (green), and 5.5\% systematics uncertainty added to the halo mass function of \citet{Tinker2008} following the prescription of \citep{Costanzi2019, Abbott2020}}. 
 \label{fig:tests_hmf}
\end{figure}

\FloatBarrier
\clearpage

\section{Constraints from neutrino summed masses to the mass of the lightest}
\label{app:neutrino}

A constraint on the summed neutrino mass will shed light on the mass of the lightest neutrino eigenstate.
Obviously, this also depends on the neutrino mass hierarchy as:

\begin{align}
\sum m_\nu = 
\begin{cases}
 m_L + \sqrt{m_L^2 + \Delta m_{21}^2} + \sqrt{m_L^2 + \Delta m_{31}^2} & \textrm{Normal Hierarchy} \\
 m_L + \sqrt{m_L^2 + \Delta m_{32}^2} + \sqrt{m_L^2 + \Delta m_{31}^2} & \textrm{Inverted Hierarchy}
\end{cases}
\end{align}
where $m_L$ is the mass of the lightest neutrino eigenstate.
We reproduce Fig.~3 in \citet{Qian2015} with updated parameters from NuFIT \citep{Esteban2020} in Fig.~\ref{fig:neutrino_hierarchy}. In the same figure, we also show with horizontal dashed lines the constraints on summed neutrino masses for the combination of \erosita with \planck CMB and lower allowed limit from neutrino oscillation experiments \citep{Tanabashi2018}.

\begin{figure}[h]
 \centering
 \includegraphics[width=0.5\textwidth]{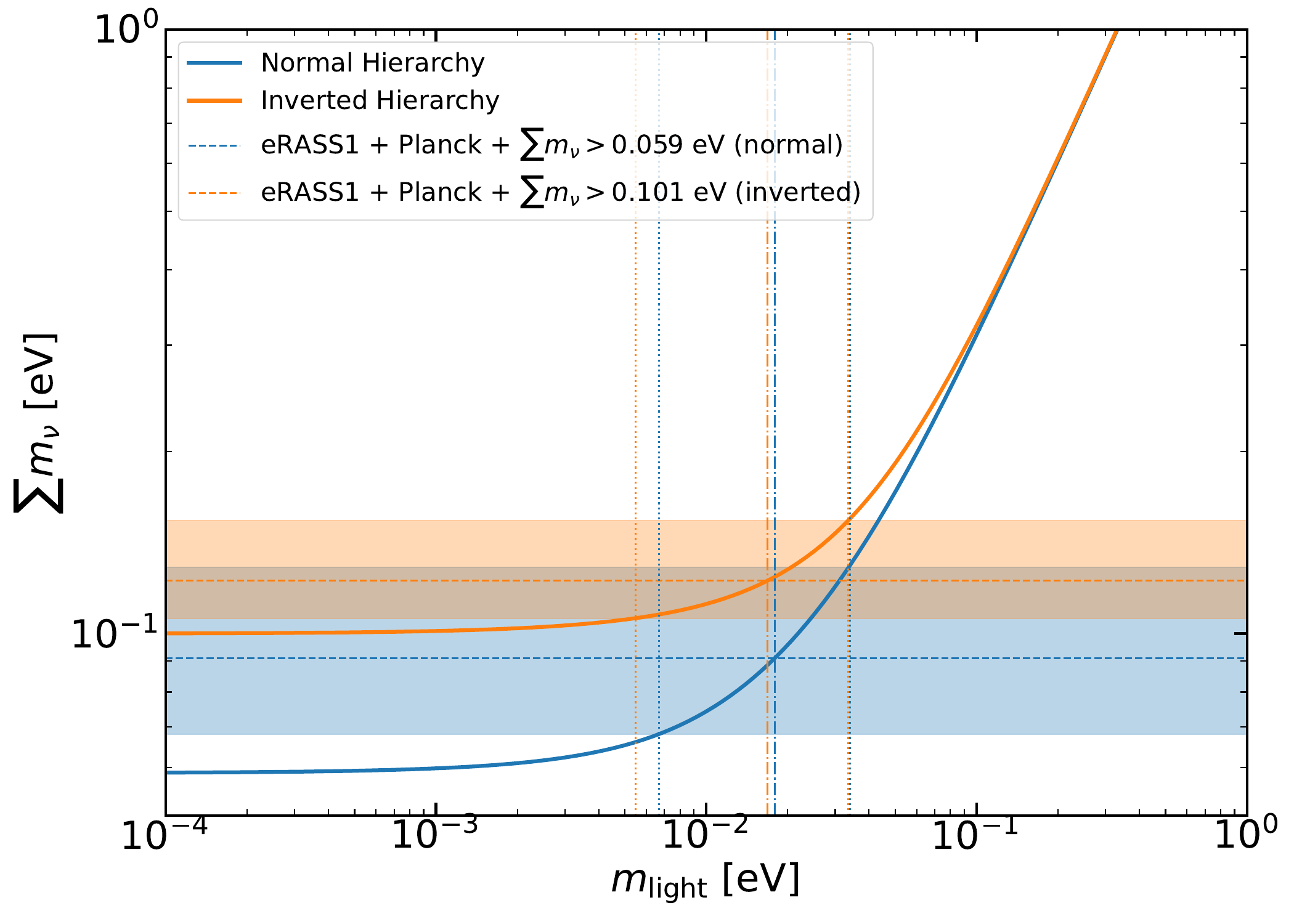}
 \caption{Current estimates of the relation between the mass of the lightest neutrino eigenstate and the summed neutrino masses (solid lines). The horizontal dashed lines and the shaded area represent the \erass and CMB combined estimate. The vertical dash-dotted line represents the lightest neutrino value corresponding to our summed mass estimate. Finally, the vertical dotted line corresponds to the $1\sigma$ uncertainty around the lightest neutrino mass estimate. We use blue and orange to distinguish between normal and inverted hierarchy. It is remarkable to show that although the difference between the models, the estimation for the lightest neutrino mass is almost identical.}
 \label{fig:neutrino_hierarchy}
\end{figure}

\FloatBarrier
\clearpage

\section{Optical and X-ray completeness} 
\label{app:Optical and X-ray completeness}

\begin{minipage}{\textwidth}
 \centering
 \includegraphics[width=\textwidth]{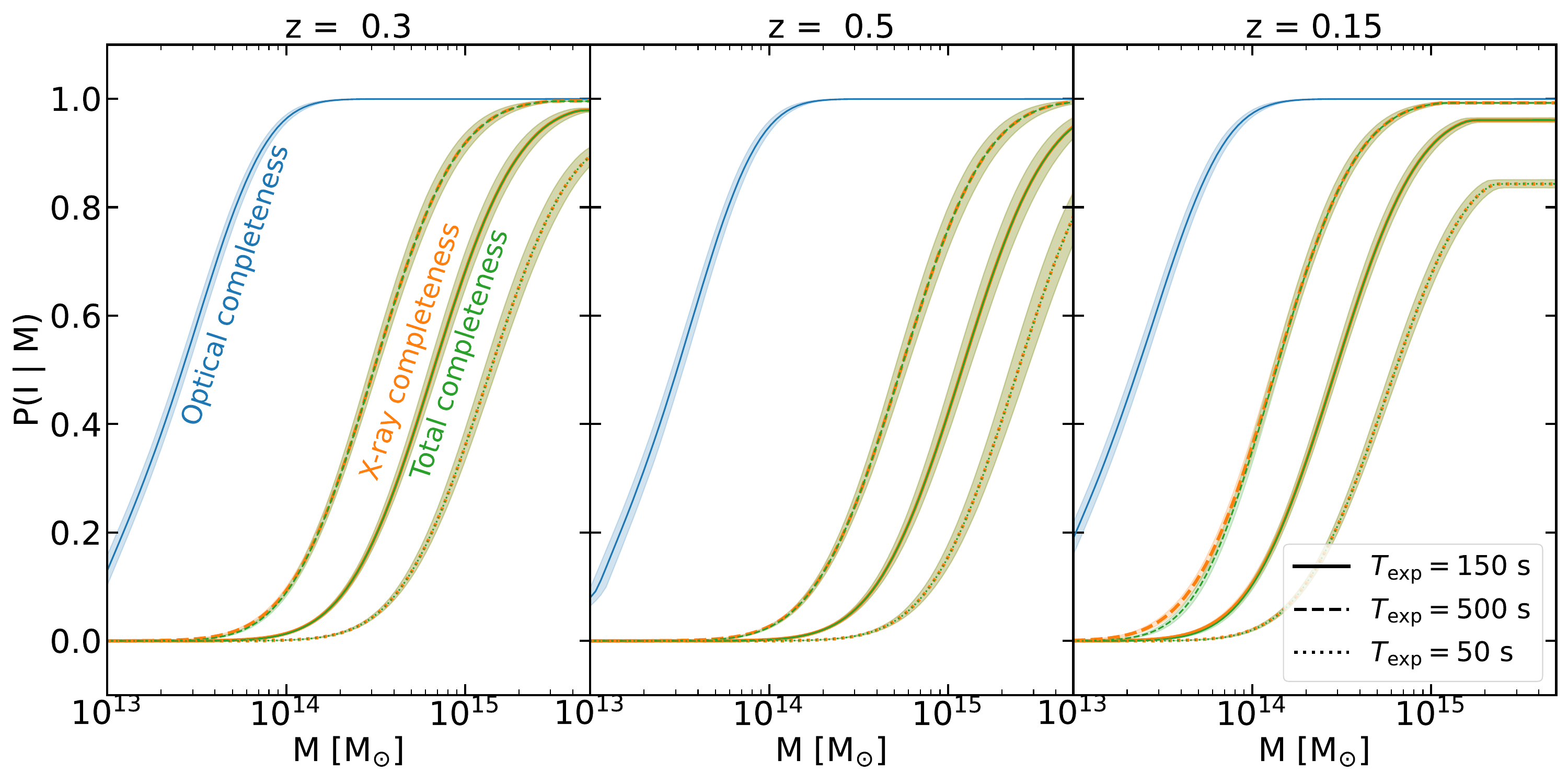}
 \captionof{figure}{Optical (blue), X-ray (orange), and total (green) completeness computed as a function of mass from the best-fit values. The three panels show different redshift, 0.3, 0.5, and 0.15, and in each panel, a solid line corresponds to a exposure of 150 s, dashed to 500 s, and dotted to 50 s. The shaded area around each line represent the 1$\sigma$ uncertainty from resampling the best fitting chain 1000 times.}
 \label{fig:completeness}
\end{minipage}
\vspace{0.5cm}

As we have chosen to pursue a forward modeling approach, the completeness is not pre-determined, but it is modeled on the fly by propagating the selection function in count rate and the selection function in richness to halo mass via the observable mass relations and their correlated scatters.
Following the approach proposed in \citet{Grandis2020} for modeling the total selection function of X-ray-selected, optically confirmed clusters, we computed the completeness curves as a function of mass and redshift for optical and X-ray selection functions.
For the optical, we have:
\begin{equation}
 P(I | M, z) = \iint 
 \Theta(\hat{\lambda} > 3) \,
 P(\hat{\lambda} | \lambda) \,
 P(\lambda | M, z) \,
 d\lambda d\hat{\lambda},
\end{equation}
while for the X-ray, we have
\begin{equation}
 P(I | M, z, \hat{\mathcal{H}}_i) = \iint 
 P(I | C_R, z, \hat{\mathcal{H}}_i) \,
 P(C_R | M, z) \,
 dC_R.
\end{equation}
We computed both of these as a function of mass and displayed them in Figure~\ref{fig:completeness}. In practice, the two incompleteness functions are allowed to correlate by the unknown correlation coefficient $\rho_{\lambda, C_\text{R}}$, which we also sample on the fly. 
As appears from the figure, the optical cut $\hat \lambda>3$ only plays a minor role for low redshift, high exposure time regions, where it leads to a minimal extra incompleteness for the few $3\times 10^{13} M_\odot$ halos that up-scatter into our sample. Computing the total number of expected objects with and without completeness indicates that this effect would reduce the total number of expected detections by 1.1\%, if not modeled.

\FloatBarrier
\clearpage

\section{$\sigma_8$}
\label{app:sigma8}
$\sigma_8$ is defined as the amplitude of linear matter fluctuations in a spherical top hat of radius $R = 8 h^{-1}$ Mpc at redshift $z = 0$, i.e., 
\begin{equation}
 \sigma_8^2 = \int_0^\infty \Delta^2(k,z=0) \left( \frac{3 j_1(kR)}{kR} \right)^2 d \log k 
\end{equation}
where $j_1(kR)$ is the spherical Bessel function of the first kind of order one, and $\Delta^2(k,z=0)$ is the power of linear matter fluctuation in logarithmic bins of wave number $k$ at redshift $z=0$,
\begin{align}
 \Delta^2(k, z=0) &= \frac{k^3 P(k, z=0)}{2\pi^2} \nonumber \\
 &= \frac{A_s}{\left( \Omega_\mathrm{m} h^2 \right)^2} \frac{4}{25} \left( \frac{k}{k_\mathrm{piv}} \right)^{n_s - 1} \left( \frac{k}{100} \right)^4 g^2(z=0) T^2(k)
\end{align}
where $g(z)$ is the growth suppression factor \citep[see Equation~10 in ][]{Huterer2023} and $T(k)$ is the linear transfer function.
It is therefore clear that $\sigma_8$ is correlated with the combination of cosmological parameters $\frac{\sqrt{A_s}}{ \Omega_\mathrm{m} h^2 }$.

\FloatBarrier
\clearpage

\section{Systematics on scaling relation modeling}
\label{app:Systematics on scaling relation modeling}
The measured intrinsic scatter in count rate is surprisingly large, $\sigma_{X} = 0.98^{+0.05}_{-0.04}$, see in Table~\ref{tab:scaling_fit}, compared with the value of $\sim$0.4 reported in WtG \citep{Mantz2016}, or the value of 0.33 reported in eFEDS \citep{Chiu2022}.
Given that our cluster sample spans a much larger range in mass and redshift than previous samples, this may indicate mass- and/or redshift-dependence of the scatter. Here, we explore a more complex model of the scatter and its impact on scaling relation and cosmological parameters. We note that this test was performed after unblinding and that we consider the model presented in Sect.~\ref{sec:scaling_relations} as our main analysis.

Thus, we model the scatter, allowing mass and redshift trends as

\begin{equation}
\sigma_X (M, z) = \sigma_{X, 0} + \sigma_{X, z} \, \log \frac{1+z}{1+z_p} + \sigma_{X, M} \, \log \frac{M}{M_p}
\end{equation}

\noindent for the X-ray count rate, and

\begin{equation}
\sigma_\lambda (M, z) = \sigma_{\lambda, 0} + \sigma_{\lambda, z} \, \log \frac{1+z}{1+z_p} + \sigma_{\lambda, M} \, \log \frac{M}{M_p}
\end{equation}

\noindent for the richness. Finally, we further unfreeze $s_{M}$ parameter, see Equation~\eqref{eq:EWL}, enabling the scatter of the WL bias to depend on mass, as it was already redshift dependent. 

We further decided to allow a mass trend in the mass slope of the count rate to mass scaling relation.
We thus modify the scaling relations as in Sect.~\ref{sec:scaling_relations}, changing Equation~\eqref{eq:xrayscaling} as follows:

\begin{equation}
\left\langle \log \frac{C_R}{C_{R,p}} \bigg| M, z \right\rangle = 
\log A_X + 
b_X(M, z) \, \log \frac{M}{M_p} + e_x(z)
,\end{equation}

\noindent where $e_x(z)$ remains as of Equation~\eqref{eq:e_xz}, while the mass-redshift dependent slope of the scaling relation is expressed by the term $b_X(M, z)$ as:

\begin{equation}
b_X(z) = B_X + F_X \, \log \frac{1+z}{1+z_p} + C_X \, \log \frac{M}{M_p}.
\end{equation}

\noindent Here, $C_X$ allows the mass-dependent slope.

It is clear that this test introduces six extra free parameters, and therefore, assesses the level of the systematics of the model choices we have made throughout this paper. 
What we find is quite interesting. In fact, we discover a degeneracy between mixture model parameters, $f_{AGN}$ and $f_{RS}$, and the scatter evolution and mass dependence parameters. 
This degeneracy can very well be explained physically. The population that occupies the low-count-rate, low-richness can be comprised of both contaminants that our detection and confirmation pipeline is not able to distinguish from a galaxy cluster or low-mass groups of galaxies that upscatter in the scaling relation by, for instance, because of the presence of AGNs in its core \citep{Bulbul2022}. 
Our simulations \citep{Comparat2020, Seppi2022} suggest that low-count-rate, low-richness entries in our catalog are due to AGN and RS contaminants. However, only a chance overlap between AGNs and clusters was introduced in \erosita digital twin, while AGNs inside group and cluster-sized halos were not introduced. However, the situation in the real extended source cosmology catalog might be somewhat different.

Allowing the scatter to change as a function of mass and redshift effectively enables the degeneracy between these two populations mentioned above, and indeed, the chains prefer a non-negligible fraction of low-mass halos that upscatter in the scaling relations. 
Even though these changes do not affect the value of $S_8 = 0.866 \pm 0.012$, and the $\Omega_\mathrm{m}$ and $\sigma_8$ change by about 1$\sigma$ but still remain fully consistent with \planck CMB \citep{Planck2020}, we remain very cautious regarding the physical interpretation of the corresponding best fitting parameters in this model.
In fact, despite the several improvements over the consistency with literature results on the value of the scatter at the pivot mass, now $0.66\pm0.04$, we have several indications that this model physically fails in determining some parameters:
\begin{itemize}
 \item The slope of the X-ray count rate scaling relation becomes significantly higher, $2.0\pm0.04$, much higher than previously reported in the literature for X-ray luminosity to mass scaling relation, see Sect.~\ref{sec:xrayscaling_relations}.
 \item The mass trend of the scatter is very strong, $-0.37\pm0.02$, with suspiciously high precision, quite higher than what is expected from numerical simulations \citep[$\sim$0.1][]{Pop2022}, and with significant implication for high-mass clusters ($>10^{15}M_\odot$) having below 10\% scatter in their luminosity to mass scaling relation, much lower than what was previously reported in dedicated studies of high-mass clusters, e.g. 40\% in WtG \citep{Mantz2016}.
 \item Mixture model parameters $f_{AGN}$ and $f_{RS}$ are forced to 0 value. This would imply that our cleaning procedure has reduced the fraction of contaminants to nearly zero. However, we know from our simulations \citep{Comparat2020, Seppi2022} and optical confirmation \citep{Kluge2024} that a remaining contamination fraction of the order of 5\% should be left in the cosmology sample.
\end{itemize}

It is thus clear that despite the intriguing results of this run, further dedicated studies are needed to physically understand what is causing the observed parameters, currently not consistent with our best understanding of the e\erosita sample.
Further multi-wavelength studies of the eROSITA cluster population are clearly necessary; for instance, if these populations are indeed AGN, they should have characteristic infrared emission \citep{Salvato2018, Salvato2022}. 
In our upcoming scaling relation paper (Pacaud et al. in preparation), we will investigate several aspects of the population modeling of galaxy clusters through scaling relation between mass and observables, with particular attention to the aspects discussed in this appendix. In our future \erosita cosmological analysis, we will use these results to update our population modeling.

\end{appendix}

\end{document}